%% file: zd2lg.tex
\documentclass[a4paper,11pt]{article}
\usepackage{a4wide}
\usepackage{amsmath,amssymb}
\usepackage{graphics,graphicx}
\usepackage{changebar}
\usepackage{color}
\usepackage{cancel}
\usepackage[all,cmtip]{xy}

\input{macrosES}

\numberwithin{equation}{section}

\begin{document}
\thispagestyle{empty}
\begin{flushright}
\end{flushright}
\begin{center}
{\LARGE\bf D-brane central charge \\ [1mm] and Landau-Ginzburg orbifolds}
\end{center}
\vspace{8mm}
\begin{center}
  {\large Johanna Knapp\footnote{{\tt
        johanna.knapp@unimelb.edu.au}}${}^{\dagger}$, Mauricio Romo\footnote{{\tt mromoj@tsinghua.edu.cn}}${}^{\ddag}$
    and Emanuel Scheidegger\footnote{{\tt esche@bicmr.pku.edu.cn}}${}^{\ast}$}
\end{center}
\vspace{3mm}
\begin{center}
${}^{\dagger}$ {\em School of Mathematics and Statistics, University of Melbourne \\\vskip 0.1 cm
  Parkville 3010 VIC, Australia}\\\vskip 0.1 cm 
$^\ddag${\em Yau Mathematical Sciences Center, Tsinghua University, Beijing, 100084, China}\\\vskip 0.1 cm 
$^\ast$ {\em Beijing International Center for Mathematical Research
(BICMR) \\ \vskip 0.1 cm
Peking University, 100871 Beijing, China}
\end{center}
\vspace{15mm}
\begin{abstract}
\noindent We propose a formula for the exact central charge of a B-type D-brane that is expected to hold in all regions of the K\"ahler moduli space of a Calabi-Yau. For Landau-Ginzburg orbifolds we propose explicit expressions for the mathematical objects that enter into the central charge formula. We show that our results are consistent with results in FJRW theory and the hemisphere partition function of the gauged linear sigma model. 
\end{abstract}
\newpage
\setcounter{tocdepth}{1}
\tableofcontents
\setcounter{footnote}{0}

\input{section1}
\input{section2}
\input{section3}
\input{section4}
\input{section5}
\input{section6}
\input{section7}
\appendix
\input{appendix}
\input{PALP}

\bibliographystyle{utphys}
\bibliography{zd2lg}
\end{document}

%% file: macrosES.tex
\usepackage{latexsym}
\usepackage{amsmath,amssymb,amsxtra,amscd,amsthm}
\usepackage{mathrsfs}
\newcommand{\scr}{\mathscr}
\makeatletter%
\@ifclassloaded{amsart}%
  {}%
  {\usepackage[nottoc,notlot,notlof]{tocbibind}}%
\makeatother%
\usepackage{epsfig}
\usepackage{pdfsync}
\usepackage{thumbpdf}
\usepackage{color}
\ifx\pdfoutput\undefined
\usepackage
[dvips,
 verbose=false,
 bookmarks=true,
 colorlinks=true,
 linkcolor=webred,
 filecolor=webbrown,
 citecolor=webgreen,
 pagecolor=webblue,
 urlcolor=webblue,
 pdftitle={},
 pdfauthor={Emanuel Scheidegger},
 pdfsubject={},
 pdfkeywords={},
 bookmarksopen=false,
 pdfpagemode=None,
 pdfview=FitH,
 pdfstartview=FitH,
 extension=pdf]{hyperref}
\else
\usepackage
[pdftex,
 verbose=false,
 bookmarks=true,
 colorlinks=true,
 linkcolor=webred,
 filecolor=webbrown,
 citecolor=webgreen,
 pagecolor=webblue,
 urlcolor=webblue,
 pdftitle={},
 pdfauthor={Emanuel Scheidegger},
 pdfsubject={},
 pdfkeywords={},
 bookmarksopen=false,
 pdfpagemode=None,
 pdfview=FitH,
 pdfstartview=FitH,
 extension=pdf]{hyperref}
\fi
\definecolor{webred}{rgb}{.8,0,0}
\definecolor{webbrown}{rgb}{.6,0,0}
\definecolor{webgreen}{rgb}{0,0.5,0}
\definecolor{webdkgreen}{rgb}{0,0.3,0}
\definecolor{webblue}{rgb}{0,0,0.5}

\usepackage{graphics,graphicx}
\usepackage{a4wide}

\def\cA{\mathcal{A}}
\def\cB{\mathcal{B}}
\def\cC{\mathcal{C}}
\def\cD{\mathcal{D}}

\def\cF{\mathcal{F}}

\def\cH{\mathcal{H}}

\def\cL{\mathcal{L}}
\def\cM{\mathcal{M}}
\def\cN{\mathcal{N}}
\def\cO{\mathcal{O}}

\def\cV{\mathcal{V}}
\def\cW{\mathcal{W}}

\def\rC{\scr{C}}

\def\rE{\scr{E}}

\def\rL{\scr{L}}

\def\fB{\mathfrak{B}}

\def\fc{\mathfrak{c}}

\def\mC{\mathbb{C}}

\def\mL{\mathbb{L}}

\def\mN{\mathbb{N}}

\def\mP{\mathbb{P}}
\def\mQ{\mathbb{Q}}
\def\mR{\mathbb{R}}

\def\mZ{\mathbb{Z}}

\def\tG{\mathrm{G}}
\def\tH{\mathrm{H}}

\def\tL{\mathrm{L}}

\def\tR{\mathrm{R}}
\def\tS{\mathrm{S}}

\def\tU{\mathrm{U}}

\def\pC{\mathsf{C}}

\def\pG{\mathsf{G}}

\def\pJ{\mathsf{J}}

\def\pM{\mathsf{M}}

\def\pQ{\mathsf{Q}}
\def\pR{\mathsf{R}}
\def\pS{\mathsf{S}}
\def\pT{\mathsf{T}}

\def\pV{\mathsf{V}}
\def\pW{\mathsf{W}}

\def\pr{\mathsf{r}}

\def\pt{\mathsf{t}}

\DeclareMathOperator{\chern}{ch}
\DeclareMathOperator{\ch}{c}
\DeclareMathOperator{\todd}{td}

\DeclareMathOperator{\rank}{rk}

\DeclareMathOperator{\str}{str}

\DeclareMathOperator{\im}{im}

\DeclareMathOperator{\ext}{ext}

\DeclareMathOperator{\Res}{Res}
\DeclareMathOperator{\End}{End}

\DeclareMathOperator{\Aut}{Aut}

\DeclareMathOperator{\eq}{=}

\renewcommand{\Re}{\mathop{\mathrm{Re}}}
\renewcommand{\mod}{\mathop{\mathrm{mod}}}

\def\bb1{\textup{\small{1}} \kern-3.8pt \textup{1}}
\newcommand{\diff}[2]{\textrm{d}^{#1}{#2}}
\newcommand{\e}[1]{\mathrm{e}^{#1}}

\def\id{\textrm{id}}

\def\diag{\textrm{diag}}

\def\conv{\mathrm{conv}}

\def\vir{\mathrm{vir}}

\newcommand{\pd}[2]{\frac{\partial #1}{\partial #2}}

\newcommand{\jbar}{\bar{\jmath}}

\newcommand{\gammabar}{\overline{\gamma}}

\newcommand{\shortexactseq}[5]{0 \longrightarrow #1
\stackrel{#2}{\longrightarrow} #3 \stackrel{#4}{\longrightarrow} #5
\longrightarrow 0}
\newcommand{\Mgn}[2]{\overline{\cM}_{ #1 , #2 } }

\def\SL2Z{\tS\tL(2,\mZ)}
\def\PSL2Z{\mP\tS\tL(2,\mZ)}

\DeclareMathOperator{\age}{age}

\def\FJRW{\mathrm{FJRW}} 

\def\inv{\mathrm{inv}}

\def\Top{\mathrm{top}}

\numberwithin{equation}{section}

\providecommand{\href}[2]{#2}

\allowdisplaybreaks

%% file: section1.tex
\section{Introduction and summary}
In the context of $\mathcal{N}=(2,2)$ superconformal field theories with boundary, the central charge $Z(B)$ of a topological B/A-type D-brane ${B}$ is defined as the correlator obtained by inserting the boundary state associated to ${B}$ on the bounding circle of an infinitely long cigar that has a Ramond-Ramond (RR-) ground state corresponding to an element of the $(a,c)/(c,c)$ ring inserted at the tip. In this work the main focus will be on the D-brane central charge of B-type D-branes in families of conformal field theories with central charge $\hat{c}\in \mathbb{Z}_{\geq 0}$. This is related to string compactifications on Calabi-Yau (CY) spaces of complex dimension $\hat{c}$.  

In practice, the central charge of a B-type D-brane $\mathcal{B}$ in a CY is difficult to compute, as it receives instanton corrections depending on the K\"ahler moduli. One difficulty is that the stringy K\"ahler moduli space $\mathcal{M}_K$ is divided into chambers, and only some of them have limiting regions that allow for a description of the CY compactification in terms of geometry. In such large volume regions the central charge of a B-brane $\mathcal{B}$ on a Calabi-Yau $X$ has the form \cite{Hosono:2000eb}
\begin{equation}
  Z^{LV}(\mathcal{B})\propto\int_X\widehat{\Gamma}_Xe^{B+\frac{1}{2\pi}\omega}\mathrm{ch}(\mathcal{B}_{LV})+\mathcal{O}(e^{-t}),
  \end{equation}
where $B$ is the $B$-field, $\omega$ is the K\"ahler class, $\mathrm{ch}(\mathcal{B}_{LV})$ is the Chern character of the brane at large volume, and $\widehat{\Gamma}_X$ is the Gamma class \cite{Iritani:2009ab,Katzarkov:2008hs}. The subleading terms are instanton corrections depending on the K\"ahler moduli $t$. The exact expression can be computed using mirror symmetry \cite{Candelas:1990rm,Brunner:1999jq}, where it can be expressed in terms of periods of the mirror CY, or directly in $X$, using supersymmetric localization \cite{Sugishita:2013jca,Honda:2013uca,Hori:2013ika}.

The main goal of this work is to propose a definition of the exact central charge that also holds in non-geometric regimes of $\mathcal{M}_K$. We argue that it has a universal form:
\begin{equation}
  \label{formula}
 Z({\mathcal{B}})=\langle \widehat{\Gamma}^{*} \circ\mathbf{J}|\mathrm{ch}({\mathcal{B}})\rangle,
  \end{equation}
where the definition of the quantities entering this formula depends on the concrete realization of the conformal field theory associated to the locus of $\mathcal{M}_K$ under consideration. The intuition for such an expression to exist comes from various directions. One of them is of course the worldsheet conformal field theory itself for which it does not play a role whether the target space has a geometric description or not. The worldsheet point of view in relation to (\ref{formula}) will be discussed in Section \ref{sec:cc-scft}. 
Another motivation are the results from supersymmetric localization \cite{Sugishita:2013jca,Honda:2013uca,Hori:2013ika} which showed that the hemisphere partition of the gauged linear sigma model (GLSM) \cite{Witten:1993yc} computes $Z(\mathcal{B})$. Since the GLSM provides a common UV description of the CFTs parametrized by $\mathcal{M}_K$, this suggests a universal structure in the conformal field theories and their observables in the IR. 

To test (\ref{formula}) we will show that it works for the case of Landau-Ginzburg orbifolds \cite{Vafa:1989xc,Intriligator:1990ua}. These are among the best-understood non-geometric descriptions of string compactifications, and the quantities entering (\ref{formula}) have been defined in the context of FJRW (Fan-Jarvis-Ruan-Witten) theory \cite{Witten:1991mk,Fan:2007ba,Chiodo:2008hv}. On the other hand, Landau-Ginzburg orbifolds arise as phases, i.e. low-energy descriptions, of GLSMs. Therefore the hemisphere partition function provides a means to test the central charge formula.

Let us summarize the main results of the article. We consider Landau-Ginzburg orbifolds specified by $(W,G,\overline{\rho}_m,\mathbb{C}_L^{\ast})$, where $W$ is the superpotential, $G$ is the orbifold group which we restrict to be abelian, $\overline{\rho}_m: G \to \tG\tL(\mC^N)$ is the matter representation, and $\mathbb{C}_L^{\ast}$ is the left R-symmetry. We denote the corresponding R-charges by $q_1,\dots,q_N$. For most of the discussion we will further assume that $G$ is admissible, which means that
\begin{equation}
\langle J\rangle\subseteq G\subseteq\mathrm{Aut}(W),\label{sec1admissible}
\end{equation}
where $J$ is the group element $\mathrm{diag}(e^{2\pi i q_{1}},\ldots,e^{2\pi i q_{N}})$. 
The condition (\ref{sec1admissible}) guarantees that the R-charges of the physical states are integral and that the theory allows for a topological A-twist. Even though some of our results are more general, we further assume that the Landau-Ginzburg orbifolds we consider arise as limiting points in the K\"ahler moduli space $\cM_K$ of a Calabi-Yau. 

We propose a general formula for the $J$-function \cite{Givental:2004ab} $\mathbf{J}$ of abelian Landau-Ginzburg orbifolds:
\begin{equation}
 \mathbf{J}_{LG}(t)=\frac{I_{LG}(u(t))}{I_{0}(u(t))},
\end{equation}
where
\begin{equation}
  I_{LG}(u) = \sum_{\gamma \in G} I_{\gamma}(u) \mathbf{e}^{(a,c)}_{\gamma },
\end{equation}
is the $I$-function. Here, $\mathbf{e}^{(a,c)}_{\gamma }$ denotes the basis elements of the narrow part $(a,c)$ chiral ring. ``Narrow'', as opposed to ``broad'', refers to those sectors of the $(a,c)$-ring that only consist of the identity element, i.e. the vacuum. Furthermore, $I_{0}$ is the component of $I_{LG}$ corresponding to $\gamma=\mathrm{id}$, and $u(t)$ denotes a change from formal variables $u$ to flat coordinates $t$ in the Landau-Ginzburg region of $\mathcal{M}_K$. Our proposal for the $I$-function for a Landau-Ginzburg orbifold with $G\cong\prod_{a=1}^h\mathbb{Z}_{d_a}$ and $h=\mathrm{dim}\mathcal{H}_{\mathrm{mar},0}$ parameters $u=(u_1,\dots,u_h)$ corresponding to the narrow marginal deformation sectors $\mathcal{H}_{\mathrm{mar},0}$ in the $(a,c)$-ring $\mathcal{H}^{(a,c)}$  is
\begin{equation}
  I_\gamma(u) =
  -\sum_{\substack{k_1,\dots,k_h \geq 0\\k'_a =
      \ell_a \mod d_a}} 
    \frac{ u^k}{\prod_{a=1}^h\Gamma(k_a+1)}
    \prod_{j=1}^N\frac{(-1)^{\langle-\sum_{a=1}^h
      k_a q_{a,h+j}+q_{j}\rangle}\Gamma(\langle\sum_{a=1}^h
      k_a q_{a,h+j}-q_{j}\rangle)}{\Gamma(1+\sum_{a=1}^h
      k_a q_{a,h+j}-q_{j})}
\end{equation}
with $\ell=(\ell_1,\dots,\ell_h)$ determined by $\gamma^{-1}J=\prod_{a=1}^hg_a^{\ell_a}$ where $g_a$ is the generator of $\mathbb{Z}_{d_a}$. 
The central piece of data is a matrix $q$ with rational entries that encodes the action of $G$ on $\mathcal{H}_{\mathrm{mar},0}$. Furthermore, $\langle x\rangle=x-\lfloor x\rfloor$. We show that $I_{LG}(u)$ satisfies a system of GKZ differential equations. 

For the Gamma class $\widehat{\Gamma}^{\ast}$ entering in (\ref{formula}) we propose the following definition for Landau-Ginzburg orbifolds in terms of the matrix $q$:
\begin{equation}
  \widehat \Gamma^*_{W,G}\mathbf{e}_{\gamma}^{(a,c)}=\widehat{\Gamma}_{\gamma}\mathbf{e}_{\gamma}^{(a,c)},\qquad \widehat{\Gamma}_{\gamma}=\prod_{j=1}^N\Gamma\left(1-\left\langle\sum_{a=1}^h
      k_a q_{a,h+j} - q_j\right\rangle \right).
\end{equation}

The information about the D-brane enters into (\ref{formula}) via the Chern character $\mathrm{ch}(\mathcal{B})\in\mathcal{H}^{(c,c)}$. B-type D-branes in Landau-Ginzburg orbifolds are matrix factorizations of $W$ \cite{Kapustin:2002bi,Brunner:2003dc}. For Landau-Ginzburg orbifolds the Chern character of a B-brane $\overline{\mathcal{B}}$ has been defined in \cite{Walcher:2004tx}.
Then (\ref{formula}) has the explicit realization in the Landau-Ginzburg orbifold setting as
\begin{equation}
  \label{eq:14}
     Z_{LG}(\overline{\cB},u) =\left\langle \textstyle\sum\nolimits_{\gamma}\widehat{\Gamma}_{\gamma}I_{\gamma}\mathbf{e}_{\gamma}^{(a,c)}\right.\left|\textstyle\sum\nolimits_{\gamma'}\mathrm{ch}(\overline{\cB})_{\gamma'}\mathbf{e}_{\gamma'}^{(c,c)}\right\rangle,
\end{equation}
where the pairing $\langle \mathbf{e}^{(a,c)}_\gamma| \mathbf{e}^{(c,c)}_{\gamma'}\rangle$ is related to the topological pairing $\langle \mathbf{e}^{(a,c)}_\gamma,\mathbf{e}^{(a,c)}_{\gamma'}\rangle$ through $\langle \mathbf{e}_\gamma^{(a,c)} |\mathbf{e}_{\gamma'}^{(c,c)} \rangle=\langle \mathbf{e}_\gamma^{(a,c)}, \mathcal{U}\circ\mathbf{e}_{\gamma'}^{(c,c)} \rangle$ where $\mathcal{U}$ is the spectral flow operator. All details can be found in Section~\ref{sec:cc-lgo}. 

The remaining sections are dedicated to testing $Z_{LG}(\overline{\cB},u)$ by various approaches. One important reference point is FJRW theory where explicit expressions for the $I$-function and the Gamma class have been given in~\cite{Chiodo:2008hv,Chiodo:2012qt}, see also~\cite{Guere:2016ab,Iritani:2016nkj}. Extending these results to orbifold groups other than $G=\langle J\rangle$ and $G=\mathrm{Aut}(W)$, and to several marginal deformations, we compute the $I$-function in FJRW theory. Our result is 
\begin{equation}
  I_{W,G}( u,z) = \sum_{\gamma' \in G} \lim_{\lambda\to0}I_{T,\gamma'}(u,z;\lambda)  e_{\gamma'}
\end{equation}
with
\begin{equation}
    \begin{aligned}
    I_{T,\gamma'}( u,z;\lambda) &= z^{1-\frac{1}{2}\deg e_{\gamma'}} 
    \sum_{\substack {\{k_\gamma \geq 0 \mid \gamma \in G^{(2)}\}\\\prod_{\gamma \in G^{(2)}} \gamma^{k_\gamma} = \gamma' }} \prod_{\gamma\in G^{(2)}}
    \frac{\left(u^{\gamma}\right)^{k_\gamma}}{k_\gamma!}  \prod_{j=1}^N \frac{\Gamma(\frac{\lambda_j}{z}-\langle \sum_{\gamma \in G^{(2)}} \theta^{\gamma}_j k_\gamma + q_j\rangle +1)}{\Gamma(\frac{\lambda_j}{z}-\sum_{\gamma \in G^{(2)}} \theta^{\gamma}_j 
      k_\gamma - q_j +1)}. \\
  \end{aligned}  
\end{equation}
Here, $e_{\gamma'}$ is the basis of the FJRW state space $\mathcal{H}_{\text{FJRW}}$ which is isomorphic to the $(a,c)$-ring. The set $G^{(2)} = \{ \gamma \in G \mid \sum_{i=1}^N\theta_i^{\gamma} = 2\}$  labels those narrow twisted sectors of $\cH_{\FJRW} = \bigoplus_{\gamma \in G} \cH_{\FJRW,\gamma}$ that correspond to the marginal deformations of the $(a,c)$-ring. Moreover, the $\theta_j^{\gamma}$, $j=1,\dots,N$, are the phases of $\gamma \in G$, and $z$ is a parameter that is arbitrary when one considers the Calabi-Yau case. We can show for a large class of Landau-Ginzburg orbifolds that
\begin{equation}
  I_{W,G}(u,-1)=I_{LG}(u).
  \end{equation}
We also propose a more general definition of the Gamma class in FJRW theory as
\begin{equation}
  \widehat{\Gamma}_{W,G}^* = \bigoplus_{\gamma \in G} \prod_{j \in I_\gamma}
  \Gamma(1- \theta_j^{\gamma^{-1}} ) \id_{\cH_{\FJRW,\gamma}}.
\end{equation}
where $I_\gamma \subset \{1,\dots,N\}$ encodes the information on the narrow sectors. This also is shown to coincide with the Landau-Ginzburg definition. The pairing on $\mathcal{H}_{\text{FJRW}}$ is, up to change of basis, the same as for the $(a,c)$-ring, the definition of the Chern character of a B-brane is also the same as in the Landau-Ginzburg case. Given all this information we can show that inserting the FJRW quantities into (\ref{formula}) yields the same result as the Landau-Ginzburg orbifold case. Details on our results in the context of FJRW theory, together with an overview of the FJRW formalism can be found in Section \ref{sec:fjrw}.

Section \ref{sec:hpf} is dedicated to the GLSM and the hemisphere partition function. The explicit form of the hemisphere partition function $Z_{D^2}(\mathcal{B},\pt')$ for an abelian GLSM with gauge group $\pG=U(1)^h$, a charge matrix $\pC=(L\ S)$ and a B-brane $\cB$ is~\cite{Hori:2013ika}:
\begin{equation}
  \begin{aligned}
  Z_{D^2}(\mathcal{B},\pt') & = \frac{1}{(2\pi)^h}\int_{\gamma}d^h\sigma\prod_{a=1}^h\Gamma\left(i\sum_{\alpha=1}^hL_{\alpha a}\sigma_{\alpha}\right)\prod_{j=1}^{N}\Gamma\left(i\sum_{\alpha=1}^hS_{\alpha j}\sigma_\alpha+{q_{j}}\right) \\
  &\phantom{=}  \cdot \e{i\sum_{a,\alpha}\sigma_\alpha L_{\alpha a}\pt'_a}\sum_{\mu=1}^{\dim \pM}e^{i\pi r^{\mu}}e^{2\pi\sum_\alpha w_{\alpha}^{\mu}\sigma_\alpha},
\end{aligned}
\end{equation}
where $\pt' \in \cM_K$ is the FI-theta parameter of the GLSM parametrizing the K\"ahler moduli space $\cM_K$. The last factor contains the information about the brane $\cB$, where $\pM$ is its Chan-Paton space and $r^{\mu}$ and $w^{\mu}_{\alpha}$ are its R- and gauge charges, respectively.

We show that the hemisphere partition function, evaluated in a Landau-Ginzburg phase, coincides with the Landau-Ginzburg central charge:
\begin{equation}
  Z_{D^2}^{LG}(\mathcal{B},\pt')=Z_{LG}(\overline{\mathcal{B}},e^{-\pt'}),
\end{equation}
where $\overline{\mathcal{B}}$ is the brane in the Landau-Ginzburg orbifold phase corresponding to $\cB$ via \cite{Herbst:2008jq}. Note that the matrix $q$ we have introduced in the Landau-Ginzburg orbifold is related to the charge matrix $\pC$ of the GLSM by
\begin{equation}
  q=L^{-1}\pC.
\end{equation}
The $h\times h$ matrix $L$ is the matrix of gauge charges of those $h$ chiral matter fields in the GLSM that get a VEV in the Landau-Ginzburg orbifold phase.

In Section \ref{sec:apps} we work out these various approaches to the D-brane central charge in several examples. The first example is the inevitable quintic, where most results can be found in the literature. For the quintic we also show that (\ref{formula}) also holds in the geometric phase of $\mathcal{M}_K$. Then we move on to two-parameter models. In one of the examples we also compute the FJRW invariants. To our knowledge, this is the first time such invariants have been computed in a multi-parameter model. Furthermore, we consider an example where the Landau-Ginzburg potential is not a Fermat polynomial. 
Finally, we discuss a four-parameter model which has broad sectors -- a case where our methods to compute the central charge do not apply.  We propose a way around this issue by introducing an alternative formulation where all the moduli are realized in terms of narrow sectors. This is the Landau-Ginzburg equivalent of a way to deal with non-torically realized moduli in geometric settings \cite{Berglund:1995gd}, and seems to apply for a well-defined class of examples with broad sectors. \\\\
{\bf Acknowledgements:} We thank Huai-Liang Chang, David Erkinger, Huijun Fan, Kentaro Hori, Hans Jockers, Maximilian Kreuzer, Wolfgang Lerche, Greg Moore, Daniel Pomerleano, Thorsten Schimannek, Yefeng Shen, Dmytro Shklyarov, and Harald Skarke for discussions and comments. JK thanks BICMR, CERN, and YMSC for hospitality. MR thanks Fields Institute, Caltech and Universidad Andres Bello for hospitality. ES thanks CERN and the University of Melbourne for hospitality. JK was partially supported by the Austrian Science Fund (FWF): [P30904-N27]. ES acknowledges support from NSFC grant No. 11431001. MR acknowledges support from the National Key Research and Development Program of China, grant No. 2020YFA0713000, and the Research Fund for International Young Scientists, NSFC grant No. 11950410500.


%% file: section2.tex
\section{D-brane central charges in $\cN=(2,2)$ SCFTs}
\label{sec:cc-scft}
In this section we will review the physics definition of the central charge of a topological D-brane. Then we will write it in terms of objects that are more familiar from a mathematical point of view. We will keep the point of view of an abstract SCFT during this section. Hence, some of the objects will not be rigorously defined for mathematical standards but when we work on specific SCFTs we will be able to relate them to geometric and/or categorical quantities.
\subsection{Worldsheet definition of D-brane central charges}
Following~\cite{Cecotti:1991me}, we start by considering an $\mathcal{N}=(2,2)$ SCFT of central charge $c$ that has two unbroken (left and right) R-symmetries which we will denote by $U(1)_{L}$ and $U(1)_{R}$. In such theories we have four supercharges $Q_{\pm}$ and $\overline{Q}_{\pm}$ and we can define four nilpotent operators, namely $Q_{A}:=\overline{Q}_{+}+Q_{-}$, $Q_{B}:=\overline{Q}_{-}+\overline{Q}_{+}$ and their corresponding conjugates $Q_{A}^{\dag}$ and $Q_{B}^{\dag}$. In flat space and with all the fermions having NS-NS boundary conditions these charges are globally defined. One can define four rings by taking the cohomology of these operators: 
\begin{eqnarray}
\mathcal{H}^{(c,c)}=H_{Q_{B}}\qquad \mathcal{H}^{(a,c)}=H_{Q_{A}}\qquad \mathcal{H}^{(c,a)}=H_{Q_{A}^{\dag}}\qquad \mathcal{H}^{(a,a)}=H_{Q_{B}^{\dag}}.
\end{eqnarray}
In the RR sector of the theory, one defines the space of Ramond vacua $\mathcal{H}_{R}$ defined by the states annihilated by $Q$ and $Q^{\dag}$ where $Q$ stands for either $Q_{A}$ or $Q_{B}$. The isomorphisms between these rings are implemented by the spectral flow operator $\mathcal{U}_{(r,\bar{r})}$, which has R-charges $(\hat{c} r,\hat{c} \bar{r})$ ($c=3\hat{c}$) \cite{Lerche:1989uy}. We summarize these isomorphisms in Table \ref{loctab}.
  \begin{table}[h]\def\arraystretch{1.5}
    \begin{center}
\begin{tabular}{|c|c|}
\hline
Locality condition & Isomorphism defined by $\mathcal{U}_{r,\bar{r}}$ \\\hline%
$q-\bar{q}\in \mathbb{Z}$& $\mathcal{U}_{\left(\frac{1}{2},\frac{1}{2}\right)}\circ \mathcal{H}_{R}\cong\mathcal{H}^{(c,c)}$\\\hline%
$q+\bar{q}\in \mathbb{Z}$& $\mathcal{U}_{\left(-\frac{1}{2},\frac{1}{2}\right)}\circ \mathcal{H}_{R}\cong\mathcal{H}^{(a,c)}$\\\hline%
$q\in \frac{1}{2}\mathbb{Z}$& $\mathcal{U}_{(1,0)}\circ \mathcal{H}^{(a,c)}\cong\mathcal{H}^{(c,c)}$\\
\hline
\end{tabular}\caption{Chiral rings and locality conditions.}\label{loctab}
\end{center}
  \end{table}
  From now on we will assume that the locality condition as indicated in the table is satisfied,  and hence $\mathcal{U}_{(r,\bar{r})}$ is well defined. In particular, we will assume that $\hat{c}\in\mathbb{Z}$ and that the charges of the physical operators are also integers.

\begin{figure}
  \begin{center}
    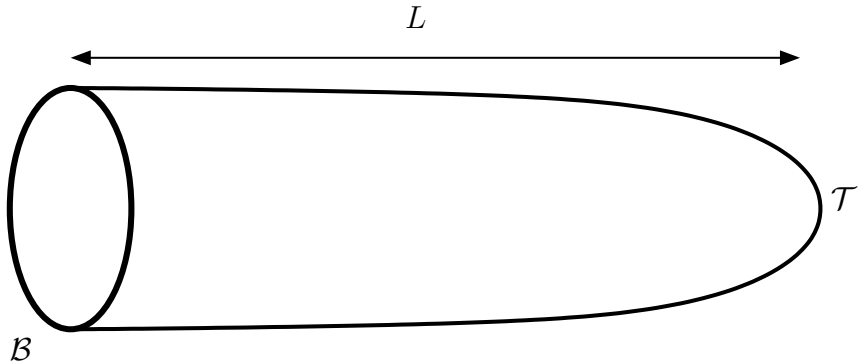\caption{D-brane central charge, where we attach a flat cylinder of length $L\rightarrow\infty$ to a hemisphere with a twist $\mathcal{T}=(\text{A},\text{B})$ and ${B}=(\text{B},\text{A})$-type boundary conditions.}\label{fig:cccv}
    \end{center}
  \end{figure}

  In order for the rings to be well defined when we put our theory on an arbitrary Riemann surface we need to perform a topological twist. We have again four options labelled as $A$, $\overline{A}$, $B$ and $\overline{B}$, depending on which supercharges become scalar\footnote{Our convention is that the A-twist corresponds to twisting by the axial R-charge $U(1)_A=U(1)_{R}-U(1)_{L}$ and the B-twist to the twist by the vector R-charge $U(1)_V=U(1)_{L}+U(1)_{R}$.}. Upon twisting, the fermions become periodic on contractible cycles and we have a natural map from operators $\phi\in\mathcal{H}^{(*,*)}$ to Ramond ground states $|\phi \rangle_{R}\in \mathcal{H}_{R}$ \cite{Cecotti:1991me}. In physical terms, this map can be described as follows: first, by the operator-state correspondence, one defines the state $|\phi\rangle$. This is a state in the NS sector. Then, one performs an appropriate topological twist, depending on whether $\phi\in \mathcal{H}^{(c,c)}$ or $\phi\in \mathcal{H}^{(a,c)}$. In the former case this is equivalent to inserting the operator $\mathcal{U}_{(-\frac{1}{2},-\frac{1}{2})}$ and in the latter corresponds to inserting $\mathcal{U}_{(\frac{1}{2},-\frac{1}{2})}$. This brings $|\phi\rangle$ to the Ramond sector. The last step is to project onto a ground state by attaching an infinitely long cylinder. More precisely, we have
\begin{eqnarray}
|\phi \rangle_{R}=\lim_{L\rightarrow\infty}e^{-LH}\mathcal{U}\circ|\phi\rangle,
\end{eqnarray}
where $L$ is the coordinate along the cylinder, $H$ is the Hamiltonian and $\mathcal{U}$ denotes the twist implemented by the spectral flow operator, as described above. This projection operation can be regarded as choosing a harmonic representative for $\phi \in H_{Q}$ \cite{Lerche:1989uy}. In order to obtain a wave function from the vector $|\phi \rangle_{R}$ we need to fix a boundary condition. We have two options, namely we can preserve either $Q_B$ (and $Q_B^\dagger$) or $Q_A$ (and $Q_A^\dagger$). The former corresponds to a B-brane and the latter to an A-brane. Since we have attached an infinitely long flat cylinder, we can put in principle any boundary condition, i.e.~either A- or B-branes. In order to obtain D-brane central charges, which are the main subject of this article, we should take boundary conditions preserving the opposite set of supercharges compared to the ones corresponding to the cohomology we insert at the tip \cite{Cecotti:1991me,Hori:2000ck}. This is an A-brane for the case of $\phi \in \mathcal{H}^{(c,c)}$ and a B-brane for the case of $\phi\in\mathcal{H}^{(a,c)}$. This coupling of A/B-branes with $(c,c)/(a,c)$-operators is actually very natural\footnote{In a nutshell, the boundary state describing the boundary CFT with A/B-boundary conditions will be a state in the $(c,c)/(a,c)$ Hilbert space, respectively. Hence A/B-boundary conditions will naturally couple to $(c,c)/(a,c)$ local operators.} \cite{Ooguri:1996ck}. We illustrate this in Figure \ref{fig:cccv}.
Therefore the central charge of an A- or B-brane ${B}$ is defined by
\begin{eqnarray}
Z({B}):=\langle {B}|\mathbf{1}_{(*,*)}\rangle,
\end{eqnarray}
where $\mathbf{1}_{(*,*)}$ is the corresponding identity operator on the $(c,c)$ or $(a,c)$ ring. If we consider specifically central charges of B-branes (as opposed to A-branes), it is known that they are subject to quantum corrections, and their integrality properties are highly nontrivial \cite{Hosono:2000eb}. If $\mathcal{M}$ is the moduli space of complex structures of a Calabi-Yau, the corresponding central charge $Z(B)$ of an A-brane is given by a period of the top holomorphic form \cite{Ooguri:1996ck}, and thus is a purely classical expression. On the other hand, if $\mathcal{M}$ is the stringy K\"ahler moduli space $\mathcal{M}_K$ of a Calabi-Yau, which decomposes into chambers, not all of which allow for a geometric description, a definition of $Z(B)$ in purely geometric terms (such as a period) will not suffice as a general expression for the D-brane central charge. 
Our main goal in this section is to propose an expression for $Z(B)$ for B-branes (even though it can be applied to A-branes as well) in terms of objects that can be defined at arbitrary loci of $\mathcal{M}_K$.

\subsection{Vacuum bundle and $J$-function}
\label{sec:vacuum-bundle}

The chiral rings admit a bi-grading given by the left and right R-charges:
\begin{eqnarray}
\mathcal{H}^{(*,*)}=\bigoplus_{q,\bar{q}}\mathcal{H}_{q,\bar{q}}^{(*,*)},
\end{eqnarray}
where $(*,*)$ stands for $(c,c)$, etc. The range of the sum $(q,\bar{q})$ depends on the ring. For the case of the $(c,c)$ ring, one has $0\leq q,\bar{q}\leq \hat{c}$, and $0\leq \bar{q}\leq \hat{c},0\leq -q\leq \hat{c}$\footnote{This bound is a consequence of the $\mathcal{N}=2$ SCFT algebra.} for the case of the $(a,c)$ ring. These rings are isomorphic as vector spaces for those theories for which the spectral flow operators are local. Each of these rings has a special
subring called the deformation subring \cite{Alim:2012gq}. This subring is finitely generated by the elements of conformal weight $(h,\bar{h})=(\frac{1}{2},\frac{1}{2})$. This corresponds to operators of charges $(1,1)$ for the case of the $(c,c)$ ring and $(-1,1)$ for the case of the $(a,c)$ ring \cite{Lerche:1989uy}. We collectively denote the corresponding subspaces $\mathcal{H}^{(c,c)}_{1,1}$ and $\mathcal{H}^{(a,c)}_{-1,1}$ by $\mathcal{H}_{\mathrm{Mar}}$ ($\mathrm{Mar}$ for ``marginal'').  
Then the deformation ring is defined by
\begin{eqnarray}
  \label{hdef}
\mathcal{H}_{\mathrm{def}}:=\langle\mathcal{H}_{\mathrm{Mar}}\rangle\qquad \mathrm{dim}\mathcal{H}_{\mathrm{Mar}}=h.
\end{eqnarray}
Because of $\mathcal{H}_{\mathrm{def}}$ being generated by operators satisfying $|q|=\bar{q}$, we denote the grading as
\begin{eqnarray}
\mathcal{H}_{\mathrm{def}}=\bigoplus_{\bar{q}}\mathcal{H}_{\mathrm{def}}^{\bar{q}}.
\end{eqnarray}
Each generator of $\mathcal{H}_{\mathrm{def}}$ can be mapped to an exactly marginal deformation of the SCFT\footnote{We remark that the dimension of $\mathcal{H}_{\mathrm{Mar}}$ (and $\mathcal{H}_{\mathrm{def}}$) is not necessarily the same for each topological ring. Usually it is not.}. If we consider $\phi_{i}\in \mathcal{H}_{\mathrm{Mar}}$, then we can write a marginal deformation
\begin{eqnarray}\label{flatcordd}
t^{i}\int_{\Sigma}\mathcal{O}^{(1,1)}_{i}\qquad i=1,\ldots,h,
\end{eqnarray}
where the operator $\mathcal{O}^{(1,1)}_{i}$  has weights $(1,1)$ and is constructed from $\phi_{i}$ as $\{\overline{Q}_{-},[Q_{+},\phi_i^{(a,c)}]\}$ or $\{Q_{-},[Q_{+},\phi_i^{(c,c)}]\}$ \cite{Cecotti:1991me}. The deformation parameters $t^{i}$, $i=1,\ldots,h$ are coordinates in a $h$-dimensional complex moduli space $\mathcal{M}$ of marginal deformations whose fibers correspond to SCFTs. The moduli space, in general, takes the form $\mathcal{M}\cong (\mathbb{C}^{*})^{h}\setminus \Delta$ where $\Delta$ is a divisor determined by the values of $t$ where the resulting theory is not well defined.\\

As shown in \cite{Cecotti:1991me} we can form a holomorphic bundle over $\mathcal{M}$ given by $\mathcal{V}=\mathcal{H}_{\mathrm{def}}\otimes \mathcal{O}_{\mathcal{M}}$, the vacuum bundle. It comes equipped with a flat connection $\nabla$ (in fact, a $\mathbb{P}^{1}$-family of connections). There is a special choice of coordinates on $\mathcal{M}$ called flat coordinates, corresponding to the $t_{i}$ \cite{Dijkgraaf:1990dj}. The ring structure of $\mathcal{H}_{\mathrm{def}}$ is given by the OPEs,
\begin{eqnarray}
\phi_{a}\cdot \phi_{b}=C_{ab}^{\ \ c }\phi_{c},
\end{eqnarray}
where the structure constants are given in terms of genus $0$ correlators:
\begin{eqnarray}\label{topm}
\qquad C_{ab}^{\ \ c }=\langle \phi_{a} \phi_{b}\phi_{d}\rangle _{S^{2}}\eta^{dc}, \qquad \eta_{ab}= \langle \phi_{a} \phi_{b}\rangle _{S^{2}}.
\end{eqnarray}
Here $\langle \ldots \rangle _{S^{2}}$ stands for the topological correlator on $S^{2}$, i.e.~the A/B-twisted correlator on $S^{2}$, for the case of the $(a,c)/(c,c)$ ring. We remark that these correlators depend holomorphically on the coordinates of $\mathcal{M}$, because we are working with a basis of states spanned by chiral operators. Such a basis is equivalent to a holomorphic basis of the $tt^{*}$-connection \cite{Cecotti:1991me}. We will work in the flat coordinates (\ref{flatcordd}). They are characterized by the fact that, in the path integral formalism, the derivative with respect to $t_{i}$ produces an insertion of the operator $\int_{\Sigma}\mathcal{O}^{(1,1)}_{i}$ \cite{Dijkgraaf:1990dj}. In such coordinates the topological metric $\eta_{ab}$ is constant, i.e.~$\partial_{a}\eta_{bc}=0$. Moreover, let us remark that there exists a hermitian metric $g_{a\bar{b}}$ on $\mathcal{M}$, usually referred to as the $tt^{*}$-metric. This metric is defined by joining two hemispheres by an infinitely long cylinder, with a $\overline{\mathcal{T}}$-twist on one hemisphere and the conjugate $\mathcal{T}$-twist  on the other one:
\begin{eqnarray}
\langle \phi_{a}| \Theta \phi_{b}\rangle = g_{a\bar{b}},
\end{eqnarray}
where $\Theta$ is a CPT conjugation operator. The flatness of the connection associated to $g_{a\bar{b}}$ is equivalent to the flatness of $\eta_{ab}$ by virtue of the $tt^{*}$-equation \cite{Dubrovin:1992yd}. Hence, when working with flat coordinates, we can set the connection to zero: $g^{-1}\partial g=0$. Using the state-operator correspondence, we chose the following frame for the bundle $\mathcal{V}$:
\begin{eqnarray}
\mathbf{e}_{a}=|\phi_{a}\rangle\qquad \mathbf{e}_{0}:=|\mathbf{1}\rangle.
\end{eqnarray}
Therefore, given a section $\mathfrak{s}=s^{a}\mathbf{e}_{a}\in\Gamma(\mathcal{M},\mathcal{V})$, in flat coordinates we have 
\begin{eqnarray}\label{nablaflat}
\nabla_{i}\mathfrak{s}=\partial_{i}s^{a}\mathbf{e}_{a}-s^{a}C_{ia}^{ \ c}\mathbf{e}_{c}.
\end{eqnarray}
This is the familiar form of the Gauss-Manin connection when acting on the D-brane central charges/boundary entropy \cite{Hori:2000ck}, written in such coordinates and in the appropriate gauge \cite{Ceresole:1992su}. There are two distinguished directions in the frame, namely $\mathbf{e}_{0}$, corresponding to the unique state with R-charges $(0,0)$ and $\mathbf{e}_{D}\in\mathcal{H}_{\hat{c},\hat{c}}^{(c,c)}$ ($\in\mathcal{H}_{-\hat{c},\hat{c}}^{(a,c)}$), $D:=\mathrm{dim}(\mathcal{H}_{\mathrm{def}})-1$, that is the unique state with maximal weight $(\hat{c}/2,\hat{c}/2)$. Now, consider the equation for flat sections:
\begin{eqnarray}\label{flateqqqq}
\nabla_{i}\mathfrak{s}=0.
\end{eqnarray}
We consider a basis of solutions $\mathfrak{s}_{(a)}$, labelled by $a=0,\ldots,\mathrm{dim}(\mathcal{H}_{\mathrm{def}})-1$ and that take the form
\begin{eqnarray}
\mathfrak{s}_{(a)}=\mathbf{e}_{a}+\ldots
\end{eqnarray}
where the terms $\ldots$ are along the directions $\mathbf{e}_{b}$ of charges $\bar{q}_{b}\geq\bar{q_{a}}$ and $b\neq a$. Explicitly,
\begin{eqnarray}\label{eqgravdesc}
\mathfrak{s}_{(a)}=\mathbf{e}_{a}+\sum_{n=1}^{\infty}\frac{1}{n!}\langle\tau_{n}(\phi_{a}),\phi_{c}\rangle_{S^{2}}\mathbf{e}^{c}\qquad \mathbf{e}^{c}:=\eta^{cb}\mathbf{e}_{b},
\end{eqnarray}
where $\tau_{n}$ denotes the $n$th gravitational descendant. Also note that the sum on the right-hand side of (\ref{eqgravdesc}) automatically contains only terms proportional to $\mathbf{e}_{b}$ with $\bar{q_{b}}>\bar{q_{a}}$ by the selection rule of $\langle\tau_{n}(\phi_{a}),\phi_{c}\rangle_{S^{2}}$, which is nonzero only if $n+\bar{q_{a}}+\bar{q_{c}}=\hat{c}$. This is a solution due to the topological recursion relations at genus $0$, satisfied by any SCFT coupled to topological gravity \cite{Witten:1989ig,Dijkgraaf:1990nc}:
\begin{eqnarray}
\langle\tau_{n}(\phi_{a}),\mathcal{O},\mathcal{O}'\rangle_{S^{2}}=n\langle\tau_{n-1}(\phi_{a}),\phi_{b}\rangle_{S^{2}}\eta^{bc}\langle\phi_{c},\mathcal{O},\mathcal{O}'\rangle_{S^{2}}.
\end{eqnarray}
Hence, we have that
\begin{eqnarray}
\partial_{i}\langle\tau_{n}(\phi_{a}),\phi_{b}\rangle_{S^{2}}=n\langle\tau_{n-1}(\phi_{a}),\phi_{c}\rangle_{S^{2}}C_{ib}^{ \ c}.
\end{eqnarray}
We define the $J$-function as the following (not necessarily flat) section of $\mathcal{V}$:
\begin{eqnarray}\label{jjjj}
\mathbf{J}:=\langle\mathfrak{s}_{(a)},\mathbf{1}\rangle_{S^{2}}\eta^{ab}\mathbf{e}_{b}.
\end{eqnarray}
As it will become useful later, let us give a more explicit expression for $\mathbf{J}$, relating it to topological invariants. First, it is easy to show that we can write
\begin{eqnarray}
\mathbf{J}=\mathbf{e}_{0}+\sum_{\bar{q_{c}}<\hat{c}}\sum_{n=1}^{\infty}\frac{1}{n!}\langle\tau_{n}(\phi_{c}),\mathbf{1}\rangle_{S^{2}}\eta^{cb}\mathbf{e}_{b}.
\end{eqnarray}
Schematically, the correlators take the following form
\begin{eqnarray}
\langle\tau_{n}(\phi_{a}),\mathbf{1}\rangle_{S^{2}}=\langle\tau_{n}(\phi_{a}),\mathbf{1},e^{t^{i}\int\mathcal{O}_{i}^{(1,1)}}\rangle\vert_{0}.
\end{eqnarray}
We use the notation $\langle\cdots\rangle\vert_{0}$ to emphasize that this has to be read as the expectation value of an operator in the theory with all marginal deformations set to zero: $t\equiv 0$. Then, we can can formally expand the exponentials
\begin{eqnarray}
\langle\tau_{n}(\phi_{a}),\mathbf{1}\rangle_{S^{2}}=\sum_{k_{1},\ldots,k_{h}\geq 0}\prod_{i=1}^{h}\frac{(t_{i})^{k_{i}}}{k_{i}!}\langle\tau_{n-1}(\phi_{a}),\prod_{j=1}^{h}(\int\mathcal{O}_{j}^{(1,1)})^{k_{j}}\rangle\vert_{0},
\end{eqnarray}
where we used the puncture equation~\cite{Witten:1989ig} (also known as string equation) to get rid of the insertion  $\mathbf{1}$. Intuition coming from the path integral formalism implies that a correlator $\langle\tau_{n-1}(\phi_{a}),\prod_{j=1}^{h}(\int\mathcal{O}_{j}^{(1,1)})^{k_{j}}\rangle\vert_{0}$ is expected to be written as an integral over the moduli space $\mathcal{M}_{\mathrm{BPS}}$ of maps\footnote{Here we are being vague. $\mathcal{M}_{\mathrm{BPS}}$ can stand for the moduli of stable maps if we are working in the A-twisted sigma model, and hence referring Gromov-Witten (GW) theory, or for the moduli of maps satisfying the Witten equation (as in FJRW theory) and so on.}  
on $S^{2}$ of a differential form determined by the operators $\phi_{j}$ whose descendants are $\mathcal{O}_{j}^{(1,1)}$\cite{Witten:1989ig}. Then, the topological invariants we are interested in are given by
\begin{eqnarray}
  \langle\tau_{n}(\phi_{a}),\phi_{j_{1}},\phi_{j_{2}},\ldots,\phi_{j_{m}}\rangle_{0,m+1}:=\frac{1}{n!}\langle\tau_{n}(\phi_{a}),\int\mathcal{O}_{j_{1}}^{(1,1)},\int\mathcal{O}_{j_{2}}^{(1,1)},\ldots,\int\mathcal{O}_{j_{m}}^{(1,1)}\rangle\vert_{0}\,.
\end{eqnarray}
 Now we can finally write $\mathbf{J}$ as
\begin{eqnarray}
\mathbf{J}=\mathbf{e}_{0}+\sum_{\bar{q}_{c}<\hat{c}}\sum_{n=1}^{\infty}\sum_{k_{1},\ldots,k_{h}\geq 0}\prod_{i=1}^{h}\frac{(t_{i})^{k_{i}}}{k_{i}!}\langle\tau_{n}(\phi_{c}),\prod_{j=1}^{h}\phi_{j}^{k_{j}}\rangle_{0,|k|+1}\eta^{cb}\mathbf{e}_{b}, 
\end{eqnarray}
where $|k|=\sum_ik_i$. Notice that the correlators $\langle\tau_{n_{1}}(\phi_{j_{1}}),\tau_{n_{2}}(\phi_{j_{2}}),\ldots,\tau_{n_{m}}(\phi_{j_{m}})\rangle_{0,m}$ vanish unless the equality
\begin{eqnarray}
\sum_{a=1}^{m}(n_{a}+\bar{q}_{j_{a}})=\hat{c}-3+m,
\end{eqnarray}
is satisfied. This can be used to simplify this expression further. In fact, for the components along $\mathbf{e}_{j}\in \mathcal{H}_{\mathrm{Mar}}$, all correlators vanish except for the ones of the form $\langle\phi_{a},\phi_{b},\mathbf{1}\rangle_{0,3}$, giving us
\begin{eqnarray}
\mathbf{J}=\mathbf{e}_{0}+\sum_{i=1}^{h}t_{i}\mathbf{e}_{i}+\sum_{\bar{q}_{c}<\hat{c}-1}\sum_{n=1}^{\infty}\sum_{k_{1},\ldots,k_{h}\geq 0}\prod_{i=1}^{h}\frac{(t_{i})^{k_{i}}}{k_{i}!}\langle\tau_{n}(\phi_{c}),\prod_{j=1}^{h}\phi_{j}^{k_{j}}\rangle_{0,|k|+1}\eta^{cb}\mathbf{e}_{b}.
\end{eqnarray}

\subsubsection*{$\hat{c}=3$: Special geometry}

In the case of $\hat{c}=3$, the holomorphic vector bundle over $\mathcal{V}\rightarrow\mathcal{M}$ enjoys an extra structure: special K\"ahler geometry. 
Let us briefly recall the properties of special K\"ahler manifolds that we will need. $\mathcal{M}$ being (locally) special K\"ahler implies, by virtue of being the moduli space of an $\mathcal{N}=(2,2)$ SCFT \cite{Strominger:1990pd}, that $\mathcal{V}$ decomposes as $\mathcal{V}\cong \mathcal{S}\oplus\overline{\mathcal{S}}$ where $\mathrm{rk}\:\mathcal{S}=h+1$. Geometrically, $\mathcal{H}_{0,0}^{(*,*)} \otimes \mathcal{O}_{\mathcal{M}}$ is a line bundle $\mathcal{L}$ over $\mathcal{M}$. Together with $\mathcal{H}_{\mathrm{Mar}}\otimes \mathcal{O}_{\mathcal{M}}$, it forms $\mathcal{S}$:
\begin{eqnarray}
\mathcal{S}=\mathcal{H}_{0,0}^{(*,*)} \otimes \mathcal{O}_{\mathcal{M}}\oplus \mathcal{H}_{\mathrm{Mar}}\otimes \mathcal{O}_{\mathcal{M}}\cong\mathcal{L}\oplus (\mathcal{L}\otimes T\mathcal{M}).
\end{eqnarray}
In our case we identify a local frame of $\mathcal{S}$ with the following states of the chiral ring:
\begin{eqnarray}
\{ \mathbf{e}_{0}, \mathbf{e}_{i} \}_{i=1}^{h}\in \mathcal{H}_{0,0}^{(*,*)} \oplus \mathcal{H}_{\mathrm{Mar}}.
\end{eqnarray}
By spectral flow arguments we can show that, as vector spaces,
\begin{eqnarray}
\mathcal{H}_{\mathrm{def}}^{2}\cong \mathcal{H}_{\mathrm{def}}^{1}.
\end{eqnarray}
Therefore, given $\phi_{j}^{(2)}\in\mathcal{H}_{\mathrm{def}}^{2}$ and $\phi^{(3)}\in \mathcal{H}_{\mathrm{def}}^{3}$, we can define
\begin{eqnarray}
\phi^{i}:=\eta^{ij}\phi_{j}^{(2)}\qquad \phi^{0}:=(\langle \mathbf{1},\phi^{(3)}\rangle_{S^{2}})^{-1}\phi^{(3)}.
\end{eqnarray}
The bundle $\overline{\mathcal{S}}$ is then spanned by $\mathbf{e}_{\bar{b}}=\mathbf{e}^{a}g_{a\bar{b}}$, $a,\bar{b}=0,\ldots, h$. 
In the frame $\{\mathbf{e}_{0},\mathbf{e}_{i},\mathbf{e}^{j},\mathbf{e}^{0}\}$, the section $\mathfrak{s}$ is given by
\begin{eqnarray}
\mathfrak{s}=\mathfrak{s}^{0}\mathbf{e}_{0}+\mathfrak{s}^{i}\mathbf{e}_{i}+\mathfrak{s}_{j}\mathbf{e}^{j}+\mathfrak{s}_{0}\mathbf{e}^{0}.
\end{eqnarray}
Then, the flatness equation (\ref{flateqqqq}) is
\begin{eqnarray}\label{systeq}
\left(
                                         \begin{array}{c}
                                           \partial_{l}\mathfrak{s}^{0} \\
                                          \partial_{l}\mathfrak{s}^{i} \\
                                          \partial_{l}\mathfrak{s}_{j}\\
                                          \partial_{l}\mathfrak{s}_{0}
                                         \end{array}\right)-
                                         \left(
                                         \begin{array}{cccc}
                                          0 & 0 & 0 & 0\\
                                          \delta_{l}^{i} & 0 & 0 & 0\\
                                          0 & C_{lij} & 0 & 0\\
                                          0 & 0 & \delta_{l}^{j} & 0\\
                                         \end{array}\right)\left(
                                         \begin{array}{c}
                                           \mathfrak{s}^{0} \\
                                          \mathfrak{s}^{i} \\
                                          \mathfrak{s}_{j}\\
                                          \mathfrak{s}_{0}
                                         \end{array}\right)=0.
\end{eqnarray}
In this particular case we can write our previously defined basis of solutions very explicitly and without need for gravitational descendants. For this purpose, we make use of the existence of the prepotential $\mathcal{F}$. In flat coordinates,
\begin{eqnarray}
  \label{eq:6}
  C_{ijk}=\frac{\partial^{2}\mathcal{F}}{\partial t^{i}\partial t^{j}\partial t^{k}}.
\end{eqnarray}
Then our basis of solutions is given by the sections:
\begin{equation}
  \begin{aligned}
\mathfrak{s}^{(0)}&=\mathbf{e}^{0}\\
\mathfrak{s}^{(k)}&=t^{k}\mathbf{e}^{0}+\mathbf{e}^{k}\\
\mathfrak{s}_{(k)}&=\frac{\partial\mathcal{F}}{\partial t^{k}}\mathbf{e}^{0}+\frac{\partial^{2}\mathcal{F}}{\partial t^{k}\partial t^{i}}\mathbf{e}^{i}+\mathbf{e}_{k}\\
\mathfrak{s}_{(0)}&=\left(\frac{\partial \mathcal{F}}{\partial t^{i}}t^{i}-2\mathcal{F}\right)\mathbf{e}^{0}+\left(\frac{\partial^{2}\mathcal{F}}{\partial t^{k}\partial t^{i}}t^{k}-\frac{\partial\mathcal{F}}{\partial t^{i}}\right)\mathbf{e}^{i}+t^{i}\mathbf{e}_{i}+\mathbf{e}_{0}.
\end{aligned}
\end{equation}
Now we can define the $J$-function: 
\begin{equation} \mathbf{J}:=\langle\mathfrak{s}^{(0)},\mathbf{1}\rangle_{S^{2}}\mathbf{e}_{0}+\langle\mathfrak{s}^{(k)},\mathbf{1}\rangle_{S^{2}} \mathbf{e}_{k}+\langle\mathfrak{s}_{(k)},\mathbf{1}\rangle_{S^{2}} \mathbf{e}^{k}+\langle\mathfrak{s}_{(0)},\mathbf{1}\rangle_{S^{2}}\mathbf{e}^{0}.
\end{equation}
In other words, we are taking the component along $\mathbf{e}^{0}$ of each section $\mathfrak{s}$. Explicitly,
\begin{equation}
  \label{eq:7}
  \mathbf{J}:=\mathbf{e}_{0}+t^{k}\mathbf{e}_{k}+\frac{\partial \mathcal{F}}{\partial t^{k}}\mathbf{e}^{k}+\left(\frac{\partial \mathcal{F}}{\partial t^{i}}t^i-2\mathcal{F}\right)\mathbf{e}^{0}.
\end{equation}
This is the familiar form, for example, for the SCFT corresponding to the IR fixed point of a Calabi-Yau sigma model, when $\mathcal{H}_{\mathrm{def}}\subset \mathcal{H}^{(a,c)}$ i.e.~for Gromov-Witten theory \cite{Cox:2000vi}.
\subsection{D-brane central charge, $\widehat \Gamma$ class, and the $J$-function}

We expect that topological D-branes $B$ and their associated boundary states $|B\rangle$ form a triangulated $A_{\infty}$-category $\mathcal{D}$ (for a review see \cite{Aspinwall:2009isa}). Well known examples are $\mathcal{D}$ corresponding to derived categories of coherent sheaves or equivariant matrix factorizations, as is the case for B-branes $\mathcal{B}$ in sigma models and Landau-Ginzburg orbifolds, respectively. Then we expect that there exists a map, the Chern character,
\begin{eqnarray}
  \mathrm{ch}:&&\mathcal{D}\to \mathcal{H}_{\mathrm{def}}\nonumber\\
  && {B}\mapsto |\phi_i\rangle \eta^{ij} \langle\phi_j|{B}\rangle.
\end{eqnarray}
The physics definition of this map is the A/B-twisted disk correlation function with boundary conditions corresponding to an object ${B}$ in $\mathcal{D}$ with the components along $\mathcal{H}_{\mathrm{def}}$ obtained by inserting the corresponding elements $\phi_i\in\mathcal{H}_{\mathrm{def}}$. For example, for B-branes in a geometric SCFT, this is the familiar Chern character \cite{Brunner:1999jq}. For Landau-Ginzburg orbifolds, the Chern character of a matrix factorization is less intuitive \cite{Walcher:2004tx,Brunner:2013ota,Kapustin:2003ga,MR2954619}. For a general dg category, a definition of the Chern character can be found in~\cite{Shklyarov:2013ab}.  

The ingredient to the central charge formula that is not obvious from an SCFT point of view is the Gamma class:
\begin{eqnarray}
\widehat{\Gamma}\in \mathrm{End}(\mathcal{H}_{\mathrm{def}}).
\end{eqnarray}
Here we abuse the notation and use the expression ``Gamma class'' a bit loosely. In principle the Gamma class is a multiplicative characteristic class, that is, a map\footnote{Here $K_{0}(\mathcal{D})$ denotes the Grothendieck group of the triangulated category $\mathcal{D}$ and can be defined in general (see \cite{neeman2005k} for a review). 
}
$\Gamma: K_{0}(\mathcal{D})\rightarrow \mathcal{H}_{\mathrm{def}}$ such that $\Gamma(E+F) = \Gamma(E)\Gamma(F)$, rather than an element of $\mathrm{End}(\mathcal{H}_{\mathrm{def}})$.
What we call $\widehat{\Gamma}$ here can be thought of as the map $K_{0}(\mathcal{D})\rightarrow \mathcal{H}_{\mathrm{def}}$ evaluated on a particular class in $K_{0}(\mathcal{D})$. For example, when $\mathcal{D}=D^{b}\mathrm{Coh}(X)$, this class is $[TX]$ and the endomorphism is given by the cup product in $H^{\bullet}(X)$. In other cases we do not have a prescription for selecting which element of $K_{0}(\mathcal{D})$ to take to obtain the desired map $\widehat{\Gamma}$, so we resort to defining it abstractly as a linear map $\widehat{\Gamma}\in \mathrm{End}(\mathcal{H}_{\mathrm{def}})$. Alternatively we can view $\widehat{\Gamma}$ as a particular element of $\mathcal{H}_{\mathrm{def}}$ acting by the ring product induced from the OPE.

One possible physics explanation for the appearance of $\widehat{\Gamma}$ are perturbative corrections. Not much is known about the Gamma class except in geometric realizations, i.e.~when the $\mathcal{N}=2$ SCFT can be interpreted as the IR fixed point of a nonlinear sigma model with target space $X$. We will now proceed to review this case and then formulate a proposal for the structure of the central charge formula for B-branes that will be used in this work.

The RR charge of a B-brane, as an element of $H^{\mathrm{even}}(X,\mathbb{Q})$ in a geometric situation, does not involve the class $\widehat{\Gamma}$, but a closely related object. In geometric cases this has been computed \cite{Green:1996dd,Cheung:1997az,Minasian:1997mm} by using the worldvolume of the D-brane as a guide. In this case, the RR charge of a brane $\mathcal{E}\in D^{b}\mathrm{Coh}(X)$ is given by $\mathrm{ch}(\mathcal{E})\sqrt{\widehat{A}(TX)}$, where $\sqrt{\widehat{A}(TX)}$ is the 'real' root (defined by the power series of the square root of $\widehat{A}$) of the characteristic class $\widehat{A}$ ($=\mathrm{Td}$ when $c_{1}(X)=0$). On the other hand, the characteristic class $\widehat{\Gamma}:=\widehat{\Gamma}(TX)\neq \sqrt{\widehat{A}(TX)}$, in general. Even though $\widehat{\Gamma}$ and $\sqrt{\widehat{A}(TX)}$ are roots of $\widehat{A}$ (in a sense to be made more precise below), one can think of the Gamma class as $\widehat{\Gamma}(TX)=\sqrt{\widehat{A}(TX)}\exp (i\widehat{\Lambda})$ where $\widehat{\Lambda}$ is some characteristic class \cite{Halverson:2013qca} such that $\widehat{\Gamma}$ respects the integrality of the open Witten index $\chi(\mathcal{E},\mathcal{F})$ for $\mathcal{E},\mathcal{F}\in\mathcal{D}$, i.e.
\begin{eqnarray}
\chi(\mathcal{E},\mathcal{F})=\langle  \widehat{\Gamma}^{*}\mathrm{ch}(\mathcal{E}^{\vee}), \widehat{\Gamma}\mathrm{ch}(\mathcal{F})\rangle_{S^{2}}\in \mathbb{Z},\label{wittenind}
\end{eqnarray}
where $ \widehat{\Gamma}^{*}$ and $\mathcal{E}^{\vee}$ are obtained from $\widehat{\Gamma}$ and $\mathcal{E}$ from an involution induced by the change of orientation of the brane\footnote{The involution for a general compact complex manifold was studied in \cite{cualduararu2005mukai}.} \cite{Hori:2000ck}. This condition is satisfied, for instance for $\mathrm{ch}(\mathcal{E})\sqrt{\widehat{A}(TX)}$, thanks to the Hirzebruch-Riemann-Roch theorem. The relation (\ref{wittenind}) imposes conditions on $\widehat{\Lambda}$ \cite{Halverson:2013qca} since it is required that (in the case $X$ is Calabi-Yau)
\begin{eqnarray}
\widehat{\Gamma}^{*}\widehat{\Gamma}=\mathrm{Td}(TX),
\end{eqnarray}
which leads to a derivation of $\widehat{\Gamma}$ in such geometric cases. Upon compactification of the type IIA superstring on a Calabi-Yau 3-fold $X$, the particles in the resulting 4d $\mathcal{N}=2$ theory associated to a an object $\mathcal{E}$ will have a central charge on their 4d supersymmetry algebra which will be given by \cite{Ceresole:1995jg}
\begin{eqnarray}
\int_{X}e^{-(B+iJ)} \widehat{\Gamma}\mathrm{ch}(\mathcal{E})+\text{instantons},
\end{eqnarray}
where $(B+iJ)$ stands for the complexified K\"ahler class of $X$. Alternatively, this formula can be written as
\begin{eqnarray}
\int_{X}e^{-(B+iJ)} \sqrt{\widehat{A}(TX)} \mathrm{ch}(\mathcal{E})+i\frac{\zeta(3)\chi(X)}{8\pi^{3}}+\text{instantons},
\end{eqnarray}
where the term $\zeta(3)\chi(X)$ comes from perturbative corrections of the nonlinear sigma model on $X$ \cite{Grisaru:1986px}. Hence, $\widehat{\Gamma}$ is a convenient way to encode the perturbative corrections to the central charge of $\mathcal{E}$. The appearance of this term in the prepotential $\cF$ in~\eqref{eq:6} (and hence in~\eqref{eq:7}) and its connection to these perturbative corrections has been already been pointed out in~\cite{Candelas:1990rm} (see also~\cite{Hosono:1994ax}).

Returning to the central charge of a general B-brane $\mathcal{B}\in\mathcal{D}$, the expectation is that, including the Gamma class, we can write
\begin{eqnarray}
Z(\mathcal{B})=\langle \widehat{\Gamma}^{*}\circ \mathbf{J}_{\mathrm{pert}}|\mathrm{ch}(\mathcal{B})\rangle+\text{instantons},
\end{eqnarray}
where $\mathbf{J}_{\mathrm{pert}}$ denotes the perturbative part of the previously defined $J$-function. For the geometric SCFTs mentioned above, this has been observed in the mathematics literature about central charges of objects in $D^{b}\mathrm{Coh}(X)$. Rigorous definitions for $\widehat{\Gamma}$ have been given \cite{libgober1999chern,Iritani:2009ab,Katzarkov:2008hs}. Similar expressions have appeared in physics \cite{Hosono:2000eb}. The exact formula is expected to take the form
\begin{eqnarray}\label{ccfinalf}
Z(\mathcal{B}):=\langle \widehat{\Gamma}^{*} \circ\mathbf{J}|\mathrm{ch}(\mathcal{B})\rangle.
\end{eqnarray}
Let us explain the pairing $\langle \cdot |\cdot \rangle$. We cannot just replace $\langle \cdot |\cdot \rangle$ by $\langle \cdot ,\cdot \rangle$ because $\widehat{\Gamma}^{*} \circ\mathbf{J}$ and $\mathrm{ch}(\mathcal{B})$ live in different chiral rings. So if, say, $\phi^{(a,c)}\in \mathcal{H}^{(a,c)}$ and $\phi^{(c,c)}\in\mathcal{H}^{(c,c)}$, thanks to the existence of the isomorphism provided by the spectral flow operator, we define
\begin{eqnarray}
  \label{eq:121}
  \langle \phi^{(a,c)} |\phi^{(c,c)} \rangle:=\langle \phi^{(a,c)}, \mathcal{U}\circ\phi^{(c,c)} \rangle_{S^{2}},
\end{eqnarray}
where the $\langle \phi^{(a,c)}, \mathcal{U}\circ\phi^{(c,c)} \rangle_{S^{2}}$ is the A-twisted two point function. 

As a final comment, we remark that the image of the map $\widehat{\Gamma}\circ \mathrm{ch}: \cD \to \cH_{\mathrm{def}}$ is a lattice in $\cH_{\mathrm{def}}$, given (\ref{wittenind}) for instance. The map $\mathbf{J} :\cM_K \to \cH_{\mathrm{def}}$ does not preserve the lattice structure when paired with $\mathrm{ch}(\mathcal{B})$ but $\widehat{\Gamma}^{*} \circ\mathbf{J}$ does. The integral local system associated to this map is $K_{\mathrm{top}}(X)$.  Further discussions on this can be found in \cite{Hosono:2000eb,Iritani:2009ab,Katzarkov:2008hs} from a physics and mathematical point of view. It would be very interesting to understand this integral structure from first principles in $\mathcal{N}=(2,2)$ SCFTs.

\subsection{Change of frame}
\label{sec:frame}

In this subsection we want to give a few remarks about the behavior of (\ref{ccfinalf}) under a change of frame of $\mathcal{V}$ (i.e.~gauge transformations). This becomes very useful, because there are many situations where one can obtain an expression for the central charge (for example, by UV computations) but it does not come in flat coordinates, and not even in the same frame as (\ref{nablaflat}). This will indeed be the case in the subsequent sections where we will deal with the so called $I$-function instead, which is equivalent to the $J$-function (\ref{jjjj}) up to a gauge transformation and a change of coordinates. Denote the Chern character of a B-brane $\mathcal{B}$, after spectral flow, by
\begin{eqnarray}
\mathcal{U}\circ\mathrm{ch}(\mathcal{B})=\mathrm{ch}(\mathcal{B})^{a}\mathbf{e}^{(a,c)}_{a}.
\end{eqnarray}
Then (\ref{ccfinalf}) takes the form
\begin{eqnarray}\label{explicitccc}
Z(\mathcal{B})=\langle \mathfrak{s}_{(a)},\mathbf{1}\rangle_{S^{2}}\eta^{ab}(\widehat{\Gamma}^{*})^{ \  c}_{b}\eta_{cd}\mathrm{ch}(\mathcal{B})^{d}.
\end{eqnarray}
Under a frame transformation $\mathbf{e}_{a}\rightarrow \mathcal{A}_{a}^{ \ b}\mathbf{e}_{a}$ all the contracted indices in (\ref{explicitccc}) remain invariant and therefore only the term $\langle \mathfrak{s}_{(a)},\mathbf{1}\rangle_{S^{2}}=\mathfrak{s}_{(a)}^{b}\eta_{b0}$ gives a nontrivial factor:
\begin{eqnarray}\label{transfgen}
Z(\mathcal{B})\rightarrow Z^{\mathcal{A}}(\mathcal{B}) =\mathcal{A}_{0}^{ \ l}\mathfrak{s}_{(a),l}\eta^{ab}(\widehat{\Gamma}^{*})^{ \  c}_{b}\eta_{cd}\mathrm{ch}(\mathcal{B})^{d}.
\end{eqnarray}
In the situations we will encounter in the following, it will be enough to consider $\mathcal{A}$ lower triangular, because we will be dealing with frame transformations that respect the filtration imposed by $\nabla$. In other words, $\mathcal{A}_{a}^{ \ b}=0$ if $\bar{q}_{b}> \bar{q}_{a}$. This means in particular that we can write $\mathcal{A}_{0}^{ \ l}=\mathcal{G}(t)\delta_{0}^{l}$ for some function $\mathcal{G}(t)$ and (\ref{transfgen}) simplifies to
\begin{eqnarray}
Z^{\mathcal{A}}(\mathcal{B})= \mathcal{G}(t)Z(\mathcal{B}).
\end{eqnarray}
This is a rather common situation, when going for example from the frame obtained from the Picard-Fuchs equations to the flat one \cite{Ceresole:1992su}.


%% file: cigar.pdf_t
\begin{picture}(0,0)%
\includegraphics{cigar.pdf}%
\end{picture}%
\setlength{\unitlength}{3315sp}%
\begingroup\makeatletter\ifx\SetFigFont\undefined%
\gdef\SetFigFont#1#2#3#4#5{%
  \reset@font\fontsize{#1}{#2pt}%
  \fontfamily{#3}\fontseries{#4}\fontshape{#5}%
  \selectfont}%
\fi\endgroup%
\begin{picture}(6120,2761)(2671,-4877)
\put(2701,-4786){\makebox(0,0)[lb]{\smash{{\SetFigFont{12}{14.4}{\familydefault}{\mddefault}{\updefault}{\color[rgb]{0,0,0}$\mathcal{B}$}%
}}}}
\put(8776,-3661){\makebox(0,0)[lb]{\smash{{\SetFigFont{12}{14.4}{\familydefault}{\mddefault}{\updefault}{\color[rgb]{0,0,0}$\mathcal{T}$}%
}}}}
\put(5626,-2311){\makebox(0,0)[lb]{\smash{{\SetFigFont{12}{14.4}{\familydefault}{\mddefault}{\updefault}{\color[rgb]{0,0,0}$L$}%
}}}}
\end{picture}%

%% file: section3.tex
\section{D-brane central charges for Landau-Ginzburg orbifolds}
\label{sec:cc-lgo}
In this section we apply the ideas of Section \ref{sec:cc-scft} to a specific class of superconformal field theories that are of particular interest in string theory: Landau-Ginzburg orbifolds. They arise as string backgrounds in non-geometric regions of the K\"ahler moduli space as originally found by  
\cite{Gepner:1987qi,Greene:1988ut}. We mainly follow the standard physics references \cite{Lerche:1989uy,Vafa:1989xc,Intriligator:1990ua}. 
      \subsection{Landau-Ginzburg orbifolds: a pr\'ecis}
      \label{sec-lgoclosed}
      We fix once and for all a basis on a vector space $\mathbb{V}$ of rank $N$ with coordinates denoted by $\phi_{j}$ $j=1,\ldots,N$. We specify a  left R-symmetry given by a $\mathbb{C}^{*}_{L}$ action on $\mathbb{V}$ with weights $q_{j}\in\mathbb{Q}\cap (0,1)$. The \emph{orbifold group} will be specified by a finite abelian group $G$ and a representation $\overline{\rho}_{m}:G\rightarrow GL(\mathbb{V})$ ($m$ stands for matter). We specify a \emph{superpotential}, that is a holomorphic, $G$-invariant function $W:\mathbb{C}^{N}\rightarrow \mathbb{C}$, $W\in \mathbb{C}[\phi_{1},\ldots,\phi_{N}]$. As an $\mathcal{N}=(2,2)$ theory, the Landau-Ginzburg orbifold is equipped with left and right R-symmetry. We will denote its generators by $F_{L}$ and $F_{R}$, and the charges of operators under them by $q$ and $\bar{q}$ respectively. The vector and axial R-symmetries are defined by
      \begin{eqnarray}
      F_{V}:=F_{L}+F_{R}\qquad F_{A}:=-F_{L}+F_{R}.
      \end{eqnarray}
      We want to have a nonanomalous vector R-symmetry, so we require $W$ to be quasi-homogeneous and of weight $1$ under a $\mathbb{C}^{*}_{L}$. Then $W(\lambda^{q_{i}}\phi_{i})=\lambda W(\phi_{i})$ \cite{Vafa:1988uu}. So, $W$ has charge $2$ under the vector R-symmetry. In addition we want to have a normalizable vacuum. We assume that a sufficient condition is that $W$ satisfies $dW^{-1}(0)=\{ 0\}$ i.e.~$W$ is called compact, or also nondegenerate in the mathematics literature.

      Quasi-homogeneity of $W$ guarantees that we always have the orbifold action $\phi_{j}\rightarrow e^{ 2 i\pi q_{j}}\phi_{j}$. 
      If $d$ denotes the lowest nonzero integer such that $dq_{j}\in \mathbb{Z}$ for all $j$, then this specifies a $\mathbb{Z}_{d}$ action we will denote by $\langle J\rangle$ where $J=\mathrm{diag}(e^{ 2i\pi q_{1}},\ldots,e^{ 2i\pi q_{N}})$.

Given $W\in \mathbb{C}[\phi_{1},\ldots,\phi_{N}]$, denote by $\mathrm{Aut}(W)$ the group of diagonal automorphisms of $W$, i.e.
\begin{eqnarray}
\mathrm{Aut}(W)=\left\{ \mathrm{diag}(e^{ 2\pi i \lambda_{1}},\ldots,e^{ 2\pi i \lambda_{N}})\in U(1)^{N}:W(e^{2\pi i\lambda_{j}}\phi_{j})=W(\phi_{j})  \right\}.
\end{eqnarray}
Using mathematical terminology, we will call an orbifold group $G$ \emph{admissible} if it satisfies
\begin{eqnarray}
\langle J\rangle\subseteq G\subseteq \mathrm{Aut}(W).
\end{eqnarray}
This condition guarantees that the left R-charges of the physical states are integral. Then the orbifold theory has spacetime supersymmetry \cite{Greene:1988ut,Vafa:1989xc}. In particular, this means that the theory is A-twistable. Even though this is a good string background \cite{Intriligator:1990ua}, in order to have a geometric interpretation in the context of string compactifications we need that the right R-charges are also integral. This is attainable by requiring $\mathrm{det}(\overline{\rho}_{m}(\gamma))=\pm 1$ for all $\gamma\in G$ \cite{Intriligator:1990ua}. In particular, in various examples that are obtained from a GLSM construction, $\overline{\rho}_{m}$ factors through\footnote{There can be situations where $\mathrm{det}(\overline{\rho}_{m}(J))=-1$ but after addition of extra massive fields, one can construct an equivalent orbifold theory where $\mathrm{det}(\overline{\rho}_{m}(J))=1$ \cite{Greene:1988ut}.} $SL(\mathbb{V})$. Note that $\mathrm{det}(\overline{\rho}_{m}(J))=\pm 1$ if and only if $\hat{c}\in \mathbb{Z}$, where $\hat{c}$ is the central charge:
\begin{eqnarray}
\hat{c}=\frac{c}{3}=\sum_{j=1}^{N}(1-2q_{j}).
\end{eqnarray}
Therefore we will be interested in Landau-Ginzburg orbifolds where $G$ is admissible. To summarize, in the following we will focus on Landau-Ginzburg orbifolds specified by the data
\begin{eqnarray}
  \label{lgdata}
(W,G,\overline{\rho}_{m},\mathbb{C}^{*}_{L})
\end{eqnarray}
satisfying:
\begin{enumerate}
 \item $W$ is quasi-homogeneous of degree $1$ with respect to the $\mathbb{C}^{*}_{L}$ action of weights $\{ q_{j}\}_{j=1}^{N}$, with $q_{j}\in \mathbb{Q}\cap (0,1)$, and $W$ is compact.
 \item $G$ is admissible.
 \item $\hat{c}\in \mathbb{Z}_{\geq 0}$.
\end{enumerate}
Let us give some more details on $W$ and the admissible orbifold groups one can associate to it. Denote the superpotential as $(\nu\geq N)$:
\begin{equation}
  W(\phi)=\sum_{\alpha=1}^{\nu}c_{\alpha}\prod_{j=1}^{N}\phi_{j}^{M_{j,\alpha}}\qquad M_{j,\alpha}\in \mathbb{Z}_{\geq 0},c_{\alpha}\in\mathbb{C}.
  \label{eq:56}
\end{equation}
We require that the exponent matrix $M:=(M_{j,\alpha})\in \mathrm{Mat}_{N\times \nu}(\mathbb{Z})$ has maximal rank, i.e.~$\mathrm{rk}(M)=N$. Quasi-homogeneous nondegenerate polynomials have been classified in \cite{Kreuzer:1992bi}. The different types are referred to as Fermat, chain and loop. While we will mostly focus on the Fermat type, most of the results presented in this section also hold for the more general cases. 

The matrix $M$ can be used to determine an explicit expression for $\mathrm{Aut}(W)$. Consider the Smith normal form of $M$:
\begin{eqnarray}
M=VSU\qquad U\in GL(\nu,\mathbb{Z}),V\in GL(N,\mathbb{Z}).
\end{eqnarray}
The matrix $S$ is zero outside the principal diagonal, whose values (the elementary divisors of $M$) are denoted by $d_{1},\ldots,d_{N}$. Since $\mathrm{rk}(M)=N$ they are nonvanishing. The matrix $S$ is unique up to permutation of its eigenvalues. Then $\mathrm{Aut}(W)\cong \mathbb{Z}_{d_{1}}\times\ldots \times\mathbb{Z}_{d_{N}}$ and is explicitly generated by
\begin{eqnarray}
g_{i}:\phi_{j}\mapsto e^{2\pi i \lambda^{(i)}_{j}}\phi_{j}\qquad \lambda^{(i)}_{j}:=d_{i}^{-1}(V^{-1})_{ij}.
\end{eqnarray}
The elements $g_{i}$ are not necessarily a minimal set of generators. There can be relations among them and/or they can be trivial (act as the identity). In the special case where $N=\nu$, i.e.~for the case of invertible polynomials, we can write the exponents $\lambda^{(i)}_{j}$ as
\begin{eqnarray}
\lambda^{(i)}_{j}=(M^{-1})_{ij}. \label{eq:minverse}
\end{eqnarray}
In general, every element of $\mathrm{Aut}(W)$ and consequently of $G$ can be written uniquely as
\begin{eqnarray}
  \label{orbdef}
\gamma=\mathrm{diag}(e^{2\pi i\theta^{\gamma}_{1}},\ldots,e^{2\pi i\theta^{\gamma}_{N}})\qquad \theta^{\gamma}_{j}\in [0,1),
\end{eqnarray}
where the $\theta^{\gamma}_{j}$ are sometimes referred to as phases. 
In the following we consider Landau-Ginzburg orbifolds with any admissible group $G\subseteq \mathrm{Aut}(W)$.

Given an orbifold specified by (\ref{lgdata}), we will describe next how to define its chiral rings, sometimes referred to as state spaces in the mathematics literature. In order to compute the $G$-invariant Hilbert space one needs to consider, for each $\gamma\in G$, the $\gamma$-twisted sectors. Let us clarify what this means in our notation. Given $G= \mathbb{Z}_{d_{1}}\times\mathbb{Z}_{d_2}\times\ldots$, a group element $\gamma\in G$ can be written as $\gamma=g_1^{k_1}g_{2}^{k_2}\ldots$ with $g_i=e^{2\pi i\lambda^{(i)}}$ and $k_i=0,\ldots,d_i-1$. In this case $\theta_j^{\gamma}=\langle k_1\lambda_j^{(1)}+k_2\lambda_j^{(2)}+\ldots \rangle$ for $j=1,\ldots,N$, where
\begin{equation}
  \label{angle}
  \langle x\rangle=x-\lfloor x\rfloor\qquad \text{for } x\in\mathbb{R}.
  \end{equation}
Given a sector labelled by $\gamma\in G$, fields in the $\gamma$-twisted sector satisfy $\phi_j(e^{2\pi i}z)=e^{2\pi i \theta_j^{\gamma}}\phi_j(z)$. If $\theta_j^{\gamma}=0$ we say that the fields $\phi_j$ satisfy untwisted boundary conditions in the $\gamma$-twisted sector. For the purpose of characterizing the chiral rings, we can restrict to zero modes as in \cite{Vafa:1989xc,Intriligator:1990ua}. 
In each sector, the $G$-invariant Hilbert space is built out of the fields satisfying untwisted boundary conditions, and one projects onto $G$-invariant states. So, schematically the $G$-invariant Hilbert space can be written as:
\begin{eqnarray}
\mathcal{H}:=\bigoplus_{\gamma\in G}PH_{\gamma}=\bigoplus_{\gamma\in G}\mathcal{H}_{\gamma},
\end{eqnarray}
where $H_{\gamma}$ is the Hilbert space of the $\gamma$-twisted sector and $P$ is the projector onto $G$-invariant states. We will focus on the $(c,c)$-ring $\mathcal{H}^{(c,c)}$ since the other rings, namely the $(a,c)$-ring and the ring of RR ground states, can be obtained from $\mathcal{H}^{(c,c)}$ by spectral flow. Write $\mathrm{Fix}(\gamma)=\{\phi_{i}:\theta_{i}^{\gamma}=0\}\subset\mathbb{C}^{N}$ and the Jacobian ring of a polynomial $F$ as
\begin{eqnarray}
\mathrm{Jac}(F)=\frac{\mathcal{R}}{(\partial F)}\qquad F\in\mathcal{R},
\end{eqnarray}
where $\mathcal{R}$ is some polynomial ring. We can write states in the unprojected $(c,c)$-ring in the $\gamma$-twisted sector as
\begin{eqnarray}
|f(\phi)\rangle_{\gamma}:=f(\phi)|0\rangle_{\gamma}\in H^{(c,c)}_{\gamma},\qquad f(\phi)\in \mathrm{Jac}(W|_{\mathrm{Fix}(\gamma)}),
\end{eqnarray}
where $|0\rangle_{\gamma}$ is the unique vacuum in the $\gamma$-twisted sector, with left/right R-charges:
\begin{eqnarray}
F_{L}|0\rangle_{\gamma}&=&\left(\mathrm{age}(\gamma)-\frac{N}{2}+\sum_{j:\theta^{\gamma}_{j}=0}q_{j}+\frac{\hat{c}}{2}\right)|0\rangle_{\gamma}\nonumber\\
F_{R}|0\rangle_{\gamma}&=&\left(-\mathrm{age}(\gamma)+\frac{N}{2}-n_{\gamma}+\sum_{j:\theta^{\gamma}_{j}=0}q_{j}+\frac{\hat{c}}{2}\right)|0\rangle_{\gamma},
\end{eqnarray}
where
\begin{eqnarray}
\mathrm{age}(\gamma)=\sum_{j=1}^{N}\theta_{j}^{\gamma}\qquad n_{\gamma}=\mathrm{dim}(\mathrm{Fix}(\gamma)).
\end{eqnarray}
The space $H^{(c,c)}_{\gamma}$ is not necessarily isomorphic to the unprojected RR ground states $H^{R}_{\gamma}$. However due to the fact that we are using admissible orbifolds, the isomorphism of vector spaces holds for the projected Hilbert spaces $\mathcal{H}_{\gamma}^{(c,c)}$ and $\mathcal{H}_{\gamma}^{R}$. The isomorphism is realized by the spectral flow operator $\mathcal{U}_{(\frac{1}{2},\frac{1}{2})}$ as (cf.~Table \ref{loctab})
\begin{eqnarray}
\mathcal{U}_{(-\frac{1}{2},-\frac{1}{2})}|0\rangle_{\gamma}=|0\rangle_{\gamma}^{R}.
\end{eqnarray}
Similarly for the $(a,c)$-ring:
\begin{eqnarray}
  \label{acccflow}
\mathcal{U}_{(-1,0)}|0\rangle_{\gamma}=|0\rangle^{(a,c)}_{\gamma J}.
\end{eqnarray}

The pairing on $\mathcal{H}_\gamma$ is given by the topological two-point function on $S^{2}$, as reviewed in Section~\ref{sec:cc-scft}. For the $(c,c)$-ring, this is given by the B-twisted correlator:
    \begin{equation}\label{pairingcccc}
   \langle-,-\rangle_\gamma : \mathcal{H}_{\gamma}^{(c,c)} \times  \mathcal{H}_{\gamma^{-1}}^{(c,c)} \to \mC.
    \end{equation}
    The pairing is symmetric and non-degenerate. The pairing on $\mathcal{H}$ is defined as $\oplus_{\gamma}\langle-,-\rangle_\gamma$.

    A situation that will be recurrent in the following is that we will focus on $\gamma$-twisted sectors $\mathcal{H}^{(c,c)}_{\gamma}$ of the $(c,c)$-ring such that $n_{\gamma}=0$, hence, they satisfy $\mathrm{dim}(\mathcal{H}^{(c,c)}_{\gamma})=1$. We refer to these sectors as \emph{narrow} sectors. All the other sectors are referred to as \emph{broad}. Note that zero-dimensional sectors, i.e. those where no state survives the projection, are also referred to as broad. The classification in terms of broad and narrow sectors is borrowed from the mathematics literature. We will say more about these sectors in Section \ref{sec:fjrw}. For now, let us remark that the pairing (\ref{pairingcccc}) takes a very simple form 
    when $\phi_\gamma,\phi_{\gamma'}$ belong to narrow sectors: 
\begin{equation}
 \eta(\phi_\gamma,\phi_{\gamma'}) :=\langle\phi_\gamma,\phi_{\gamma'}\rangle = \frac{1}{|G|}\delta_{\gamma , \gamma'^{-1}}.\label{lgpairing}
\end{equation}
We will define $\mathcal{H}_{narrow}$ as:
\begin{equation}
 \mathcal{H}_{narrow}:=\bigoplus_{\gamma\in G_{0}}\mathcal{H}_{\gamma},
\end{equation}
with $G_{0}=\{\gamma\in G|n_{\gamma}=0\}$. We will denote the corresponding subring of the deformation ring (\ref{hdef}) as
\begin{equation}
\mathcal{H}_{\mathrm{def},0}:= \langle\mathcal{H}_{\mathrm{Mar}}\cap\mathcal{H}_{narrow} \rangle.
\end{equation}
To conclude this subsection we remark that, whenever we have a large volume point, corresponding to a smooth geometry $X$, in the space of marginal deformations of a Landau-Ginzburg orbifold, we have an isomorphism of vector spaces
\begin{equation}
H^{\hat{c}-q,\bar{q}}(X)\cong \mathcal{H}_{q,\bar{q}}^{(c,c)}.
\end{equation}
\subsection{D-branes in Landau-Ginzburg orbifolds}
\label{sec-lgbranes}
B-type D-branes in (topological) Landau-Ginzburg models are characterized in terms of matrix factorizations of the Landau-Ginzburg superpotential \cite{Kapustin:2002bi,Brunner:2003dc}. A matrix factorization is defined by the set of data 
\begin{equation}\label{mfbbb}
\overline{\mathcal{B}}=(\overline{M},\sigma,\overline{Q},\overline{\rho},\overline{R})
\end{equation}
where $\overline{M}$ (the Chan-Paton space) is a free $\mathbb{C}[\phi_{1},\ldots,\phi_{N}]$-module, 
 $\sigma$ is an involution on $\overline{M}$, inducing a $\mathbb{Z}_2$-grading (so we can write $\overline{M}=\overline{M}_{0}\oplus \overline{M}_{1}$, with $\sigma \overline{M}_{i}=(-1)^{i}\overline{M}_{i}$) and $\overline{Q}(\phi)$ is a $\mathbb{Z}_2$-odd endomorphism on $\overline{M}$ satisfying
\begin{equation}
  \label{lg-mf}
  {\overline{Q}}^2=W\cdot\mathrm {id}_{\overline{M}}.
\end{equation}
This definition can be extended by various gradings. For Landau-Ginzburg orbifolds with orbifold group $G$ these are the vector R-charge and the $G$-grading. This has been defined in \cite{Walcher:2004tx}, see also \cite{Jockers:2006sm,Herbst:2008jq}.  Under the vector R-charge $F_{V}$, $W$ has charge $2$: $W(\lambda^{2q_j}\phi_j)=\lambda^2W(\phi_j)$ with the charges $q_j$ of the left R-symmetry as in Section \ref{sec-lgoclosed}. Therefore, by (\ref{lg-mf}), $\overline{Q}$ must have vector R-charge $1$. This defines a representation $\overline{R}:U(1)_{V}\rightarrow GL(\overline{M})$ of the vector R-symmetry satisfying
\begin{equation}
  \overline{R}(\lambda)\overline{Q}(\lambda^{2q_j}\phi_j)\overline{R}^{-1}(\lambda)=\lambda \overline{Q}(\phi_j),
\end{equation}
as well as another representation of $G$, $\overline{\rho}:G\rightarrow GL(\overline{M})$, satisfying
\begin{equation}
  \label{def-orbmat}
  \overline{\rho}(\gamma)^{-1}\overline{Q}(e^{2\pi i\theta_{j}^{\gamma}}\phi_j)\overline{\rho}(\gamma)=\overline{Q}(\phi_{j}),
\end{equation}
and compatible with $\overline{R}$. 

We denote the category of matrix factorizations 
as $MF_{G}(W)$ and its objects are identified with the $\overline{\mathcal{B}}$ (defined as in (\ref{mfbbb})).
In \cite{Walcher:2004tx} the RR-charge of an Landau-Ginzburg brane has been defined\footnote{See also \cite{Brunner:2013ota} for a first principle derivation from orbifold defects.}.

Pick a Landau-Ginzburg B-brane $\overline{\mathcal{B}}$, $\gamma\in G$, and a Ramond ground state in the $\gamma$-twisted sector: $|\gamma,\alpha\rangle\in\mathcal{H}_{\gamma}^{R}$. 
We should think of this state, as we described in Section \ref{sec:cc-scft}, as coming from an element $\phi_{(\alpha,\gamma)}\in \mathcal{H}^{(c,c)}_{\gamma}$. Let $|\overline{\mathcal{B}}\rangle$ be the boundary state characterized by the brane. Then the Chern character of $\overline{\mathcal{B}}$, is given by
\begin{eqnarray}\label{chernlg}
\mathrm{ch}:{MF}_{G}(W)\rightarrow \bigoplus_{\gamma\in G}\mathcal{H}_{\gamma}^{R}\qquad \overline{\mathcal{B}}\mapsto \langle\gamma,\alpha|\overline{\mathcal{B}}\rangle\eta^{(\gamma,\alpha),(\gamma',\alpha')} \langle \gamma',\alpha'|
\end{eqnarray} 
where $\langle\gamma,\alpha|\overline{\mathcal{B}}\rangle$ is given by the following bulk-boundary two-point function on the disk computed by the residue integral
\begin{eqnarray}
  \label{rr-walcher}
  \langle\gamma,\alpha|\overline{\mathcal{B}}\rangle &=&\frac{1}{n_{\gamma}!}\mathrm{Res}_{W_{\gamma}}\left(\alpha\cdot \mathrm{str}\left[\overline{\rho}_{\gamma}(\partial \overline{Q}_{\gamma})^{\wedge n_{\gamma}}\right] \right)\nonumber\\
  &=&\frac{1}{n_{\gamma}!}\oint\frac{\alpha\cdot \mathrm{str}\left[\overline{\rho}_{\gamma}(\partial \overline{Q}_{\gamma})^{\wedge n_{\gamma}}\right]}{\prod_{l\in I^{\gamma}}\partial_{l}W_{\gamma}},
\end{eqnarray}
where $\mathrm{str}(\cdot)=\mathrm{Tr}_{\overline{M}}(\sigma\cdot)$, $\overline{Q}_{\gamma}=\overline{Q}|_{\mathrm{Fix}(\gamma)}$, $I^{\gamma}$ are the labels of the coordinates in $\mathrm{Fix}(\gamma)$ and $(\partial \overline{Q}_{\gamma})^{\wedge n_{\gamma}}$ denotes the antisymmetrized product of derivatives of $\overline{Q}_{\gamma}$ with respect to the untwisted fields in the $\gamma$-twisted sector.

Note that the RR-charge vanishes trivially whenever $n_{\gamma}$ is odd. In the special case where $n_{\gamma}=0$, i.e.~when we have a single RR ground state in that sector, the expression (\ref{rr-walcher}) reduces to
\begin{equation}
  ^{R}_{\gamma}\langle 0|\overline{\mathcal{B}}\rangle=\mathrm{str}(\overline{\rho}(\gamma)).
  \label{eq:49}
\end{equation}
As we will see in Sections \ref{sec:hpf} and \ref{sec:apps}, this is precisely what one gets when one evaluates the brane factor of the hemisphere partition function of the associated GLSM in the Landau-Ginzburg phase.

\subsection{Marginal deformations}
\label{sec:deformations}

So far, we have only discussed Landau-Ginzburg orbifolds that we view as located at specific points in the stringy K\"ahler moduli space. However, many properties described here cannot be defined just considering Landau-Ginzburg orbifolds on their own. In order to define the exact central charge of a B-type D-brane, and in particular the $I$-function (and subsequently, the $J$-function) entering the proposed formula, we also have to take into account deformations away from the Landau-Ginzburg point.

To a Landau-Ginzburg orbifold $(W,G,\overline{\rho}_{m},\mathbb{C}_L^{\ast})$ and its deformations we associate a rational matrix $q$ which plays a
central role in the definition of the $I$-function of FJRW theory, and in the gauged linear sigma model associated to $(W,G,\overline{\rho}_{m},\mathbb{C}_L^{\ast})$.

To motivate $q$, recall \cite{Vafa:1989ih} that there is an action of $G$ on the chiral ring $\mathcal{H}$, known as quantum symmetry. This symmetry acts via the dual group $G^{\ast}=\mathrm{Hom}(G,\mathbb{C}^{\ast})$ of $G$ by multiplication with a character of $G$:
\begin{equation}
  \gamma\cdot \alpha=\chi_{\gamma}(\gamma')\alpha,\qquad \alpha\in\mathcal{H}_{\gamma'},\:  \chi_{\gamma}\in G^{\ast}.
\end{equation}
In particular it acts on the space of marginal deformations $\mathcal{H}_{\mathrm{Mar}}$. We wish to reformulate this $G$-action in terms of the original $G$-action on the chiral fields $\phi_j\in\mathbb{C}^{N}$. In essence, we want to infer the action of $G$ on the $(a,c)$-ring, for which currently no explicit description is known, from the action of $G$ on the $(c,c)$-ring. This is possible because of spectral flow/mirror symmetry. The interplay between quantum symmetry and mirror symmetry has been studied in \cite{Kreuzer:1995yi}. We can express this reformulation in terms of a linear map $\mathcal{H}_{\mathrm{Mar}}\to\mathcal{H}_{\mathrm{Mar}}\times \mathbb{C}^N$. The associated $h\times(h+N)$ matrix is the matrix $q$.

To determine the matrix $q$ we proceed as follows. Consider the $(-1,1)$-operators in the $(a,c)$-ring of $(W,G,\overline{\rho}_{m},\mathbb{C}_L^{\ast})$. We denote them by $\mathcal{O}_{\gamma,\mu}\in\mathcal{H}^{(a,c)}_{\gamma, (-1,1)} $ where $\gamma$ labels the twisted sector they belong to and $\mu=1,\ldots,\dim \cH^{(a,c)}_{\gamma, (-1,1)}$. 
Since we are considering only admissible orbifolds, the operators $\mathcal{O}_{\gamma,\mu}$ can be represented by spectral flow of operators in the $(c,c)$-ring of the form:
\begin{equation}
  \label{eq:118}
  \mathcal{O}_{\gamma,\mu}=\mathcal{U}_{(-1,0)}\circ f_{\mu}(\phi)|0\rangle^{(c,c)}_{\gamma J^{-1}}\qquad f_{\mu}(\phi)\in \mathrm{Jac}(W|_{\mathrm{Fix}(\gamma J^{-1})})
\end{equation}
for some monomial $f_{\mu}(\phi)$. We will be interested in operators that belong to narrow sectors, i.e. whenever $\mathrm{Fix}(\gamma J^{-1})=\{ 0\}$. Suppose that we have $h$ operators spanning the subspace of $\mathcal{H}^{(a,c)}_{(-1,1)}$ consisting only of narrow sectors. In such a case we consider $\mathcal{O}_{\gamma_a}=\mathcal{U}_{(-1,0)}\circ |0\rangle^{(c,c)}_{\gamma_{a} J^{-1}}$, $a=1,\dots,h$. Then we define the matrix $q^{\mathrm{LG}}\in\mathrm{Mat}_{h\times (h+N)}(\mathbb{Q})$ as
\begin{equation}
  \label{qmardef}
q^{\mathrm{LG}}_{a,b}=\delta_{a,b}, \qquad q^{\mathrm{LG}}_{a,h+j}=-\theta^{\gamma^{-1}_{a}}_{j}\qquad  \text{ \ for \ } \left\{\begin{array}{c}a,b=1,\ldots,h\\ j=1,\ldots,N. \end{array} \right. 
\end{equation}
The notation $q^{\mathrm{LG}}$ is to emphasize that this matrix only depends on the $(a,c)$-operators of charges $(-1,1)$ which are in one-to-one correspondence with exactly marginal deformations. We will see in the following that there are further definitions of the matrix $q$ which give equivalent $I$-functions. 

To understand the subtleties in the definition of the matrix $q$, we will outline a mirror interpretation which will become important in subsequent sections when comparing different ways to obtain the central charges for B-type D-branes.

Since we have assumed $\mathrm{rk}({M})=N$, i.e.~the rank of the matrix of exponents of $W$ is maximal, we can always go to a point in the complex structure moduli space (i.e.~a choice in variables $c_{\alpha}$) where $W$ is invertible i.e.~$\nu=N$. From now on, we will assume
\begin{equation}
  M\in \mathrm{Mat}_{N\times N}(\mathbb{Z}_{\geq 0}) \qquad \mathrm{det}(M)\neq 0.
\end{equation}
In order to distinguish columns and rows of $M$ we will write $M_{j,\alpha}$ when referring to its components, where $j,\alpha=1,\ldots,N$. We will also assume 
$G\subseteq SL(\mathbb{V})$.
We remark that, when thinking of the Landau-Ginzburg orbifold as a particular SCFT at a point in $\mathcal{M}_{K}$, the $I$-function we are about to define is expected to depend only on deformations along $\mathcal{M}_{K}$, and is independent of the choice of $c_{\alpha}$, as long as $W$ still satisfies the conditions of nondegeneracy at such a point.
 By the invertibility of $W$, we can describe the generators of $\mathrm{Aut}(W)$ explicitly in terms of the matrix elements $(M^{-1})_{\alpha,j}$. Then any element $\gamma\in \mathrm{Aut}(W)$ takes the form
 \begin{equation}
\gamma\cdot\phi_{j}=e^{2\pi i (n^{T}M^{-1})_{j}}\phi_{j} \qquad n\in \mathbb{Z}^{N}.
\end{equation}
More precisely, the integer vector $n$ takes values in the quotient $\mathbb{Z}^{N}/\{v\in \mathbb{Z}^{N}:(v^{T}M^{-1})_{j}\in\mathbb{Z}\}$ and is determined by $\gamma$. We choose a set of generators $\{ g_{1},\ldots, g_{N}\}$ of $\mathrm{Aut}(W)$ by
\begin{equation}
  \label{phiaut}
g_{\alpha}\cdot\phi_{j}=e^{-2\pi i M^{-1}_{\alpha,j}}\phi_{j}. 
\end{equation}
The fact that the $g_{\alpha}$ generate $\mathrm{Aut}(W)$ is shown in \cite{Krawitz:2010ab}. Now define the transpose potential $W^{T}$ by
\begin{equation}
W^{T}:=\sum_{j=1}^{N}\prod_{\alpha=1}^{N}y_{\alpha}^{M^{T}_{\alpha,j}}.
\end{equation}
One can show that $\mathrm{Aut}(W) \cong \mathrm{Aut}(W^{T})$ \cite{Kreuzer:2002uu}. Denote the generators of $\mathrm{Aut}(W^{T})$ as $\bar{g}_{j}$ acting on the $y$ variables as
\begin{equation}
\bar{g}_{j}\cdot y_{\alpha}=e^{-2\pi i M^{-T}_{j,\alpha}}y_{\alpha}. 
\end{equation}


Let $\iota: G \hookrightarrow \Aut(W)$ be the embedding of $G$
into $\Aut(W)$. This embedding can be described explicitly by using (\ref{phiaut}). 
Then any element on the image of $\iota$ takes the form
\begin{equation}
 \iota(g)\cdot\phi_{j}=e^{2\pi i (n^{T}M^{-1})_{j}}\phi_{j}=\overline{\rho}_{m}(g)\cdot \phi_{j}\qquad n\in \mathbb{Z}^{N},g\in G.
\end{equation}
The integer vector $n$ is determined by $g\in G$. There is an embedding
\begin{equation}
\iota\spcheck:\mathrm{Aut}(W^{T})\hookrightarrow G^{\ast}
\end{equation}
given explicitly by\footnote{Here we view $\tilde{v}^{T}M^{-T}$, $\tilde{v}\in\mathbb{Z}^{N}$ as an element of $\Aut(W^{T})$, acting on $y_{\alpha}$ as $y_{\alpha}\mapsto \exp(2\pi i (\tilde{v}^{T}M^{-T})_{\alpha})y_{\alpha}$. Then, we can define an element $\varphi(\tilde{v}^{T}M^{-T})\in G^{\ast}=\mathrm{Hom}(G,\mathbb{C}^{*})$ through the embedding of $G$ into $\Aut(W)$ by mapping $g \in G$ to $v^{T}M^{-1}$ (for some $v\in\mathbb{Z}$) and then to $\exp(2\pi i v^{T}M^{-1}\tilde{v})\in \mathbb{C}^{*}$.} $\exp(2\pi i v^{T}M^{-1}\tilde{v})$ for $v,\tilde{v}\in \mathbb{Z}^{N}$. Then we set
\begin{equation}
  G\spcheck := \ker \iota\spcheck=\{\tilde{v}\in\mathbb{Z}^N\mid\tilde{v}^TM^{-T}v\in\mathbb{Z},\:\forall v\in G\}.
  \label{eq:57}
\end{equation}
In the following, by abusing notation, we will think of the elements $g\spcheck\in G\spcheck$ as vectors $\tilde v$ in an appropriate lattice $\mathbb{Z}^{N}$ with a scalar product defined by $M^{-T}$ and write $g\spcheck \cdot v = \tilde v M^{-T} v$. Among the generators of $G\spcheck$ there is the distinguished generator $J\spcheck$ satisfying $J\spcheck \cdot (1,\ldots,1)^T = 1$.
Now we define the following set: 
\begin{equation}
  \cA_{\mathrm{ext}}:=\{ v\in (\mZ_{\geq 0})^N \mid J\spcheck \cdot v = 1, \;g \spcheck \cdot v = 0 \mod \mZ,\; \forall g\spcheck \in G\spcheck \} 
  \label{eq:52}
\end{equation}
The condition on $J\spcheck$ guarantees that the potential marginal deformations $v$ have vector R-charge $2$. 
Note that it suffices to verify these conditions on the generators of $G\spcheck$.  Then the elements of $\cA_{\mathrm{ext}}$ are vectors $v\in (\mZ_{\geq 0})^N$ defining invariant monomials in $\mathbb{C}[y_{1},\ldots,y_{N}]$ under the action of $G\spcheck$. We claim that this set characterizes the space of marginal deformations of the $(a,c)$-ring, however with the ambiguity that some of these deformations can be related by field redefinitions. A detailed  discussion of this interpretation will be given below. 

Clearly, the row vectors of the matrix $M$ belong to $\cA_{\mathrm{ext}}$. We arrange the vectors $v \in \cA_{\mathrm{ext}}$ as columns of a matrix $M\spcheck = (M'\; M^T)$ where the columns of the matrix $M'$ contain the solutions $v$ that are not row
vectors of $M$. Note that $\rank M\spcheck = \rank M=N$. 
The linear relations among the vectors $v$ will correspond to marginal deformations. To obtain them we choose a particular representative of the kernel of the matrix $M\spcheck$. This is encoded in the matrix $q^{\mathrm{ext}}$: 
\begin{equation}
  \label{eq:120}
  q^{\mathrm{ext}}\in \mathrm{Mat}_{\hat{h} \times (\hat{h}+N)}(\mathbb{Q}),\qquad M\spcheck (q^{\mathrm{ext}})^{T}=0,\qquad q^{\mathrm{ext}}_{\hat{a},\hat{b}}=\delta_{\hat{a},\hat{b}} \text{ \ for \ }\hat{a},\hat{b}=1,\ldots,\hat{h} \,. 
\end{equation}
The row vectors of $q^{\mathrm{ext}}$ span $\ker M\spcheck$ and the column vectors corresponding to the columns of $M'$ are the standard basis vectors $e_i\in\mathbb{R}^{\hat{h}}$, $i= 1,\dots,\rank M'=\hat{h}$. We have $\hat{h}=\rank q^{\mathrm{ext}} \geq \dim (\mathcal{H}_{\mathrm{Mar},0}^{(a,c)})$. For an alternative derivation of $q^{\ext}$ in the language of~\cite{Kreuzer:1994np} see Appendix~\ref{app-qq}.

Excluding from $\cA_{\mathrm{ext}}$ the vectors $v$ that have exactly one component which is $0$ corresponds to restricting to a subset of linearly independent marginal deformations. On the (geometric) mirror, this condition amounts to modding out by non-linear automorphisms on the toric ambient space preserving $W$ \cite{Hosono:1993qy,MR1253648,Berglund:1995gd}. This yields the set
\begin{equation}
  \label{eq:116}
  \cA_{\mathrm{geom}}:=  \cA_{\mathrm{ext}} \setminus \{ v\in (\mZ_{\geq 0})^N \mid v\text{ \ has a single\ }0\text{\ entry} \}.
\end{equation}
We can repeat the same procedure using $\cA_{\mathrm{geom}}$ and denote the resulting matrix $q^{\mathrm{geom}}\in \mathrm{Mat}_{h \times (h+N)}(\mathbb{Q})$. Similarly, we can define the set
\begin{equation}
  \label{eq:135}
  \cA_{\mathrm{LG}}:=  \cA_{\mathrm{ext}} \setminus\{ v \in (\mZ_{\geq 0})^N \mid \prod_{j=1}^N y_j^{v_j} = 0 \in \mathrm{Jac}(W^T) \}.
\end{equation}
We will show in Appendix~\ref{app-qq} that the matrix  $q^{\mathrm{LG}}$ obtained as the kernel of matrix of the vectors in $\cA_{\mathrm{LG}}$ agrees with the matrix defined in~\eqref{qmardef}. The purpose of $q^{\mathrm{ext}}$ is that we can obtain the matrices $q^{\mathrm{LG}}$ and $q^{\mathrm{geom}}$ from it by removing rows and columns.

Let us give some physics intuition of the necessity for $q^{\mathrm{ext}}$ when taking into account the global structure of $\mathcal{M}_{K}$. We are considering limiting points in $\mathcal{M}_K$ that have some concrete realization of the worldsheet CFT, e.g~ in terms of a Landau-Ginzburg orbifold or a non-linear sigma model. The states corresponding to marginal deformations arise from the cohomology of some BRST operator in the respective theory. A priori, these are completely different theories and one should not expect a simple relation between the elements of the different deformation spaces and their representatives. When comparing Landau-Ginzburg and large volume points in $\mathcal{M}_K$ the deformations coming from narrow sectors are characterised, via mirror symmetry, in terms of monomials. However, a monomial representative in geometry may be not necessarily be a ``good'' representative in the Landau-Ginzburg theory and vice versa. We claim that the matrix $q^{\mathrm{ext}}$ captures enough information to accommodate for all loci in $\mathcal{M}_K$ and that the reduction to $q^{\mathrm{LG}}$ or $q^{\mathrm{geom}}$ then accounts for the ``natural'' set of representatives at the respective locus of $\mathcal{M}_K$. We further claim that the choices of $q^{\mathrm{LG}}$ or $q^{\mathrm{geom}}$ lead to  equivalent descriptions in the following sense. The corresponding monomial representatives are related by non-linear field redefinitions. The corresponding $I$-functions are related by rational functions of the parameters they depend on and by a change of frame (gauge). This will be discussed in Section~\ref{sec:gkz}.

Let us illustrate this by a simple example. Consider the Fermat polynomial $W^T=y_1^8+y_2^8+y_3^8+y_4^8+y_5^2$. Interpreting this equation as the mirror of the degree $8$ hypersurface in $\mathbb{P}(11114)$ with $(h^{1,1}=1,h^{1,2}=149)$, the single complex structure deformation of the model can be represented by $m_1=y_1y_2y_3y_4y_5$. This corresponds to the interior point in the associated $N$-lattice polytope. On the other hand, interpreting $W^T$ in the Landau-Ginzburg setting, $m_1$ is not in the chiral ring but $m_2=y_1^2y_2^2y_3^2y_4^2$ is. From the geometric viewpoint $m_2$ is a point in a facet of the polytope and hence would be excluded \cite{MR1253648}. Obviously, $m_1$ and $m_2$ are related by a field redefinition $y_5\mapsto y_5+\alpha y_1y_2y_3y_4$ for some suitable $\alpha$. In our prescription $q^{\mathrm{ext}}$ would capture both, $m_1$ and $m_2$, while $q^{\mathrm{geom}}$ would take into account deformations by $m_1$ and $q^{\mathrm{LG}}$ would take into account deformations by $m_2$. Concretely, the matrices read
\begin{equation}
  \label{eq:119}
  \begin{aligned}
    q^{\ext} &=
    \begin{pmatrix}
      1 & 0 & -\frac{1}{8} & -\frac{1}{8} &-\frac{1}{8} &-\frac{1}{8} &-\frac{1}{2} \\
      0 & 1 & -\frac{1}{4} & -\frac{1}{4} &-\frac{1}{4} &-\frac{1}{4} & 0 \\
    \end{pmatrix},
    \\
    q^{\mathrm{geom}} &=
    \begin{pmatrix}
      1 & -\frac{1}{8} & -\frac{1}{8} &-\frac{1}{8} &-\frac{1}{8} &-\frac{1}{2} \\
    \end{pmatrix},
    \\
    q^{\mathrm{LG}} &=
    \begin{pmatrix}
      1 & -\frac{1}{4} & -\frac{1}{4} &-\frac{1}{4} &-\frac{1}{4} & 0 \\
    \end{pmatrix} \,.
  \end{aligned}
\end{equation}
For further examples we refer to Section~\ref{sec:apps}.

To summarize, we have gone a long way round to define the matrix $q$ encoding the action of $G$ on $\cH_{\mathrm{Mar}}$. Lacking a direct description of the marginal deformations in the $(a,c)$-ring (A-model after twist), we have used mirror symmetry to get a description of these deformations in terms of the $(c,c)$-ring of the mirror. Looking at~\eqref{eq:57} and~\eqref{eq:52}, the data encoding the deformations does not really depend on mirror symmetry as it only involves $W$, via $M$, and $G$. This suggests that a direct description of the marginal deformations in the $(a,c)$-ring should be possible. In particular, the nonlinear automorphisms of $W^T$ should have a counterpart in terms of additional symmetries among the unprojected twisted sectors $H_\gamma$ of the $(a,c)$-ring. It would be interesting to find such a description. 


In the following we refer to any choice of $q^{\ext}, q^{\mathrm{LG}}$, or $q^{\mathrm{geom}}$ by just $q$ and its rank by $h$. To conclude this section, we note two simple properties of $q$. First, multiplying the condition $M\spcheck q = 0$ by $M^{-T}$ from the right, the condition $J\spcheck v = 1$ in~\eqref{eq:52} implies the following relation:
\begin{equation}
  \label{eq:48}
  \sum_{j=1}^N q_{a,h+j} = -1, \quad a=1,\dots,h.
\end{equation}
Second, since $\overline{\rho}_{m}$ factors through $\tS\tL(\mathbb{V})$ then the action of, $g\spcheck\in G\spcheck$ on $y$ variables, also has determinant $1$. Therefore $v=(1,\ldots,1)^{T}\in \cA_{\mathrm{ext}}$ and so, for some $i_0$
\begin{equation}
  q_{i_0,h+j} = -q_{j},  \quad j=1,\dots,N. 
  \label{eq:39}
\end{equation}
We will take $i_0=1$.  

\subsection{$I$-function and Gamma class}
\label{sec:lggamma}

Given a Landau-Ginzburg orbifold $(W,G,\overline{\rho}_{m},\mathbb{C}_L^{\ast})$, we will propose a formula for the $I$-function and the Gamma class in terms of the matrix $q$. The formula is more general than the explicit expressions that can be found in the mathematics literature, as it covers the case with more than one K\"ahler parameter, and the orbifold group can be more general than $G=\langle J\rangle$ or $G=\mathrm{Aut}(W)$.

To define the $I$-function and the Gamma class, we will need another matrix, $L$, which encodes a certain lattice of periodicities determined by the group $G$. Based on this matrix, we can further define an integral matrix $\pC=(L\:\; S)$ such that $L^{-1}\pC=q$. If the Landau-Ginzburg orbifold arises as a phase of an abelian GLSM then the matrix $\pC$ is nothing but the matrix of U(1) charges, and the prescription to obtain $\pC$ is in some sense the ``inverse'' of an algorithm derived in \cite{Clarke:2010ep}, based on a criterion formulated in \cite{Herbst:2008jq}, to find Landau-Ginzburg phases of abelian GLSMs. See Section~\ref{sec:hpf} for further details about this algorithm.

Recall from the previous section that the matrix $q$ encodes the action of $G$ on the chiral ring $\cH$. We can reconstruct a group that is isomorphic to $G$ from it in the following way. Take all the $h\times h$ submatrices $B$ of $q$ and take any $B$ for which $|\mathrm{det}(B)|=\frac{1}{|G|}$. Then we set $L=(B^{-1})^T$. Now consider the Smith normal form of the matrix $L^T$: There are matrices $U,  V \in \tG\tL(h,\mZ)$ such that
\begin{equation}
  U L^T V  = D=\diag(d_1.\dots,d_h),
  \label{eq:5}
\end{equation}
where the $d_i$ satisfy $d_{i+1} \mid d_{i}, i=1,\dots,h-1$. The choice of $U,V$ is not unique, but will lead to equivalent results. 
The abstract way of thinking about the Smith normal form is that it yields a presentation of the group
\begin{equation}
  \label{eq:122}
  G_{\mathrm{orb}} := \mZ_{d_1} \times \dots \times  \mZ_{d_h}
\end{equation}
in terms of free abelian groups
\begin{equation}
  \label{eq:123}
  \shortexactseq{\mZ^h}{L^T}{\mZ^h}{}{G_{\mathrm{orb}}}.
\end{equation}
By construction of the matrix $q$ there is a choice of $L$ such that there exists an isomorphism of abelian groups
\begin{equation}
  \label{eq:124}
  F: G_{\mathrm{orb}} \cong G.
\end{equation}
In this way, we have reconstructed $G$ from $q$, up to isomorphism. We fix an isomorphism $F$ once and for all. The matrix $L$ will play a central role in the following. For practical purposes, we will choose an ordering of the elementary divisors such that the first factor $\mZ_{d_1} \subset G_{\mathrm{orb}}$ is identified under $F$ with the subgroup of $G$ generated by $J$, i.e.~$\langle J \rangle \subset G$. In all the examples we discuss, this is of order $d_1=d$, where $d$ is the degree of $W$ and hence coincides with our conventions of the Smith normal form. From now on, we will assume that this holds. 

We have denoted this group in~\eqref{eq:122} by $G_{\mathrm{orb}}$ because it is related to the unbroken gauge group of the associated GLSM with U(1) charge matrix $\pC$ given by $L^{-1}\pC=q$ in the corresponding Landau-Ginzburg orbifold phase.
We will see in examples that often different choices of $L$ are related by an integral change of basis, and hence are equivalent.
From a physics point of view, obtaining the GLSM from a Landau-Ginzburg model is highly non-trivial. In particular, one cannot expect the prescription to be unique, since the same Landau-Ginzburg model could arise from different GLSMs. In other words, an IR theory can have different UV completions. The conditions on the minors of $q$ ensures that the resulting GLSM does not have any gauge group elements that act trivially on the fields. The latter happens for Landau-Ginzburg orbifolds with massive fields, i.e.~when $n$ of the $\phi_j$ have $q_j=\frac{1}{2}$, and the matrix $q^{\mathrm{LG}}$. In this case, one may consider relaxing the condition on the minors $B$ above to $|\mathrm{det}(B)|=\frac{2^n}{|G|}$. One such example is the $q^{\mathrm{LG}}$ given in (\ref{eq:119}). Then the prescription below to determine the $I$-function and the Gamma class still works, but the meaningfulness of the GLSM needs further study\footnote{See Footnote~\ref{fn:LG-GLSM}.}. 

We can write~\eqref{eq:123} more explicitly as
\begin{equation}
  \label{eq:125}
  G_{\mathrm{orb}} = \mZ^h / L^T\mZ^h = \{k \in \mZ^h \}/\{ k\sim k+L^{T}m,\forall m\in \mZ^h \}.
\end{equation}
We can define an action of $G_{\mathrm{orb}}$ on $\mC^N$ by defining a representation $\widetilde \rho_m : G_{\mathrm{orb}} \to \mC^N$ by 
\begin{equation}
  \label{eq:126}
  \begin{aligned}
    \widetilde \rho_m( [k] ) &= \diag( \e{2\pi i (k^{T}q)_{1}},\dots,  \e{2\pi i (k^{T}q)_{N}}), \\
    (k^{T}q)_{j}&:=\sum_{a=1}^{h}k_{a}q_{a,h+j}. \qquad j= 1,\dots,N.
\end{aligned}
\end{equation}
The isomorphism $F$ then implies that
\begin{equation}
  \label{eq:128}
  \widetilde \rho_m = \overline{\rho}_m \circ F.
\end{equation}
Let $g_a$ be a (canonical) generator of $\mZ_{d_a}$, i.e.~$d_a$ is the smallest positive integer such that $g_a^{d_a} =1$. Then we can write an arbitrary element $\gamma \in G_{\mathrm{orb}}$ as
\begin{equation}
  \gamma = \prod_{m=1}^h g_a^{\ell_a}, \qquad (\ell_1,\dots,\ell_h) \in \cF.
  \label{eq:18}
\end{equation}
Here $\cF \subset \mZ^h$ is a fundamental domain for $L^T \mZ^h \subseteq \mZ^h$ in~\eqref{eq:123}, isomorphic as a set to $G_{\mathrm{orb}}$, so that
\begin{equation}
  k'_a \equiv \ell_a \mod d_a, \qquad a = 1,\dots, h \,.
  \label{eq:20}
\end{equation}
where we use the matrix $U$ to change to the basis of $\mZ^h$ in which the $G_{\mathrm{orb}}$-action is diagonal:
\begin{equation}
  \label{eq:127}
  k':=Uk\in\mathbb{Z}^{h}.
\end{equation}
In the examples we will make the following choice for $\cF$:
\begin{equation}
  \cF = \big\{ \ell=(\ell_1,\dots,\ell_h) \mid \ell_1 \in  \{ 1,\dots, d\}, \ell_a \in  \{ 0,1,\dots,d_a-1\} , a=2,\dots, h \big\}.
  \label{eq:19}
\end{equation}
We will often identify $\gamma = F([k]) \in G$ with $\ell=(\ell_1,\dots,\ell_h) \in \cF$ and write $\ell$ as an index instead of $\gamma$. Note in particular that the summation of $\ell_1$ labeling the elements of $\langle J\rangle$ is chosen to start with $1$. Under these conventions, the labelling $(\ell_1,\dots,\ell_h)$ coincides with the labels of the twisted sectors $\cH_{\gamma^{-1}}^{(c,c)}$ or, by~(\ref{acccflow}), of $\cH_{\gamma^{-1}J}^{(a,c)}$. This will be practical in relation to the twisted sectors of the state space of FJRW theory in Section~\ref{sec:some-details} and the examples in Section \ref{sec:apps}.

Consider a narrow sector $\mathcal{H}_{\gamma}^{(a,c)}$, i.e. $n_{J^{-1}\gamma} = 0$, and recall that this implies $\mathrm{dim}(\mathcal{H}_{\gamma}^{(a,c)})=1$. 
This is the only situation that will be considered in the following. 
Then $\mathcal{H}^{(a,c)}_\gamma$ is canonically generated by a vector $\mathbf{e}^{(a,c)}_\gamma$. We choose coordinates $u=(u_1,\dots,u_h)$ on the space of narrow marginal deformations $\cH^{(a,c)}_{-1,1} \cap \cH_{narrow} \cong \mC^h$.  We introduce the function $ I_{\gamma^{-1}J}(u) = I_{\ell}(u)$ 
with 
\begin{equation}
  I_{\ell}(u) = -\sum_{\substack{k_1,\dots,k_h \geq 0\\k'_a \equiv
      \ell_a \mod d_a}} 
    \frac{ u^k}{\prod_{a=1}^h\Gamma(k_a+1)}
    \prod_{j=1}^N\frac{(-1)^{\langle-\sum_{a=1}^h
      k_a q_{a,h+j}+q_{j}\rangle}\Gamma(\langle\sum_{a=1}^h
      k_a q_{a,h+j}-q_{j}\rangle)}{\Gamma(1+\sum_{a=1}^h
      k_a q_{a,h+j}-q_{j})}.
    \label{eq:21}
\end{equation}
The sum in $I_{\ell}(u)$ can be understood as a sum over all nonnegative $h$-tuples of integers $k\in(\mathbb{Z}_{\geq 0})^{h}$ such that the class $[k']=[Uk]\in G_{\mathrm{orb}}$ satisfies $[k'_{a}]=[\ell_a]$. 
We have set $u^k = \prod_{a=1}^h u_a^{k_a}$. We will view $\mC^h$ as a local coordinate neighborhood in $\mathcal{M}_K$ near the Landau-Ginzburg point. We will relate them to the marginal deformation parameters in Section~\ref{sec:flat}. We define
\begin{equation}
  \label{eq:22}
  I_{LG}(u) = \sum_{\gamma \in G} I_{\gamma}(u) \mathbf{e}^{(a,c)}_{\gamma }.
\end{equation}
This is our proposal for the $I$-function of a Landau-Ginzburg orbifold $(W,G,\overline{\rho}_{m},\mathbb{C}_L^{\ast})$. The $I$-function satisfies a system of GKZ differential equations and the $I_{\ell}(u)$ transform diagonally under the $G$-action. For details, see Section \ref{sec:gkz}.

In Section \ref{sec:fjrw} we will see that this is consistent with the $I$-function defined in FJRW theory. In Section \ref{sec:hpf} we will recover it from the hemisphere partition function of the associated GLSM. In order to make contact with these results, it is useful to rewrite this expression. We apply the reflection formula for the Gamma functions in the numerator of~\eqref{eq:21} in the form
\begin{equation}
  \label{eq:9}
  \Gamma(z) = \frac{2\pi i \e{-i\pi z}}{1-\e{-2\pi i z}}\frac{1}{\Gamma(1-z)}
\end{equation}
and obtain
\begin{equation}
  \label{eq:10}
  \begin{aligned}
  I_\ell(u) &= -(2\pi i)^N 
    \sum_{\substack{k_1,\dots,k_h \geq 0\\k'_a \equiv
      \ell_a \mod d_a}} 
    \frac{u^k}{\prod_{a=1}^h\Gamma(k_a+1)} \prod_{j=1}^N
    \frac{1}{\Gamma(1+\sum_{a=1}^h
      q_{a,h+j}k_a-q_{j})} \\
    &\phantom{=} \cdot \frac{1}{1-\e{-2\pi i (
      \sum_{a=1}^hq_{a,h+j}k_a-q_{j})}}\frac{1}{\Gamma(1-\langle
      \sum_{a=1}^hq_{a,h+j}k_a-q_{j}\rangle) } \,.
  \end{aligned}
\end{equation}
Note that the expression $\left\langle\sum_{a=1}^h
     k_a q_{a,h+j}-q_{j}\right\rangle=\left\langle
     (k^{T} q)_{j}-q_{j}\right\rangle$ 
depends on $k$ only through $\ell$ since by~\eqref{eq:5} we can write $k=U^{-1}\ell +U^{-1}Dm$ for some $m\in\mathbb{Z}^{h}$. Then the expression inside the angle brackets equals $((U^{-1}\ell)^Tq)_j \mod \mZ$. 

To finalize this section, we define a provisional Gamma class for Landau-Ginzburg orbifolds. It is defined by the operator
\begin{equation}
  \widehat \Gamma^*_{W,G} : \mathcal{H}^{(a,c)} \to \mathcal{H}^{(a,c)} \label{eq:40},
\end{equation}
where $\widehat \Gamma^*_{W,G}$ acts diagonally and its eigenvalue on $\mathcal{H}^{(a,c)}_\gamma$ is 
\begin{equation}
  \widehat \Gamma^*_{W,G}\mathbf{e}_{\gamma}^{(a,c)}=\widehat{\Gamma}_{\gamma}\mathbf{e}_{\gamma}^{(a,c)}\qquad \hat{\Gamma}_{\gamma}=\prod_{j=1}^N\Gamma\left(1-\left\langle\sum_{a=1}^h
      k_a q_{a,h+j} - q_j\right\rangle \right).\label{lggamma}
\end{equation}
By the above reasoning the eigenvalue only depends on $\gamma = F([k])$. As for the $I$-function, we will often write $\widehat\Gamma_{\gamma^{-1}J} = \widehat\Gamma_{\ell}$. We will argue in Section~\ref{sec:central-charge-fjrw} that this operator represents the action of the Gamma class, as discussed in Section~\ref{sec:cc-scft}. Then we can apply this operator to $I_{LG}$ and obtain
\begin{equation}
  \label{eq:12}
  \begin{aligned}
  \widehat \Gamma^*_{W,G} (I_{LG})(u) &= -(2\pi i)^N \sum_{\gamma \in G}
    \sum_{\substack{k_1,\dots, k_h \geq 0\\ k_a' \equiv \ell_a \mod d_a}} 
    \frac{u^k}{\prod_{a=1}^h\Gamma(k_a+1)}
    \prod_{j=1}^N \frac{1}{\Gamma(1+\sum_{a=1}^h
      q_{a,h+j}k_a-q_{j})} \\
    &\phantom{=} \cdot \frac{1}{1-\e{-2\pi i (
      \sum_{a=1}^hq_{a,h+j}k_a-q_{j})}} \mathbf{e}^{(a,c)}_{\gamma}.
  \end{aligned}
\end{equation}

\subsection{The central charge formula}
\label{sec:centr-charge-form}

According to our proposal~\eqref{ccfinalf} in Section~\ref{sec:cc-scft}, provided $I_{LG}(u)$ is related to the $J$-function by a change of frame and coordinates (to be made precise in Section \ref{sec:flat}), the central charge function in Landau-Ginzburg orbifolds should have the form 
\begin{equation}
  \label{eq:41}
      Z_{LG}(\overline{\cB},u) = \langle \widehat \Gamma^*_{W,G}(I_{LG}(u))|
    \chern(\overline{\cB}) \rangle
  \end{equation}
with the pairing defined in~\eqref{eq:121}, and where $\chern(\overline{\cB})$ and $\widehat \Gamma^*_{W,G}(I_{LG}(u))$ are given in~\eqref{chernlg} and~\eqref{eq:12}, respectively. Substituting these expressions yields
\begin{equation}
  \begin{aligned}
    Z_{LG}(\overline{\cB},u) 
    &= \sum_{\gamma,\gamma' \in G} \widehat{\Gamma}_{\gamma} I_\gamma(u)
    \chern_{\gamma'}(\overline{\cB}) \langle \mathcal{U}_{(1,0)}\circ \mathbf{e}^{(a,c)}_\gamma, \mathbf{e}^{(c,c)}_{(\gamma')^{-1}} \rangle_{S^{2}}\\
    &= \frac{1}{|G|}\sum_{\gamma \in G} \widehat{\Gamma}_{\gamma} I_{\gamma}(u)
    \chern_{\gamma J^{-1}}(\overline{\cB}) \\
    &= \frac{1}{|G|}\sum_{\ell \in \cF} \widehat{\Gamma}_{\ell} I_{\ell}(u) \str_{\overline M}\overline\rho(J^{\ell_1-1} \prod_{a=2}^{h}
    g_a^{\ell_a}) \\
    &= -\frac{(2\pi i)^N}{|G|} \sum_{\ell\in \cF} \sum_{\substack{k_1,\dots,k_h \geq 0\\k'_a =
      \ell_a \mod d_a}} 
    \frac{u^k}{\prod_{a=1}^h\Gamma(k_a+1)}
    \prod_{j=1}^N\frac{1}{\Gamma(1+\sum_{a=1}^h
      k_a q_{a,h+j}-q_{j})} \\
    &\phantom{=} \cdot \frac{1}{1-\e{-2\pi i (
      \sum_{a=1}^hq_{a,h+j}k_a-q_{j})}} \str_{\overline M}\overline\rho(J^{\ell_1-1} \prod_{a=2}^{h}
    g_a^{\ell_a}). 
  \end{aligned}
 \label{eq:26}
\end{equation}
Here we used the identification through $F$ of the factor $\mZ_{d_1}$ in $G_{\mathrm{orb}}$ with the subgroup $\langle J\rangle$ of $G$, and exhibited the corresponding contribution to the supertrace.
If we use the fact that $\gamma = J^{\ell_1}\prod_{a=2}^{h} g_a^{\ell_a} =
J^{k'_1}\prod_{a=2}^{h} g_a^{k_a'}$ by~(\ref{eq:20}) to write the supertrace in terms of the $k_a'$, we can combine both
sums into a single sum over $k_i$ and obtain
\begin{equation}
  \begin{aligned}
    Z_{LG}(\overline{\cB},u)   &= -\frac{(2\pi i)^N}{|G|} \sum_{k_1,\dots,k_h \geq 0}
    \frac{u^k}{\prod_{a=1}^h\Gamma(k_a+1)}
    \prod_{j=1}^N\frac{1}{\Gamma(1+\sum_{a=1}^h
      k_a q_{a,h+j}-q_{j})}\\
    &\phantom{=} \cdot \frac{1}{1-\e{-2\pi i (
      \sum_{a=1}^hq_{a,h+j}k_a-q_{j})}} \str_{\overline M}\overline\rho(J^{k'_1-1} \prod_{a=2}^{h}
    g_a^{k'_a}). 
  \end{aligned}
  \label{eq:25}
\end{equation}
\subsection{Flat coordinates and $J$-function}
\label{sec:flat}

So far we have defined the function $I_{LG}(u)$, starting from the Landau-Ginzburg orbifold, which depends holomorphically in some formal coordinates $u$. We want to conjecture that the central charge function $Z_{LG}(\overline{\cB},u)$ we defined in terms of $I_{LG}(u)$ differs from the central charge $Z(\mathcal{B})$ (\ref{ccfinalf}) defined in terms of the $J$-function by a change of coordinates and frame as discussed in Section \ref{sec:frame}. In the following we will define such a change. For this, first the consider all the narrow sectors which correspond to marginal deformations in the $(a,c)$-ring, i.e.~the sectors $\mathcal{H}_{\gamma}^{(a,c)}$ that satisfy 
\begin{equation}
n_{\gamma J^{-1}}=0, \qquad -F_{L}(\mathcal{H}_{\gamma}^{(a,c)})=F_{R}(\mathcal{H}_{\gamma}^{(a,c)})=1.
\end{equation}
Denote these elements by $\phi^{(a,c)}_{a}$, $a=1,\ldots,h$, or their corresponding states by $\mathbf{e}^{(a,c)}_{a}$. Next, note that the state $\phi^{(a,c)}_{0}:=\phi^{(a,c)}_{\mathrm{id}}$ (or $\mathbf{e}^{(a,c)}_{0}$) is always narrow since $n_{J^{-1}}=0$ (no field has zero R-charge) and is the unique state with lowest charges $(q,\bar{q})=(0,0)$ in the $(a,c)$-ring. So we expect that $I_{LG}(u)$ always has a nonzero component $I_{0}\mathbf{e}_{0}^{(a,c)}$. We use this to define the functions
\begin{equation}
  \label{eq:flatcoord}
  t_{a}:=\frac{I_{\gamma_{a}}}{I_{0}}.
\end{equation}
These will be our flat coordinates, i.e.~these are the coordinates that will be identified with the exactly marginal deformations of the IR SCFT:
\begin{equation}
 t_{a}\int \mathcal{O}^{(1,1)}_{a} \qquad \mathcal{O}^{(1,1)}_{a}=\{ \overline{Q}_{-},[Q_{+},\phi^{(a,c)}_{a}]\}.
\end{equation}
The relations~(\ref{eq:flatcoord}) are expected to be invertible and hence we can define $u(t)$. Then, the $J$-function is defined, in terms of $I_{LG}(u)$, by
\begin{equation}
 \mathbf{J}_{LG}(t)=\frac{I_{LG}(u(t))}{I_{0}(u(t))}.
\end{equation}

\subsection{GKZ differential equations and monodromy}
\label{sec:gkz}
Now we show that the $I$-function satisfies a system of GKZ differential equations and explain its dependence on the various choices of the matrix $q$ introduced in Section~\ref{sec:deformations}. Furthermore we discuss its behavior under Landau-Ginzburg monodromy.

First, we give another interpretation of the data $q$ and $\cA$ defined in Section~\ref{sec:deformations}. The sets $\cA_{\mathrm{ext}}$, $\cA_{\mathrm{geom}}$ and $\cA_{\mathrm{LG}}$ are sets of integral points $\cA=\{v_1,\dots,v_p\}$ in $\mR^N$, where $p=h+N, h=\rank q$. Until further specification $q$ stands for any of $q^{\mathrm{ext}}, q^{\mathrm{geom}}$, or $q^{\mathrm{LG}}$. By construction these vectors span the lattice $\mZ^{N}$ and satisfy $v \cdot J\spcheck =1$ for all $v \in \cA$.  Moreover, the rows of the matrix $q$ generate the lattice $\mL \subset \mZ^h$ of linear relations among the elements of $\cA$,
\begin{equation}
  \label{eq:15}
  \mL := \{ l = (l_1,\dots,l_p) \mid \sum_{i=1}^p l_i v_i = 0\}.
\end{equation}
In fact, $\mL$ is isomorphic the lattice $L^T\mZ^h$ from~\eqref{eq:123}. The GKZ system with parameter $\beta=(\beta_1,\dots,\beta_N) \in \mC^N$ is a system of partial differential equations for functions $\Psi(u), u \in \mC^p$ given by the following equations~\cite{Gelfand:1989ab} (see also~\cite{Hosono:1993qy,Horja:1999ab}):
\begin{equation}
  \label{eq:36}
  \begin{aligned}
    \left(-\beta + \sum_{i=1}^p v_i u_i \pd{}{u_i} \right) \Psi &=0, \qquad \text{for } v_i \in \cA,\\
    \left( \prod_{i:l_i>0} \left(\pd{}{u_i}\right)^{l_i} -  \prod_{i:l_i<0} \left(\pd{}{u_i}\right)^{-l_i} \right) \Psi &= 0 , \qquad \text{for } l \in \mL\,.
\end{aligned}
\end{equation}
To show that $I_{\ell}(u)$ in (\ref{eq:10}) satisfies the GKZ system for the set $\cA$ with parameter $\beta=0$ we proceed as follows. First, since the expression in angle brackets appearing in $I_{\ell}(u)$ only depends on $\ell$ we can collect all the functions containing this expression into an overall constant $f_{\ell}$. Next, since the isomorphism $\mL \cong L^T\mZ^h$ is explicitly given by $l_i = \sum_aq_{ai}k_a$, the sum over $k$ can be written as a sum over $l\in\mL$. We shift $k_1 \mapsto k_1-1$ to remove the $-q_j$. Then, we can write $I_{\ell}(u_1,\dots,u_h) = \widetilde I_{\ell}(u_1,\dots,u_p)|_{u_{h+1}=\dots=u_{h+N}=1}$ with
\begin{equation}
  \label{eq:37}
  \begin{aligned}
    \widetilde I_{\ell}(u_1,\dots,u_p)
     &= f_\ell
     \sum_{\substack{k_1,\dots,k_h \geq 0\\k'_i \equiv
       \ell_i \mod d_i}}\prod_{a=1}^h 
     \frac{u_a^{\sum_{b} q_{ba}k_b}}{\Gamma(1+\sum_{b=1}^h q_{ba}k_b)} \prod_{j=1}^N
     \frac{u_{h+j}^{\sum_{b} q_{b,h+j}k_b}}{\Gamma(1+\sum_{b=1}^h
      q_{b,h+j}k_b)} \\
     &= f_\ell
    \sum_{l \in \mL \cap \cC}\prod_{i=1}^p 
    \frac{u^{l_i+\gamma_i}}{\Gamma(1+l_i+\gamma_i)}\,,
  \end{aligned}
\end{equation}
where $\gamma=0$ and $\cC \subset \mL_{\mR}$ is the cone corresponding to $(\mR_{\geq 0})^h\subset \mR^h$. By~\cite{Gelfand:1989ab}, this is a solution of~\eqref{eq:36}. This proves the claim about $I_{\ell}(u)$. It would be interesting to show that the $I$-function also satisfies a system of Picard-Fuchs differential equations.

The choice of the matrix $L$ in Section~\ref{sec:lggamma} corresponds to a choice of a regular triangulation of the convex hull $\conv\cA \subset \mR^N$ of $\cA$. Equivalently, this corresponds to a choice of a maximal cone $\cC$ in the secondary fan of the polytope $\conv\cA$. This cone will play a role in describing the Landau-Ginzburg phase of a GLSM in Section~\ref{sec:glsm-hemisph-part}.

Choosing $\mathcal{A}_{\mathrm{ext}}$ as the set to define the GKZ system, one also gets additional differential operators compared to those coming from $\mathcal{A}_{\mathrm{LG}}$, or $\mathcal{A}_{\mathrm{geom}}$. Besides those coming from the additional vectors $v_i$, there are additional first order differential operators of the form
\begin{equation}
 \label{eq:137}
  \left(\sum_{i_1,i_2=1}^p C^b_{i_1,i_2} u_{i_1}\pd{}{u_{i_2}} \right) \Psi =0, \qquad \text{for } b=1,\dots,p - N- h,\\
\end{equation}
that encode polynomial relations among the variables $u$ associated to the set $\mathcal{A}_{\mathrm{ext}}$. Here $p = |\cA_{\mathrm{ext}}|$, but $h = \rank q^{\mathrm{LG}} = \rank q^{\mathrm{geom}}$. If the subsets  $\mathcal{A}_{\mathrm{LG}}$ and $\mathcal{A}_{\mathrm{geom}}$ are different, then these relations correspond to polynomial relations among the monomial deformation parameters $u_a^{\mathrm{geom}}$ and $u_a^{\mathrm{LG}}$, respectively. These yield rational functions $u_a^{\mathrm{geom}} = u_a^{\mathrm{geom}} \big(u_1^{\mathrm{LG}},\dots, u_h^{\mathrm{LG}}\big)$, $a=1,\dots,h$.  In the context of periods in the geometric mirror B-model, these additional differential operators have been introduced and studied in~\cite{Hosono:1993qy,Hosono:1995bm}. As a consequence, the $I$-functions obtained from $q^{\mathrm{geom}}$ and $q^{\mathrm{LG}}$ will be related schematically as follows:
\begin{equation}
  \label{eq:136}
  I_{LG}( u^{\mathrm{LG}};  q^{\mathrm{LG}} ) = \frac{ I_{LG}( u^{\mathrm{geom}}( u^{\mathrm{LG}} );  q^{\mathrm{geom}} ) } { H(u^{\mathrm{LG}})},
\end{equation}
where $H(u)=H(u_1,\dots,u_h)$ is an invertible holomorphic function, and we have exhibited the dependence of the $I$-function on the matrix $q$ from which it is constructed. Hence, this leads to a change of frame as discussed in Section~\ref{sec:frame}. This is the fundamental reason to introduce the extended set of vectors $\cA_{\mathrm{ext}}$ and the corresponding matrix $q^{\mathrm{ext}}$. In the example given in (\ref{eq:119}) one finds $u^{\mathrm{LG}}=\frac{1}{4}\left(u^{\mathrm{geom}}\right)^2$ and $H=1$.

Finally, let us discuss the monodromy properties of $I_{LG}(u)$. Since $u$ are coordinates on $\cH^{(a,c)}_{-1,.1} \cap \cH_{narrow} = \bigoplus_{a=1}^h \cH^{(a,c)}_{\gamma_a}$, the quantum symmetry~\eqref{eq:118} induces an action of $G$ on them: $u_{\gamma_a} \mapsto \chi_{\gamma'}(\gamma_a ^{-1}) u_{\gamma_a}$. Under this action, we have $u^k \mapsto \chi_{\gamma'}(\prod_{a=1}^h \gamma_a^{ -k_a}) u^k$. In the sum~\eqref{eq:21}, such a term contributes to $I_{\gamma^{-1}J}(u)$ if $\prod_{a=1}^h \gamma_a^{ k_a} = \gamma$. Therefore, the action of $G$ induces an action $I_{\gamma^{-1}J}(u) \mapsto \chi_{\gamma'}(\gamma^{-1} ) I_{\gamma^{-1}J}(u)$. Moreover, $I_{LG}(u)$ in~\eqref{eq:22} is invariant. Therefore, in this basis of $\cH^{(a,c)}$, the action of the local monodromy about the Landau-Ginzburg point in $\cM_K$  on the $I$-function is diagonal. This is closely related to the Galois action discussed in~\cite{Chiodo:2012qt}.


%% file: section4.tex
\section{FJRW theory}
\label{sec:fjrw}
Fan-Jarvis-Ruan-Witten (FJRW) theory is the mathematical analog for Landau-Ginzburg orbifolds of what Gromov-Witten theory is for nonlinear sigma models with target space an almost complex, symplectic manifold. The essential ideas have been formulated in physics by Witten in~\cite{Witten:1991mk,Witten:1993ab} in the one variable case\footnote{See Sections 2 and 6 of~\cite{Dubrovin:1994hc} and Section 2.4 of~\cite{Dijkgraaf:1991qh} for a short review.}, and the corresponding mathematical theory has been worked out in~\cite{Fan:2007ba,Fan:2007vi} in the general case, again following ideas of Witten. The actual computation of the FJRW invariants for a Landau-Ginzburg orbifold with $\widehat{c}=3$ that corresponds to a compact Calabi-Yau threefold has been performed in~\cite{Chiodo:2008hv} for the quintic in $\mP^4$ and generalized to other cases with one K\"ahler parameter in~\cite{Chiodo:2012qt}. These authors have also shown that the FJRW invariants contain the same information as the Gromov-Witten invariants, as is expected from by Landau-Ginzburg/Calabi-Yau correspondence~\cite{Witten:1993yc}. For reviews of FJRW theory see also~\cite{Ruan:2012ab,Francis:2015ab}. 

\subsection{$W$-spin structures}
\label{sec:w-spin-structures}

Gromov-Witten theory is the description of topologically nontrivial holomorphic maps $\phi: C \to X$ from a Riemann surface $C$ to a symplectic manifold $X$. In physics, these are worldsheet instantons in a nonlinear sigma model with target $X$, i.e. they are solutions to the equations
\begin{equation}
  \label{eq:58}
  \bar \partial \phi_i = 0
\end{equation}
in local coordinates on $X$.

In a Landau-Ginzburg model with potential $W: \mC^N \to \mC$ whose critical points are non--degenerate the analogous topologically nontrivial field configurations are given by soliton solutions to the BPS equations~\cite{Fendley:1990zj,Cecotti:1992rm}
\begin{equation}
  \label{eq:59}
  \bar \partial \phi_i + \alpha \overline{\partial_{\phi_i}W(\phi)} = 0
\end{equation}
in coordinates $(\phi_1,\dots,\phi_N)$ on $\mC^N$ and with $|\alpha|=1$. In the present situation the critical point of $W$, however, is very degenerate.

Moreover, in the presence of a finite group $G$ acting on $\mC^N$ such that $(W,G)$ defines a Landau-Ginzburg orbifold the Riemann surface $C$ must carry an orbifold structure, too. Roughly speaking, the required orbifold structure is a $d$-spin structure, or more generally a $W$-spin structure, to be described below. In~\cite{Witten:1991mk} the solutions to~\eqref{eq:59} for Riemann surfaces with a $d$-spin structure 
and the potential $W=x^d$ have been studied. After generalization to  $W$-spin structures, this leads to a moduli space of ``maps'' from a Riemann surface $C$ of genus $g$ equipped with a $W$-spin structure to $\mC^N$ satisfying~\eqref{eq:59} and compatible with the action of $G$. More precisely, these ``maps'' will be sections of certain line bundles over $C$. For this purpose, we first need to understand the moduli space $\cW_{g,n}(W,G)$ of $W$-spin structures on $C$. The goal of this subsection is to give a brief description of $\cW_{g,n}(W,G)$. 

The basic idea is as follows:\footnote{For the sake of exposition we ignore many further technical details in the description below.} Let $\cM_{g,n}$ be the moduli space of complex Riemann surfaces (or complex algebraic curves) of genus $g$ with $n$ marked points $\sigma_1, \dots, \sigma_n$, and let $\Mgn{g}{n}$ be its Deligne-Mumford compactification obtained by adjoining singular curves with at most double points. This is the (compactified) moduli space of stable curves where stable means that the maps only admit finite automorphism groups. In Gromov-Witten theory one considers maps of such stable curves into a symplectic manifold or an algebraic variety $X$~\cite{Witten:1990hr}. 

In FJRW theory one instead starts with orbicurves. These are stable curves $\cC$ for which the marked points and the nodes (and only those) are allowed to be orbifold points. If the orbifold groups at these points are all subgroups of $\mZ_d$ for some $d$, $\cC$ is called $d$-stable. Such a curve is canonically equipped with the sheaf $\omega_{\cC,\log}$ which is the sheaf of logarithmic differential forms on $\cC$ with simple poles only at the marked points and the nodes. The next datum one needs is a $d$-spin structure on $\cC$ which, roughly speaking, is a $d$-th root of its (suitably twisted) canonical bundle. More precisely, it is an orbifold line bundle $\cL \to \cC$ together with an isomorphism $\varphi:\cL^{\otimes d} \cong \omega_{\cC,\log}$. If $d=2$ and $n=0$ this is an ordinary spin structure on the curve $\cC$. For $\cL$ to have integer degree, such a bundle only exists if $2g-2+n$ is divisible by $d$. When this condition is met, there are $d^{2g}$ choices of pairs $(\cL,\varphi)$ on $\cC$. The choice of an isomorphism class of $\cL$ determines locally a cover of $\Mgn{g}{n}$. But since the pair $(\cL,\varphi)$ has $\mZ_d$ as its isomorphism group, it may not exist globally over $\Mgn{g}{n}$. The reason is that it can happen that $\varphi$ ceases to be an isomorphism at a node. Indeed, it is argued in~\cite{Witten:1991mk} that globally the cover is ramified over the boundary of $\cM_{g,n}$ in $\Mgn{g}{n}$.

Since we allow the marked points to be orbifold points, the orbifold line bundle $\cL$ can have nontrivial monodromy by $\gamma(i) \in \mZ_d$ around each of the marked points $\sigma_i$, $i=1,\dots,n$, i.e. $\mZ_d$ acts on the fiber $\cL_{\sigma_i}$ by $M_{\sigma_i}(\cL_{\sigma_i}) = \gamma(i)\cL_{\sigma_i}$. Therefore, we need to specify a collection of elements $(\gamma(1),\dots,\gamma(n)) \in (\mZ_d)^n$, or if we write $\gamma(i) = \exp(2\pi i \frac{m_i}{d})$, a collection of integers $0 \leq m_i \leq d-1$, $i=1,\dots,n$. We will see in Section~\ref{sec:brief-guide-fjrw} that the theory has a very different behaviour depending on whether $\gamma(i) \not = 1$ for all $i$ or $\gamma(i)=1$ for some $i$. 

A reformulation of the previous discussion is that this data can be used to define a map from the $d$-stable curve to the (topologically twisted) Landau-Ginzburg orbifold with potential $W(\phi) = \phi^d$ and orbifold group $G=\mZ_d$. The relation to the choice of the pair $(\cL,\varphi)$ is roughly as follows. Before the topological twist, the field $\phi$ is a scalar with $U(1)_V$ R-charge $q=\frac{2}{d}$. After the A-twist, it becomes a section of $\cL = \omega_{\cC}^q$ and hence $W(\phi)$ a section of $\omega_{\cC}$, where $\omega_{\cC}$ is the canonical sheaf. Therefore, formally replacing $\phi$ by $\cL$ in $W(\phi)$ yields a line bundle $W(\cL)$ on $\cC$, satisfying $W(\cL) \cong \omega_{\cC}$. Taking into account the marked points, one expects an isomorphism $\varphi: W(\cL) = \cL^d \cong \omega_{\cC,\log}$ to the sheaf of logarithmic differential forms.
In~\cite{Witten:1991mk} the equivalent descriptions in terms of a topologically twisted $\cN=2$ $SU(2)/U(1)$ Kazama-Suzuki model or in terms of a topologically twisted gauged WZW model on $SU(2)$ with gauge group $U(1)$, either of them coupled to topological gravity, are considered. In the context of the gauged WZW model, if $\cC$ is a smooth curve, the bundle carrying the $U(1)$ connection is the line bundle $\cL$. The choice of $\gamma(i)$ describes the choice of a flat connection on $\cL$, up to the action of $G$.  Ultimately, one is interested in the correlation functions with insertions of chiral primary fields $\alpha_i$ at the marked points $\sigma_i$. From the point of view of the Landau-Ginzburg orbifold, the choice of $\gamma(i)$ corresponds to the choice of a twisted sector in the chiral ring $\cH^{(a,c)}$ for the chiral primary $\alpha_i$, which is labelled by $\gamma(i)$.

The latter formulation is the starting point for a generalization to a large class of Landau-Ginzburg orbifolds.
For a general Landau-Ginzburg potential, i.e. a non-degenerate, quasihomogeneous polynomial $W \in \mC[x_1,\dots,x_N]$ as in Section~\ref{sec-lgoclosed} (cf.~\eqref{eq:56}),
\begin{equation}
  W(x_1,\dots,x_N) = \sum_{\alpha=1}^\nu c_\alpha W_\alpha = \sum_{\alpha=1}^\nu c_\alpha \prod_{j=1}^N x_j^{M_{j\alpha}} \,,
  \label{eq:65}
\end{equation}
we choose a line bundle $\cL_j$ on $\cC$ for each variable $x_j$, $j=1,\dots,N$ and an isomorphism
\begin{equation}
  \varphi_\alpha : W_\alpha(\cL_1,\dots,\cL_N) := \bigotimes_{j=1}^N \cL_j^{\otimes M_{j\alpha}} \cong \omega_{\cC,\log} \,
\end{equation}
for each monomial $W_\alpha$, $\alpha=1,\dots,\nu$.  This collection of line bundles and isomorphisms $(\cL_j,\varphi_\alpha)_{1 \leq j \leq N, 1 \leq \alpha \leq \nu}$ is called a $W$-spin structure on $\cC$ if the $\varphi_\alpha$ satisfy certain compatibility conditions depending on the choice of the group $G$. These will be described shortly.

Again, we have to specify the monodromy $\gamma_j(i)$ of the line bundle $\cL_j$ at the marked point $\sigma_i$ for all $j=1,\dots,N$ and $i=1,\dots,n$. A priori, we only know that $\gamma_j(i) \in \mZ_m$ for some $m\in \mN$. It is an amazing fact~\cite{Fan:2007ba} that $\gamma(i) = \diag(\gamma_1(i),\dots,\gamma_N(i))$ acting on $\mC^N$ defines an automorphism of $W$. In fact, the orbifold structure of $\bigoplus_{j=1}^N \cL_j$ is completely parametrized by $\Aut(W)$, the group of all diagonal automorphisms of $W$, cf.~\eqref{orbdef}. Therefore we are actually working with a Landau-Ginzburg model orbifolded by $\Aut(W)$. Recall from Section~\ref{sec-lgoclosed} that we have the canonical grading element $J = \diag(\exp(2\pi i q_1),\dots,\exp(2\pi i q_N)) \in \Aut(W)$, where $(q_1,\dots,q_N)$ are the (normalized) weights of $(x_1,\dots,x_N)$, making $W$ quasihomogeneous.
It turns out that one can define new $W$-spin structures by restricting to any subgroup $G \subset \Aut(W)$, as long as $G$ contains the grading element $J$. The reason for this condition will be given below. However, this implies that the isomorphisms $\varphi_\alpha$ must satisfy certain compatibility conditions as alluded to above. These compatibility conditions are obtained as follows. One adds to $W$ any polynomial $W_G$ such that $W+W_G$ is a non-degenerate, quasihomogeneous polynomial of the same weights as $W$ and that $\Aut(W+W_G) \cong G$. Then the compatibility conditions are $W'(\cL_1,\dots,\cL_N) \cong \omega_{\cC,\log}$ for all monomials $W'$ in $W_G$. By~\cite{Fan:2007ba}, the resulting $W$-spin structure is independent of the choice of $W_G$. As an example, consider $G= \langle J\rangle$. In this case, a $d$-spin structure $(\cC,\cL,\varphi)$ gives rise to a $W$-spin structure by setting $\cL_j = \cL^{\otimes q_jd}$, $j=1,\dots, N$, see~\cite{Chiodo:2012qt}. In general, a $W$-spin structure does not necessarily come from a $d$-spin structure in this way.

We set $\gammabar= (\gamma(1),\dots,\gamma(n)) \in G^n$. Given the data $W,G,g,n$ and $\gammabar$, the moduli space (more accurately, moduli stack) of $W$-spin structures is defined in~\cite{Fan:2007ba} as:
\begin{equation}
  \begin{aligned}
    \cW_{g,n}(W,G)(\gammabar) &= \Big\{ (\cC;\sigma_1,\dots,\sigma_n;\cL_1,\dots,\cL_N;\varphi_1,\dots,\varphi_{\nu_G}) \Big| \\
    &\varphi_\alpha : W_\alpha(\cL_1,\dots,\cL_N) \cong \omega_{\cC,\log} \text{ for every } \alpha = 1,\dots, \nu_G, \\
    &M_{\sigma_i}(\cL_j) = \gamma_j(i) \text{ for all } j=1,\dots,N, i=1,\dots,n \Big\} / \sim \,,
\end{aligned}
\end{equation}
where $ M_{\sigma_i}(\cL_j)$ is the monodromy of the line bundle $\cL_j$ at the marked point $\sigma_i$ and $\nu_G$ is the number of monomials of $W+W_G$ with $W_G$ as above. We also set
\begin{equation}
   \cW_{g,n}(W,G) = \bigsqcup_{\gammabar \in G^n} \cW_{g,n}(W,G)(\gammabar) \,.
\end{equation}
This moduli space comes with a natural map to the moduli space of stable curves
\begin{equation}
  \mathrm{st}: \cW_{g,n}(W,G) \to \overline{\cM}_{g,n} 
  \label{eq:75}
\end{equation}
which forgets the data $(\cL_j,\varphi_\alpha)_{1 \leq j \leq N, 1 \leq \alpha \leq \nu}$ of the $W$-spin structure. 
In~\cite{Fan:2007ba} it is shown that $\cW_{g,n}(W,G)$ is a finite cover of $\Mgn{g}{n}$ if $2g-2+n>0$. In particular, it is smooth and compact. This cover is the one of~\cite{Witten:1991mk} reviewed at the beginning of this subsection. For the compactness, it is essential that the grading element $J$ is contained in $G$. An explicit description of $\cW_{g,n}(W,G)$ for the Fermat quintic $W= \sum_{i=1}^5 x_i^5$ and $G=\langle J\rangle$ has been worked out in~\cite{Chiodo:2008hv}. 

Note that for the underlying physical theory to admit a topological A-twist, the $\tU(1)_V$ symmetry must be preserved.  This leads to a selection rule which translates into a condition for $\cW_{g,n}(W,G)(\gammabar)$ to be non-empty. It is empty unless
\begin{equation}
  \label{eq:76}
  \gamma(1)\cdot\dots\cdot\gamma(n) = J^{2g-2+n}.
\end{equation}
This condition is equivalent to the requirement that the degree of $|\cL_j|$,  $\deg|\cL_j| = q_j(2g-2+n) - \sum_{i=1}^n \theta_j^{\gamma(i)}$, is an integer. Here $|\cL_j|$ is the pushforward of $\cL_j$ on the orbicurve $\cC$ to the underlying coarse curve $C$. There is also a selection rule coming from the absence of $\tU(1)_A$ anomaly which will be reviewed in the next subsection.

\subsection{A brief guide to FJRW theory}
\label{sec:brief-guide-fjrw}

The moduli spaces $\cW_{g,n}(W,G)$ are the analogs for Landau-Ginzburg orbifolds $(W,G)$ of the moduli spaces of stable maps $\overline{\cM}_{g,n}(X,\beta)$ in Gromov-Witten theory where $(W,G)$ plays the role of $X$ while there is no analog for $\beta \in \tH_2(X,\mZ)$. The fact that $\cW_{g,n}(W,G)$ is smooth and a finite cover of $\overline{\cM}_{g,n}$ makes it much more tractable than $\overline{\cM}_{g,n}(X,\beta)$. There is a slight difference, though. While the points in $\overline{\cM}_{g,n}(X,\beta)$ automatically satisfy~\eqref{eq:58}, we still need to impose~\eqref{eq:59} on sections of the $\cL_j$ corresponding to points in $\cW_{g,n}(W,G)$. 
For this reason, the evaluation of the correlation functions is still hard.

In Gromov-Witten theory the correlation functions are obtained by integrating certain cohomology classes against the fundamental class of $\overline{\cM}_{g,n}(X,\beta)$. Since this is not smooth in general, the ordinary fundamental class does not exist, and has to be replaced by a so-called ``virtual'' fundamental class. In the present case, even though $\cW_{g,n}(W,G)$ is smooth, the presence of $\cL_j $ leads to an obstruction and one still requires an analog of this virtual class in order to define the analogs of the Gromov-Witten classes,~i.e. the cohomology classes over which one has to integrate to get the invariants. For the moduli space of $d$-spin structures, such an analog has been constructed in~\cite{Jarvis:2001ab} and~\cite{Polishchuk:2001bc}, following a formal argument (for $W(x) = x^d, G=\mZ_d$) formulated by Witten in~\cite{Witten:1993ab}.  

Based on this argument, an analytic construction of the virtual cycle  $[\cW_{g,n}(W,G)]^{\vir} \in \tH_*(\cW_{g,n}(W,G))$ was given in~\cite{Fan:2007ba} for a general polynomial and group. It satisfies certain key properties and axioms such that the set of correlations functions defines a cohomological field theory, called FJRW theory, in the sense of Kontsevich and Manin~\cite{Kontsevich:1994qz} (cf. also~\cite{Francis:2015ab}), on the space $\cH_{\FJRW}(W,G)$ of chiral primary fields of the Landau-Ginzburg orbifold $(W,G)$. FJRW theory is then intersection theory on $\cW_{g,n}(W,G)$, generalizing the case of topological gravity~\cite{Witten:1990hr}. 
In the following, we are going to outline some of the main ideas.

We wish to compute the genus $g$ correlation functions with $n$ insertions
\begin{equation}
  \label{eq:67}
  \langle \tau_{a_1}(\alpha_1) \tau_{a_2}(\alpha_2) \dots \tau_{a_n}(\alpha_n) \rangle_{g,n} \,,
\end{equation}
where $\alpha_i$ are chiral primary fields in $\cH^{(a,c)}$, which is the state space $\cH_{\FJRW}$ in FJRW theory, to be discussed in Section~\ref{sec:some-details}. We recall that the operators $\tau_{a_i}$ for the gravitational descendants~\cite{Witten:1989ig} are defined as follows. Associated with each marked point $\sigma_i$, there is a natural line bundle $L_i$ on $\Mgn{g}{n}$ whose fiber over the point $(C,\sigma_1,\dots,\sigma_n)$ is the cotangent space to $C$ at $\sigma_i$. Its first Chern class is usually denoted by $\psi_i \in \tH^2(\Mgn{g}{n})$. Pulling these back to $\cW_{g,n}(W,G)$ by the natural map~\eqref{eq:75} we get classes in $\tH^2(\cW_{g,n}(W,G))$ which we will also denote by $\psi_i$. The gravitational descendants play an essential role in the definition of the $J$-function (cf. Section~\ref{sec:cc-scft}). The properties of the $J$-function then allow for a computation of the correlation functions in genus zero, as we will discuss in Section~\ref{sec:i-j-function}. 

In~\cite{Witten:1991mk} an argument to compute (\ref{eq:67}) was given for the topologically twisted gauged WZW model corresponding to $W(\phi)= \phi^d$. There should be a formulation in terms of A-twisted Landau-Ginzburg theory for an arbitrary superpotential $W(\phi_1,\dots,\phi_N)$ coupled to topological gravity. To our knowledge, this has not been done and we will not perform a detailed discussion here. We restrict ourselves to outlining the required steps to compute correlation functions since this will motivate the results of FJRW theory from a physics perspective. After the A-twist the fermions (not to be confused with the gravitational descendants $\psi_i$) take values in
\begin{equation}
  \begin{aligned}
  \psi_{+}^j&\in C^\infty (\mathcal{C},\cL_j\otimes\overline{\omega}_{\mathcal{C},\log}), &{\overline\psi}_{+}^{\jbar}&\in C^\infty (\mathcal{C},\overline {\cL_j}), \\
  \psi_{-}^j&\in C^\infty (\mathcal{C},\cL_j ), &{\overline\psi}^{\jbar}_{-}&\in C^\infty (\mathcal{C},\overline{\cL_j}\otimes{\omega}_{\mathcal{C},\log}),
\end{aligned}
\end{equation}
for $j,\jbar = 1,\dots,N$. 
This assignment is consistent with terms in the action of the form $\psi_+^j\psi_-^k\partial_j\partial_kW$ which is necessary for making a Landau-Ginzburg orbifold A-twistable \cite{Guffin:2008kt}. The coupling to gravity should work by making use of the standard Noether procedure. One expects to end up with an action $S[\phi_j,\psi_{\pm}^j,\overline\psi_{\pm}^{\jbar},h_{\alpha\beta},\chi_{\alpha\beta}]$, where $h_{\alpha\beta}$ is the metric on $\cC$ and $\chi_{\alpha\beta}$ is the gravitino. Following standard arguments \cite{Witten:1991mk,Cordes:1994fc,Hori:2003ic}, the path integral with insertions of gravitational descendants $\psi_i$ and states $\alpha_i$  is expected to reduce to
\begin{equation}
  \label{eq:61}
  \int_{\cM_{\textrm{BPS}}}  \ch_{\mathrm{top}}(\rE) \prod_{i=1}^n \psi_i^{a_i} \prod_{i=1}^n \alpha_i.
\end{equation}
The integral over $\cM_{\textrm{BPS}}$ is determined by the fact the path integral  localizes on the fermionic zero modes. The vanishing locus of the variation of the fermions, $Q\psi_{\pm}^j=0$, is the space of solutions to the BPS equation~\eqref{eq:59}. By general arguments \cite{Cordes:1994fc}, we expect that $\ch_{\mathrm{top}}(\rE)$ represents the (generalized) Euler class of the (generalized) bundle of fermionic zero modes that arises from evaluating the path integral of the two- and four fermion terms in the action. A mathematical definition will be given below. In general, it is not known how to evaluate this path integral since a representation of the states $\alpha_i$ in the terms of the fields $\phi_j$ is lacking. In the special case $W=x^d, G=\mZ_d$, the path integral could be evaluated in~\cite{Witten:1991mk} using the equivalent formulation in terms of the topologically twisted gauged WZW model.

Now we give a description of $\cM_{\mathrm{BPS}}$ in the framework of $W$-spin structures. Given $(\cC,\sigma_i,\cL_j,\varphi_k) \in \cW_{g,n}(W,G)(\gammabar)$, the BPS equation~\eqref{eq:59} describing the fixed point of the fermionic symmetry is viewed as a system of PDEs for smooth sections $s_j \in C^\infty(\cC,\cL_j)$ of $\cL_j$, $j=1,\dots,N$ (we choose $\alpha=1$ here):
\begin{equation}
  \label{eq:90}
  \bar\partial s_j + \overline{\partial_j W(s_1,\dots,s_N)} = 0, \qquad j=1,\dots,N\,.
\end{equation}
These equations, called Witten equation in~~\cite{Fan:2007ba}, make sense under the isomorphisms $\varphi_k$ and a suitable choice of a Hermitian metric on $\cL_j$. The latter is needed since $\bar\partial s_j \in \Omega^{0,1}(\cL_j)$ while $\overline{\partial_j W} \in C^\infty(\cC,\overline{\omega}_{\cC,\log}\otimes\overline{\cL}^{-1}_j)$ and these two spaces can be identified via such a metric. The Witten equation should be viewed as the counterpart in LG theory to the Cauchy-Riemann equation $\bar\partial_J u = 0$ for a smooth map $u:C \to X$ in Gromov-Witten theory. 

When all the marked points $\sigma_i$ correspond to the narrow sector, i.e. when $\gamma_j(i) \not= 1$ for all $i$ and $j$, the zero sections are the only solutions to the BPS equations. In a broad sector, however, we have fields other than the vacuum satisfying untwisted boundary conditions $\phi_j(e^{2\pi i}z) = \gamma(i)\phi_j(z)$ with $ \gamma(i)=1$ (see Section~\ref{sec-lgoclosed}). After perturbing $W$ into a holomorphic Morse function, i.e.~such that its critical points become nondegenerate, the vacua corresponding to these untwisted fields allow for nontrivial solitonic solutions~\cite{Fan:2007vi} of the type studied in~\cite{Cecotti:1992rm}.

In~\cite{Fan:2007ba} the moduli space of solutions to the BPS equations (or Witten equations) for a fixed point in $\mathcal{W}_{g,n}$ was defined as 
\begin{equation}
 \{ (s_1,\dots,s_N) \in C^\infty(\cC,\bigoplus_{j=1}^N \cL_j) \mid \bar\partial s_j + \overline{\partial_j W(s_1,\dots,s_N)} = 0, \; j=1,\dots,N \} /\sim\,,
\end{equation}
where $\sim$ takes into account automorphisms of the sections. This describes a fiber in the moduli space $\mathcal{M}_{\textrm{BPS}}$ in (\ref{eq:61}).

It was shown that, after perturbing $W$ into a holomorphic Morse function, the space of solutions defines a homology class
\begin{equation}
  [\cW_{g,n}(W,G)(\gammabar)]^{\vir} \in\tH_*(\cW_{g,n}(W,G)(\gammabar)) \otimes \bigotimes_i \left(\cH_{\gamma(i)}\right)\spcheck \,,
  \label{eq:92}
\end{equation}
called the virtual cycle, where $\cH_{\gamma(i)}$ are the twisted sectors, cf.~Section \ref{sec:some-details}. This class is independent of the perturbation and satisfies a number of key axioms that we will not spell out here. The real dimension of this cycle is
\begin{equation}
  \label{eq:77}
  2D(\gammabar) = 6(g-1) + 2n + 2 \sum_{j=1}^N\chi(|\cL_j|), 
\end{equation}
where $\chi(|\cL_j|)$ is the holomorphic Euler characteristic of $|\cL_j|$. 
Here the notion of a $W$-spin structure bears fruit. It provides a natural setting for studying the solutions of the BPS equations compatible with the action of the orbifold group $G$ on $W$. 
The fact that $\cW_{g,n}(W,G)$ is a finite cover of $\Mgn{g}{n}$ can be used to push forward the Poincar\'e dual of the virtual cycle to $\Mgn{g}{n}$. 
Given $\alpha_i \in \cH$, $i=1,\dots,n$, this procedure yields a cohomology class on $\Mgn{g}{n}$ defined as
\begin{equation}
  \Lambda^{W,G}_{g,n}(\alpha_1,\dots,\alpha_n) = \frac{|G|^g}{\deg \textrm{st}} \textrm{PD } \textrm{st}_* \left(  [\cW_{g,n}(W,G)(\gammabar)]^{\vir}  \cap \prod_{i=1}^n \alpha_i\right) \quad \in \tH^*(\cM_{g,n}),
  \label{eq:66}
\end{equation}
where PD stands for Poincar\'e dual. This class is nonzero only if $\alpha_i$ lies in $\cH_{\gamma(i)}$ and its real dimension is $2D(\gammabar) - \sum_{i=1}^N d^{\gamma(i)}$ where $d^{\gamma(i)}$ the dimension of the fixed point locus $(\mC^N)^{\gamma(i)}$ of $\gamma(i)$. The axioms of the virtual cycle then guarantee that the collection of classes $\Lambda_{g,n}^{W,G}$ satisfy the axioms of a cohomological field theory \cite{Fan:2007ba}. Ideally, one would like to have 
$[\cW_{g,n}(W,G)(\gammabar)]^{\vir} = \ch_{\mathrm{top}}(\rE) \cap [\cW_{g,n}(W,G)(\gammabar)]$ where is $\rE \to \cW_{g,n}$ is a vector bundle naturally associated to the line bundles $\cL_j$.

We remark that while the original description~\cite{Fan:2007ba} reviewed here was of analytic nature, there have been algebraic constructions of $[\cW_{g,n}(W,G)]^{\vir}$ in~\cite{Polishchuk:2016ab,Chang:2015ab,Kiem:2018ab} which are shown to be equivalent for narrow insertions. To our knowledge, the equivalence for insertions from broad sectors is known for the ADE potentials and still open in general.

Now, we turn to the description of $\ch_{\mathrm{top}}(\rE)$. This is a certain ``top Chern class'' of the push-forward by $\pi$ of the obstruction complex $\rE^\bullet$. The latter is a two-term complex $[\rE^0 \to \rE^1]$ of coherent sheaves over $\cW_{g,n}(W,G)$ built out of the universal line bundles $\rL_j \to \cW_{g,n}(W,G)$. Here, we denote the universal curve by $\pi: \rC \to \cW_{g,n}(W,G)$. Let $p = (\cC;\sigma_i,\cL_j,\varphi_k) \in \cW_{g,n}(W,G)$. Then the fiber of $\rC$ over $p$ is the curve $\rC_p = \pi^{-1}(p)=\cC$ and the fiber of $\rL_j$ over $p$ is $\rL_{j,p} = \cL_j$. The coherent sheaves $\rE^i$ are then defined as $\rE^i=\bigoplus_{j=1}^N \tR^i\pi_*\rL_j$ for $i=0,1$, where $\left(\tR^i\pi_*\rL_j\right)_p=\tH^i(\pi^{-1}(p),\rL_{j,p}) = \tH^i(\cC,\cL_j)$.  
The vector space $\tH^1(\cC,\cL_j)$ corresponds to the fermionic zero modes in the path integral and was denoted by $V$ in~\cite{Witten:1989ig}. In general, it can happen that these zero modes are not independent. The map in the complex describes the relations among them. More precisely, the first term comes from $\psi_+^i,\psi_-^{\overline{i}}$ zero modes, the second comes from $\psi_+^{\overline{i}},\psi_-^i$ zero modes.

In general, this ``top Chern class''  is difficult to construct. Since $\rE^\bullet$ is in general not a sheaf, but only a two-term complex of sheaves, the construction of an analog of the top Chern class, sketched in~\cite{Witten:1991mk} in terms of an index-theoretic construction, has not yet been formulated in general mathematically. There are two notable situations in which an effective method for computation of this class has been developed. These apply to narrow sectors of Fermat and chain polynomials, respectively, and will be summarized shortly.

We first discuss the special case when $\tR^\bullet\pi_*\cL_j$ is concave in genus $g$, i.e. when $\tH^0(\cC,\cL_j) = 0$ for every genus $g$ $W$-spin curve $\cC$. One can show~\cite{Chiodo:2012qt} that if $W = \sum_{j=1}^N x_{j}^{\frac{1}{q_j}}$ is a Fermat polynomial and $\gammabar$ consists only of narrow sectors, then $\tR^0\pi_*\rL_j = 0$ for every $j$.  Moreover, in this case $\tR^1\pi_*\rL_j$ is a vector bundle and the virtual class becomes
\begin{equation}
  [\cW_{0,n}(W,G)]^{\vir} = (-1)^N\prod_{j=1}^N \ch_{\Top}(\tR^1\pi_*\rL_j) \cap  [\cW_{0,n}(W,G)] \,.
  \label{eq:72}
\end{equation}
The concave situation is in fact used as an axiom that the virtual class has to satisfy~\cite{Fan:2007ba}. 

In general, even though neither of the terms in the complex $\rE^\bullet$ is a vector bundle, $\rE^\bullet$ can be replaced by a complex of vector bundles $[A \to B]$ that is quasi-isomorphic to $\rE^\bullet$~\cite{Polishchuk:2016ab}. This complex is not unique. The na{\"i}ve idea would be to define $\ch_{\Top}(\rE^\bullet) \sideset{"}{\!"}\eq \frac{\ch_{\Top}(B)}{\ch_{\Top}(A)}$, but this does not work at first since $\ch_{\Top}$ is not an invertible class in general. 
Instead, the following characteristic class is considered~\cite{Guere:2016ab}. Let $V \to X$ be a complex vector bundle of rank $r$ over a complex manifold $X$ with Chern roots $\alpha_1,\dots,\alpha_r$. Then we set
\begin{equation}
  \label{eq:98}
  \fc_t(V) = \prod_{k=1}^r \frac{\e{\alpha_k}-t}{\e{\alpha_k}-1}\alpha_k  \in \tH^*(X)[t], \qquad t\not =1. 
\end{equation}
Note that $\lim_{t\to 1} \fc_t(V) = \ch_{\Top}(V)$. More generally, for a K-theory class $V=[V_0 - V_1] \in K(X)$ we set $\fc_t(V) = \frac{\fc_t(V_0)}{\fc_t(V_1)}$, and the limit $t \to 1$ generally diverges. In the present context, viewing the complex $[A \to B]$ as $[B-A] \in K(\cW_{g,n}(W,G))$, the idea then is to set
\begin{equation}
  \label{eq:99}
  [\cW_{g,n}(W,G)]^{\vir} = \lim_{t\to 1} \fc_t([B - A]) \cap  [\cW_{g,n}(W,G)] \,.
\end{equation}
By means of the isomorphisms $\varphi_k: W_k(\rL_j) \cong \omega_{\rC,\log}$, this idea can be made precise for chain polynomials~\cite{Guere:2016ab}. The author shows, in the algebraic formalism of~\cite{Polishchuk:2016ab}, that the limit exists and the resulting virtual class does not depend on the choice of the complex $[A \to B]$. Therefore, one can define
\begin{equation}
  \label{nonconc}
  \ch_{\Top}(\rE^\bullet) := \lim_{t\to 1} \fc_t(\rE^\bullet) = \lim_{t\to 1} \fc_t([B - A]).
  \end{equation}
To summarize, the mathematical formulation of~\eqref{eq:61} is as follows. The correlation functions~\eqref{eq:67} are understood as multilinear maps from the state space $\cH(W,G)$, i.e. the chiral ring $\cH^{(a,c)}$, to the cohomology of the moduli space $\overline{\cM}_{g,n}$:
\begin{equation}
  \label{eq:55}
  \begin{aligned}
    \Lambda^{W,G}_{g,n} : \cH(W,G)^{\otimes n} &\to \tH^*(\cM_{g,n},\mC)\\
    (\alpha_1,\dots,\alpha_n) &\mapsto \Lambda_{g,n}(\alpha_1 \otimes \dots \otimes \alpha_n) = \langle \tau_{a_1}(\alpha_1) , \dots, \tau_{a_n}(\alpha_n) \rangle_{g,n}
\end{aligned}
\end{equation}
In general, given the virtual class $\Lambda_{g,n}^{\vir}$ in~\eqref{eq:66}, the FJRW invariants are defined as
\begin{equation}
  \label{eq:54}
  \langle \tau_{a_1}(\alpha_1) , \dots, \tau_{a_n}(\alpha_n) \rangle_{g,n}
  = \int_{\Mgn{g}{n}}  \Lambda_{g,n}^{W,G}(\alpha_1,\dots,\alpha_n) \cap \prod_{i=1}^n  \psi_i^{a_i}  \,.
\end{equation}
The absence of the $\tU(1)_A$ anomaly implies that
\begin{equation}
  \langle \tau_{a_1}(\alpha_1) , \dots, \tau_{a_n}(\alpha_n) \rangle_{g,n} \not = 0 \quad \text{ only if } \quad D(\gammabar) - \frac{1}{2} \sum_{i=1}^N d^{\gamma(i)} =  \sum_{i=1}^n a_i.
  \label{eq:93}
\end{equation}
These invariants ``count'' the number of solutions to the Witten equation~\eqref{eq:90} in a similar way as the Gromov-Witten invariants ``count'' the number of solutions to the Cauchy-Riemann equations, i.e. the number of holomorphic maps. In favorable circumstances, these can be computed explicitly. 

Equivalently, for a general CohFT with state space $\cH$, equipped with a symmetric nondegenerate bilinear form $(\cdot,\cdot)$ and a distinguished nonzero element $e_1$, we can define a generating function as
\begin{equation}
  \label{eq:68}
  F_g(t) = \sum_{n=0}^\infty \sum_{\substack{a_1,\dots,a_n\geq 0\\0 \leq k_1,\dots,k_n
      \leq M}} \langle \tau_{a_1}(e_{k_1}) \dots \tau_{a_n}(e_{k_n})
  \rangle_{g,n}  \frac{t_{a_1}^{k_1} \dots t_{a_n}^{k_n} }{n!},
\end{equation}
where $M = \dim\cH$ and $e_{1}, \dots, e_{M}$ is a basis for $\cH$ such that $e_1$ is the identity of the ring structure on $\cH$. The superscripts in $t_{a_i}^{k_i}$ are indices and not powers. 

The generating function $F_g$ of all the FJRW invariants in~\eqref{eq:54} satisfies the WDVV equation, the dilaton equation, the string equation and the topological recursion relations~\cite{Fan:2007ba}. Since the string equation will play a role in several places, we reproduce it here for completeness. With $t_a = \sum_{k=1}^M t_a^k e_k$ it reads
\begin{equation}
  \label{eq:100}
  \pd{F_g}{t^1_0} = \frac{1}{2} (t_0,t_0) + \sum_{b=0}^\infty \sum_{k=1}^M t_{b+1}^k \pd{F_g}{t_b^k} \,.
\end{equation}
The generating function $F_g$ hence has the same structure as topological gravity and Gromov-Witten theory~\cite{Witten:1990hr}. Therefore, much of the formalism developed for Gromov-Witten theory, such as Givental's symplectic formalism, applies to FJRW theory, as we will summarize in Section~\ref{sec:i-j-function}. While~\eqref{eq:54} is still abstract, Givental's symplectic formalism in particular allows for explicit computations of the FJRW correlation functions in genus zero.

\subsection{The state space and the chiral ring of LG orbifolds}
\label{sec:some-details}

In this subsection, we relate the definitions and properties of the state space $\cH_{\FJRW}$ in FJRW theory to the chiral ring $\cH^{(a,c)}$ in physics.

Given an admissible group $G$ and an element $\gamma \in G$, let
$(\mC^N)^\gamma \subseteq \mC^N$ denote the subspace of $\mC^N$ of
$\gamma$-invariants, i.e. the subspace of elements that are fixed by $\gamma$:
\begin{equation}
  (\mC^N)^\gamma = \{ (x_1,\dots,x_N) \in \mC^N \mid \gamma(x_1,\dots,x_N) =
  (x_1,\dots,x_N) \}.
\end{equation}
We denote the set of fixed indices by (writing $\gamma$ as in~\eqref{orbdef})
\begin{equation}
  I^\gamma = \{ j \in \{1,\dots,N\} \mid \theta^\gamma_j = 0 \}
\end{equation}
and we write $d^\gamma = |I^\gamma| = \dim (\mC^N)^\gamma$. For the
complement, we set $I_\gamma = \{1,\dots, N\} \setminus I^\gamma$ and
$d_\gamma = |I_\gamma| = N - d^\gamma$.

The state space for the LG orbifold $(W, G)$ is defined as the vector space (cf. Section~\ref{sec-lgoclosed})
\begin{equation}
    \cH_{\FJRW}(W,G) = \bigoplus_{\gamma\in G} \cH_{\FJRW,\gamma} = \bigoplus_{\gamma\in G} \left(\mathrm{Jac}(W_\gamma) \otimes \diff{}{x}_\gamma\right)^G,
\end{equation}
where $W_\gamma = W|_{(\mC^N)^\gamma}$ is the $\gamma$-invariant part of the polynomial
$W$, $\mathrm{Jac}(W_\gamma)$ is its Jacobian ring, the differential form
$\diff{}{x}_\gamma$ is $\bigwedge_{j\in I^\gamma} \diff{}{x_j}$,
and the superscript $G$ stands for the $G$-invariant part. Since $W$ and $G$ are fixed, we will just write $\cH_{\FJRW}$ for $\cH_{\FJRW}(W,G)$ for notational ease.

There are alternative, isomorphic definitions in terms of relative Chen-Ruan cohomology~\cite{Fan:2007ba}, and Hochschild homology of the category of $G$-equivariant matrix factorizations of $W$~\cite{Polishchuk:2016ab}. In the former case, we have $\cH_{\FJRW,\gamma} \cong \tH^{d^\gamma}((\mC^N)^{\gamma}, W^{+\infty}_\gamma;\mC)^G$. In particular, the dual space $(\cH_{\FJRW,\gamma})\spcheck$ can be identified with the relative homology $\tH_{d^\gamma}((\mC^N)^{\gamma}, W^{+\infty}_\gamma;\mC)^G$ so that the virtual class in~\eqref{eq:92} can really be thought of as a homology class of degree $d^\gamma$.

For any $\gamma \in \Aut(W)$, the set of broad variables with respect
to $\gamma$ is $\fB_\gamma = \{x_j \mid j \in I^\gamma \}$. In
physics, these variables are called untwisted fields in the $\gamma$-twisted sector~\cite{Intriligator:1990ua,Walcher:2004tx}.
The direct summand $\cH_{\FJRW,\gamma}$ when $d^\gamma = 0$ or, equivalently,
when $\fB_\gamma = \emptyset$ is called a narrow sector and a broad sector otherwise. Note that a narrow sector satisfies $\dim \cH_{\FJRW,\gamma} = 1$. This coincides with the notion of broad and narrow that we have used in the previous sections. We would like to point out that nontrivial broad sectors can appear in a Fermat polynomial of Calabi-Yau type such as in the example discussed in Section~\ref{sec:4-parameter-example}. 

Next, we explain a number of additional structures on the state
space $\cH_{\FJRW}$ known from Section~\ref{sec-lgoclosed}. It carries a bigrading, a nondegenerate pairing, and a product.

Recall the natural $\mQ$ grading on the Jacobian ring $\mathrm{Jac}(W_\gamma)$,
defined by the weights $q_1,\dots,q_N$. This gives a $\mQ$ grading on $\mathrm{Jac}(W_\gamma) \otimes \diff{}{x_\gamma}$
defined by
\begin{equation}
  \deg_W(m \diff{}{x_\gamma}) = \deg_{\mC[x]}(m) + \sum_{j \in I^\gamma} q_j \ .
\end{equation}
In other words, if $m = \prod_{j \in I^\gamma} x_j^{v_j}$, then
$\deg_{W}(m \diff{}{x_\gamma}) = \sum_{j \in I^\gamma} (v_j+1)q_j$.
Recall that $\sum_{j=1}^N q_j = \age(J)$. Following~\cite{Chiodo:2011ab,Chiodo:2008hv,Francis:2015ab} we define the bigrading on $\cH_{\FJRW,\gamma}$ as follows: For $\alpha \in \cH_{\FJRW,\gamma}$ we set
\begin{equation}
  (\deg_{+}\alpha, \deg_{-}\alpha) =\left ( d^\gamma -
    \deg_{W} \alpha + \age(\gamma) - \age(J), \deg_{W} \alpha
    +\age(\gamma) -\age(J)\right).
\end{equation}
and for the total degree
\begin{equation}
  \label{eq:87}
 \deg \alpha = \deg_{+}\alpha + \deg_{-}\alpha \,.
\end{equation}
In fact, we have a decomposition
\begin{equation}
  \begin{aligned}
      \cH_{\FJRW,\gamma} &= \bigoplus_{p+q = d^\gamma + 2\age(\gamma)-2\age(J)}
      \cH_{\FJRW,\gamma}^{p,q}, \\
      \cH_{\FJRW,\gamma}^{p,q} &= \{ \alpha \in \cH_{\FJRW,\gamma} \mid
      \deg_+\alpha = p, \deg_-\alpha = q \}.
  \end{aligned}
\end{equation}
The state space of FJRW theory is then equipped with the bigrading
\begin{equation}
  \cH_{\FJRW}^{p,q}(W,G) = \bigoplus_{\gamma \in G} \cH_{\FJRW,\gamma}^{p,q}.
\end{equation}
As a bigraded vector space, $\cH_{\FJRW}(W,G)$ is determined only by
the weights $q_1,\dots,q_N$ and the action of the group $G$ on $\mC^N$~\cite{Francis:2015ab}. Note that the degree of the class $\Lambda_{g,n}^{W,G}(\alpha_1,\dots,\alpha_n)$ in~\eqref{eq:66} can now be rewritten as
\begin{equation}
  D(\gammabar) - \frac{1}{2} \sum_{i=1}^n d^{\gamma(i)} = (\widehat c -3)(1-g) + n - \frac{1}{2} \sum_{i=1}^n \deg \alpha_i \,,
  \label{eq:97}
\end{equation}
if $\alpha_i \in \cH_{\FJRW,\gamma(i)}$ for $i=1,\dots,n$.

The comparison to the original bigrading defined in~\cite{Intriligator:1990ua} (cf. Section 2) for the $\cH^{(c,c)}$ is
\begin{equation}
  \label{eq:64}
  \cH_{\FJRW,\gamma}^{p,q} \cong \cH^{(c,c)}_{\gamma}{}^{q,\widehat{c} - p} \cong \cH^{(c,c)}_{\gamma^{-1}}{}^{\widehat{c} - p,q}.
\end{equation}
To see this, let $q_+$ be the charge of $F_L$ and $q_-$ be the charge
of $F_R$. Then the left and right $\tU(1)$ charges of an element $\alpha =
m\diff{}{x_\gamma} \in \cH_{\FJRW,\gamma} = \cH^{(c,c)}_\gamma$ are
\begin{equation}
  q_{\pm} = \deg_{\mC[x]} m \pm \sum_{j \in I_\gamma} \left(\theta^\gamma_j - \lfloor
    \theta^\gamma_j \rfloor - \tfrac{1}{2}\right) + \sum_{j\in
    I^\gamma} \left(q_j -\tfrac{1}{2} \right) + \tfrac{\widehat{c}}{2}.
\end{equation}
Assuming that $0 \leq \theta^\gamma_j < 1$ we can drop the
term $\lfloor\theta^\gamma_j \rfloor$. Moreover, since
$\theta^\gamma_j = 0$ for $j \in I^\gamma$, we can write
\begin{equation}
  q_{\pm} = \deg_{\mC[x]} m \pm \left( \age(\gamma) -\tfrac{1}{2} (N - d^\gamma) \right) + \sum_{j\in
    I^\gamma} q_j  -\tfrac{1}{2} d^\gamma + \tfrac{\widehat{c}}{2}.
\end{equation}
Finally, using $\deg_{W}\alpha = \deg_{\mC[x]}m + \sum_{j\in I^\gamma}
q_j$ and $\widehat{c} = N - 2\age(J)$ we obtain
\begin{equation}
  \begin{aligned}
    q_+ &= \deg_{W} \alpha +\age(\gamma) -\age(J) &= \deg_-\alpha,\\
    \widehat{c} - q_- &=  d^\gamma - \deg_{W} \alpha + \age(\gamma) - \age(J) &= \deg_+\alpha. 
  \end{aligned}
\end{equation}
From Section~\ref{sec-lgoclosed} we therefore get the isomorphism
\begin{equation}
  \label{eq:78}
  \begin{aligned}
    \cH_{\FJRW,\gamma}^{p,q} &\cong \cH_{J\gamma^{-1}}^{(a,c)} \\
    e_\gamma & \mapsto {\bf e}_{J\gamma^{-1}}^{(a,c)}.
\end{aligned}
\end{equation}
Even though we work in the A-model and hence with $(a,c)$-rings, for the FJRW formalism and explicit calculations it is often more convenient to work in the FRJW/B-model basis, which we will do in the following and in section \ref{sec:apps}. 

Note that in the Calabi-Yau case, we actually have a $\mZ$ bigrading. There is also a coarse $\mZ/2\mZ$
grading given by the total degree mod 2, i.e. by $d^\gamma \mod
2$. We will call a sector $\cH_\gamma$ of even degree if $d^\gamma \mod 2 = 0$ and of odd degree otherwise.

Since $(\mC^N)^{\gamma} = (\mC^N)^{\gamma^{-1}}$, i.e. $\gamma$ and
$\gamma^{-1}$ have the same fixed point set, there is an obvious
isomorphism $\varepsilon : \cH_{\FJRW,\gamma} \to \cH_{\FJRW,\gamma^{-1}}$. The
residue pairing on $\mathrm{Jac}(W_\gamma) \otimes \diff{}{x_\gamma}$ induces a
pairing $\langle-,-\rangle_\gamma : \cH_{\FJRW,\gamma} \otimes  \cH_{\FJRW,\gamma^{-1}} \to
\mC, (f,g) \mapsto \langle f,g\rangle_\gamma = \langle f,\varepsilon^*g\rangle$ which is symmetric and non-degenerate.
The pairing on $\cH_{\FJRW}$ is defined as the direct sum of the pairings $\langle-,-\rangle_\gamma$  on $\cH_{\FJRW,\gamma}$.
Fixing a basis for $\cH_{\FJRW}$, we denote the pairing by a matrix $\eta_{\alpha\beta} = \langle\alpha,\beta\rangle$, with inverse $\eta^{\alpha\beta}$. The restriction of the pairing to the narrow sectors then takes the
following form (see (\ref{lgpairing}))
\begin{equation}
 \eta_{\gamma,{\gamma'}} = \eta_{e_\gamma,e_{\gamma'}} = \frac{1}{|G|}\delta_{\gamma, \gamma'{}^{-1}}, \qquad \gamma,\gamma' \in G \,.
\end{equation}

There are two types of twisted sectors of special interest. The sector $\cH_{\FJRW,J}$ is always
narrow, hence has a canonical generator which we call $1_J$. By the
Calabi-Yau condition $\sum q_j = d$, we have $\deg 1_J = 0$. The
second case is when $\sum q_j = d$ and $\left(\deg_{+}(\alpha),
  \deg_{-}(\alpha)\right) = (1,1)$. This is the case if and only if
$d_\gamma = 0 \mod 2$ and $\age(\gamma) =
2-\frac{1}{2}d^\gamma$. These are the sectors that contain the
deformation classes discussed in Section~\ref{sec:deformations}. In particular, note that there can be broad sectors that are of even degree and therefore contribute to the deformation classes. An example is given in Section~\ref{sec:4-parameter-example}. 

From the general discussion in Section~\ref{sec:cc-scft} we expect that $\cH_{\FJRW}(W,G)$ can be equipped with a product to make it into a ring. To define the product $\ast:\cH_{\FJRW} \times \cH_{\FJRW} \to \cH_{\FJRW}$, we need the FJRW invariants. For $\alpha_1,\alpha_2 \in \cH_{\FJRW}$ we set
\begin{equation}
  \label{eq:102}
  \alpha_1 \ast \alpha_2 = \sum_{k,k'=1}^M \langle \alpha_1,\alpha_2, e_k\rangle_{0,3} \eta^{e_k,e_{k'}} e_{k'},
\end{equation}
where $e_1,\dots,e_M$ is a fixed basis of $\cH_{\FJRW}(W,G)$. The state $e_1 := 1_J$ is the unit of this product. The product can only be determined once we have a prescription for computing the FJRW invariants in genus zero.

We define the narrow part of $\cH_{\FJRW}$ by $\cH_{narrow} = \bigoplus_{\gamma : d^\gamma = 0} \cH_{\FJRW,\gamma}$ and the broad part by $\cH_{broad} = \bigoplus_{\gamma : d^\gamma > 0} \cH_{\FJRW,\gamma}$.
The decomposition $\cH_{\FJRW} = \cH_{narrow} \oplus
\cH_{broad}$ corresponds under the isomorphism $\cH_{\FJRW} \cong \tH^*(X,\mC)$
(Landau-Ginzburg/Calabi-Yau correspondence) to the decomposition
$\tH^*(X,\mC) = \tH^*_{amb}(X,\mC) \oplus \tH^*_{prim}(X,\mC)$ of the cohomology of a Calabi-Yau hypersurface $X$.The two spaces in this decomposition are defined starting from the embedding $\iota: X \to \mP(\Sigma)$ into a (smooth) toric varietry $\mP(\Sigma)$ given by a fan $\Sigma$. Then the ambient cohomology is $\tH^*_{amb}(X,\mC) = \im(\iota^*: \tH^*(\mP(\Sigma),\mC) \to \tH^*(X,\mC))$ and the primitive cohomology is $\tH^*_{prim}(X,\mC) = \ker(\iota_*: \tH^*(X,\mC) \to \tH^*(\mP(\Sigma),\mC))$. The correspondence was shown in~\cite{Chiodo:2011ab} for Calabi-Yau hypersurfaces in weighted projective spaces. We expect it to hold more generally for any Calabi-Yau hypersurface $X$ in a toric variety that has a Landau-Ginzburg phase. We would like to point out that the map $\iota^*$ restricted to the even cohomology need not be surjective. An example is discussed in Section~\ref{sec:4-parameter-example}. We will return to this point after we have introduced the $J$-function.

\subsection{$I$- and $J$-functions}
\label{sec:i-j-function}

In this subsection we review how Givental's symplectic formalism~\cite{Givental:2004ab} can be applied to FJRW theory in order to compute the genus zero FJRW invariants in the narrow sectors. This has been done for one-parameter families of Landau-Ginzburg orbifolds of Calabi-Yau type with $G=\langle J\rangle$ in~\cite{Chiodo:2008hv,Chiodo:2012qt} and $G=\Aut(W)$ in~\cite{Guere:2016ab}. Here, we generalize it to multiparameter families with general $G\subset\mathrm{Aut}(W)$. From now on we will restrict ourselves to genus zero. The genus zero descendant potential $F_0$ in~\eqref{eq:68} can be recovered from the so-called $J$-function of finitely many variables via a reconstruction theorem~\cite{Givental:2004ab} essentially due to~\cite{Dijkgraaf:1990nc,Dubrovin:1994hc}. It turns out that there exists a family of $J$-functions parametrized by a set of variables $s=(s_0,s_1,\dots)$ that interpolates between the (rescaled) invariants of $\overline{\cM}_{g,n}$ and certain equivariant FJRW invariants. The actual FJRW invariants are then obtained in the non-equivariant limit. In the following we review this procedure and apply it to the case of multiparameter LG orbifolds. 

For computational purposes, the authors of~\cite{Chiodo:2008hv,Chiodo:2012qt} have made two modifications, referred to as ``extension'' and ``twist'', to the description of FJRW theory given so far. This defines new invariants that are different from the invariants of the full theory, but still are a natural and computable extension of the narrow sector invariants.

Let us first define the extended invariants. For this purpose, we define the extended (narrow) state space replacing every broad sector by a one-dimensional auxiliary space $\mC e_\gamma$, thereby effectively making it narrow: 
\begin{equation}
  \cH_{\FJRW}^{\ext}(W,G) = \bigoplus_{\gamma \in G} \mC e_{\gamma} =
  \cH_{narrow} \oplus \bigoplus_{d^\gamma > 0} \mC e_{\gamma} \,.
  \label{eq:74}
\end{equation}
The grading~\eqref{eq:87} on $\cH^{\ext}_{\FJRW}(W,G)$ is modified in the way to include the new sectors by setting
\begin{equation}
  \label{eq:101}
  \deg \alpha = 2 d^\gamma + 2 \age(\gamma) - 2 \age(J).
\end{equation}
This extension is introduced for practical purposes. The new states are
only a computational tool and play the role of placeholders in the
theory. In this way, we can work with all twisted sectors
on equal footing, without paying attention to those which are absent, or
broad. The disadvantage of this modification is that we are ignoring contributions from the broad sectors. This is because no computational description is known so far. Moreover, this modification can introduce unphysical states, see e.g. the discussion in Section~\ref{sec:central-charge-fjrw}.

In order to include the extended sectors properly, the moduli stack
$\cW_{0,n}(W,G)(\gammabar)$ has to be modified accordingly: The essential idea is to undo the monodromy of $\rL_j$ at
$\sigma_i$ by twisting $\rL_j$ to $\widetilde \rL_j =\rL_j(-\sum_{i=1}^n\sigma_i)$. This procedure will guarantee that the new invariants involving classes from the extended sectors vanish.

We define the extended FJRW invariants to be
\begin{equation}
  \label{eq:105}
  \langle \tau_{a_1}(\alpha_1) \dots \tau_{a_n}(\alpha_n)
  \rangle^{\ext}_{0,n} = \int_{\cW_{0,n}(W,G)(\gammabar) }
  c_{\Top}(\bigoplus_{j=1}^N \tR^1\pi_*\widetilde\rL_j) \cap
  \prod_{i=1}^n \psi_i^{a_i} \cap \prod_{i=1}^n \alpha_i
\end{equation}
for $\alpha_i\in \cH^{\ext}_{\FJRW,\gamma(i)}$, $i=1,\dots,n$.
It is shown in \cite{Chiodo:2008hv,Chiodo:2012qt} that  the extended invariants vanish if one of
the entries $\alpha_j$ does not belong to $\cH_{narrow}$. Otherwise
\begin{equation}
    \langle \tau_{a_1}(\alpha_1) \dots \tau_{a_n}(\alpha_n)
  \rangle^{\ext}_{0,n} = \langle \tau_{a_1}(\alpha_1) \dots \tau_{a_n}(\alpha_n)
  \rangle_{0,n} \,.
\end{equation}

The second modification concerns the Euler class of the obstruction complex. 
Rescaling the fiber of each line bundle $\rL_j$, defines a $T=(\mC^\ast)^N$ action on a $W$-spin structure given by with character $-\lambda_j \in \tH_T^2(\mathrm{pt})$, $j=1,\dots,N$.  This induces an action on $\cW_{0,n}(W,G)$ and on the extended obstruction bundle $\rE^1 = \bigoplus_{j=1}^N \tR^1\pi_*\widetilde \rL_j$. 
Then the $T$-equivariant Euler class $e_T$ of $\rE^1 $
is given by
\begin{equation}
  e_T\left( \rE^1 \right) = \prod_{j=1}^N \sum_{\ell=0}^{r_j} \lambda_j^{r_j-\ell} \chern_\ell(\tR^1\pi_*\widetilde \rL_j),
\end{equation}
with $r_j = \rank \tR^1\pi_*\widetilde \rL_j$. Note that there is an explicit formula for expressing  $\chern_k(\tR^1\pi_*\widetilde\rL_j)$ in terms of the tautological classes in $\tH^*(\overline{\cW}_{g,n}(W,G))$~\cite{Chiodo:2008ab}. In the non-equivariant limit $\lambda_j \to 0 $ we have
\begin{equation}
  \lim_{\lambda_j\to 0} e_T(\rE^1) = \ch_{\Top}(\rE^1) \,.
  \label{eq:112}
\end{equation}
More generally, we may express an invertible multiplicative characteristic class of $\rE^1$ as 
\begin{equation}
  e(s)(\rE^1) = \exp\left( \sum_{j=1}^N \sum_{\ell
      \geq 0} s^{(j)}_\ell \chern_\ell(\tR^1\pi_*\widetilde\rL_j) \right) \in
  \tH^*(\cW_{0,n}(W,G),\mC) \otimes_{\mC} \mC[[s]],
\end{equation}
where we write $s=(s_\ell^{(j)})_{\ell\in \mZ_{\geq 0}, 1\leq j \leq N}$ with $\exp(s_0^{(j)}), s_\ell ^{(j)} \in \mC$ for $1 \leq j \leq N,\ell > 0$. These variables can be collected into the generating functions
\begin{equation}
  \label{eq:71}
   s^{(j)}(x) = \sum_{\ell \geq 0} s_\ell^{(j)} \frac{x^\ell}{\ell!}, \qquad j=1,\dots,N \,.
 \end{equation}
Following~\cite{Chiodo:2008hv,Chiodo:2012qt} we consider two specializations. On one hand, setting $s_\ell^{(j)} = 0$ for all $j$ and $\ell$ yields $e(0) = 1$, and the virtual class becomes the ordinary fundamental class. We will see below that in this case the corresponding correlators can be computed explicitly. On the other hand, consider the specialization
\begin{equation}
  s_\ell^{(j)} =
  \begin{cases}
    -\log \lambda_j & \ell=0\\
    (\ell-1)! (-\lambda_j)^{-\ell} & \ell>0,
  \end{cases}
  \label{eq:73}
\end{equation}
which we will abbreviate by $s=\lambda$. This specialization yields $\exp(-s^{(j)}(x)) = x + \lambda_j$ and therefore recovers the $T$-equivariant Euler class
\begin{equation}
  \label{eq:70}
  e(s)(\rE^1)|_{s=\lambda} = e_T(\rE^1) \,.
\end{equation}
In the limit $\lambda_j \to 0$ we obtain the virtual class. In particular, the variation by $s$ {\em interpolates} between integrals of $\psi$-classes over $\cW_{g,n}(W,G)$ (which are generically  $|G|$-fold covers of $\Mgn{0}{n}$ and hence are easy to compute) and the FJRW invariants (which we want to compute).  Equations (\ref{eq:73}) and (\ref{eq:70}) are valid for the concave case. For the non-concave case \cite{Guere:2016ab} one replaces $c_{\Top}$ in (\ref{eq:105}) by (\ref{nonconc}). Furthermore $\lambda_j$ no longer has an interpretation as an equivariant parameter, but formally the derivation is the same. 

Given these modifications we define the $s$-twisted virtual class on $\cW_{0,n}(W,G)$ as the
class
\begin{equation}
  \left[ \cW_{0,n}(W,G) (\overline \gamma) \right]^{\vir,s} = e(s)(
  \rE^1 ) \cap \left[ \cW_{0,n}(W,G) (\overline \gamma)\right]
\end{equation}
and the twisted invariants
\begin{equation}
  \langle \tau_{a_1}(\alpha_1) \dots \tau_{a_n}(\alpha_n)
  \rangle^{\ext,s}_{0,n} = \int_{\cW_{0,n}(W,G) (\overline \gamma) }
  e(s)(\rE^1) \cup \prod_{i=1}^n \alpha_i \cup \prod_{i=1}^n \psi_i^{a_i}.
\end{equation}
There is an $s$-twisted pairing $\eta^{\ext,s}: \cH^{\ext}_{\FJRW}(W,G)
\otimes_{\mC} \mC[[s]] \times \cH ^{\ext}_{\FJRW}(W,G)
\otimes_{\mC} \mC[[s]] \to \mC[[s]]$ given as follows: For any
$e_\gamma, e_{\gamma'} \in \cH ^{\ext}_{\FJRW}(W,G)$ we set 
\begin{equation}
  \eta ^{\ext,s} (e_\gamma,e_{\gamma'}) = \frac{1}{|G|} \prod_{j\in I^\gamma}
  \exp(-s_0^{(j)}) \delta_{\gamma ,(\gamma')^{-1}}
\end{equation}
and then extend it by linearity. As in the unmodified case in Section~\ref{sec:brief-guide-fjrw}, we can define a generating
function for the invariants in~\eqref{eq:54} as
\begin{equation}
  F_0^s(t) = \sum_{n=0}^\infty \sum_{\substack{a_1,\dots,a_n\geq 0\\\gamma(1),\dots,\gamma(n) \in G}} \langle \tau_{a_1}(e_{\gamma(1)}) \dots \tau_{a_n}(e_{\gamma(n)})
  \rangle^{\ext,s}_{0,n}  \frac{t_{a_1}^{\gamma(i)} \dots t_{a_n}^{\gamma(n)} }{n!},
\end{equation}
where $M = \dim\cH^{\ext}_{\FJRW}(W,G)$ and $e_{\gamma}$ is the generator of $\cH_\gamma$ (cf.~\eqref{eq:74}). We denote its specialization to~\eqref{eq:71} by $F_0^T(t) = F_0^s(t)|_{s=\lambda}$. In the non-equivariant limit, this becomes
\begin{equation}
  \label{eq:69}
  \lim_{\lambda \to 0} F_0^T = F_0.
\end{equation}
where we set $\lambda_j = -q_j\lambda$, $j=1,\dots,N$, and then take the limit $\lambda \to 0$.

We briefly return to the specialization $s=0$. In this case, some of these modified invariants can be explicitly determined by reducing them to integrals over $\Mgn{g}{n}$ which are explicitly known in many cases, see e.g.~\cite{Faber:1999ab}. In particular for $g=0$, the string equation~\eqref{eq:100} implies that~\cite{Chiodo:2008hv}
\begin{equation}
  \begin{aligned}
    \langle \tau_{a_1}(e_1),\dots,
    \tau_{a_n}(e_n)\rangle_{0,n}^{s=0} &=
    \frac{1}{|G|} \int_{\overline{\cM}_{0,n}} \prod_{i=1}^n
    \psi_i^{a_i} \\
    &= 
    \begin{cases}
      \frac{1}{|G|} \frac{\left( \sum_{i=1}^n a_i\right)!}{a_1!\cdots
        a_n!} & \text{if }  n-3 = \sum_{i=1}^n a_i \text{ and }
      \gamma(1)\dots\gamma(n) = J^{n-2}\\
      0 & \text{otherwise}
    \end{cases}
  \end{aligned}
  \label{eq:60} 
\end{equation}
The vanishing conditions follow from the non-emptiness of $\cW_{0,n}(W,G)(\overline{\gamma})$ in~\eqref{eq:76} and the absence of $\tU(1)_A$ anomaly~\eqref{eq:93}.

Finally, we come to Givental's symplectic formalism~\cite{Givental:2004ab}. In this formalism a new variable $z$ is introduced and one considers the symplectic vector space of formal Laurent series  $\cV^s = \cH_{\FJRW}^{\ext}(W,G) \otimes_{\mC} \mC[[s]] \otimes_{\mC} \mC[z][[z^{-1}]]$ with symplectic form $\Omega(f,g) = \Res_{z=0}(\eta^{\ext,s}(f(z), g(-z)))$. The variable $z$ can be identified with the parameter $z$ in the $tt^*$ or Dubrovin connection~\cite{Cecotti:1992rm, Dubrovin:1994hc}. 
Introducing Darboux coordinates $(q^\gamma_a, p_{\gamma,b})$ dual to the basis of $\cV^s$ given by $(e_\gamma z^a, \eta_{\ext,s}^{\gamma\gamma'}e_{\gamma'}(-z)^{-1-b})$, $\gamma\in G$, $a,b \in \mZ_{\geq 0}$, the important point is that after the change of variables $t_a^\gamma = q_a^\gamma - \delta^{\gamma}_J\delta_{a}^1$ the generating function $F_0^s$ becomes a function of $q_a^\gamma $. Its graph therefore defines a Lagrangian subspace $\cL^s = \{ (q^\gamma_a, p_\gamma^b) \in \cV^s \mid  p_\gamma^a = \pd{F_0^s}{q_\gamma^a} \} \subset \cV^s$ which is a cone and has further very special geometric properties that encode the dilaton equation, the string equation and the topological recursion relations satisfied by $F^s_0$.

This cone has an alternative characterization in terms of the twisted $J$-function. The twisted $J$-function, $J^s : \cH ^{\ext,s}_{\FJRW}(W,G) \to \cV^s, t\mapsto J^s(t,z)$, is a family of points in $\cV^s$ parametrized by $t = \sum_{\gamma \in G} t_0^\gamma e_\gamma \in \cH^{\ext,s}_{\FJRW}(W,G)$. From now on, we will omit the index $0$ and write $t^\gamma =
t_0^\gamma$. The function $J^s$ is defined as
\begin{equation}
  J^s(t,z) = z 1_J + t + \sum_{n=2}^\infty \sum_{b=0}^\infty
  \sum_{\gamma,\gamma' \in G} \frac{1}{n!} \langle
  t,\dots,t,\tau_b(e_\gamma)\rangle^{\ext,s}_{0,n+1}
  \eta_s^{\gamma\gamma'} z^{-1-b}e_{\gamma'}.
\end{equation}
One can show that $J^s(t,-z) \in\cL^s\subset \cV^s$ and that it is the unique such function of the form
\begin{equation}
  J^s(t,-z) = - z 1_J + t + O(z^{-1}).
  \label{eq:96}
\end{equation}
It follows that the cone $\cL^s$ is uniquely determined by the image of $J^s(t,-z)$ via the string equation~\eqref{eq:100}~\cite{Givental:2004ab}. 

We evaluate the untwisted $(s=0)$ correlators using~\eqref{eq:60} and find
\begin{equation}
  \label{eq:79}
     J^0(t,z) = \sum_{ \{k_\gamma \geq
     0 \mid \gamma \in G
     \}}\prod_{\gamma\in G}
 \frac{\left(t^{\gamma}\right)^{k_\gamma}}{k_\gamma!} z^{1-|k|} 
 e_{\widehat \gamma(k)}\,,
\end{equation}
where we have set $\widehat \gamma (k) = \prod_{\gamma \in G}\gamma^{k_\gamma}$ and $|k| = \sum_{\gamma \in G} k_\gamma$. We write this function as
\begin{equation}
  J^0(t,z) = \sum_{ \{k_\gamma \geq
     0 \mid \gamma \in G
     \}} J^0_{\{k_\gamma\}}(t,z)
  \label{eq:80}
\end{equation}
with coefficients
\begin{equation}
  J^0_{\{k_\gamma\}}(t,z) =\prod_{\gamma\in G}
 \frac{\left(t^{\gamma}\right)^{k_\gamma}}{k_\gamma!} z^{1-|k|} 
 e_{\widehat \gamma(k)}\,.
\end{equation}

To determine $J^s$ one introduces another function $I^s:  \cH ^{\ext,s}_{\FJRW}(W,G) \to \cV^s$ which is obtained from $J^0$ by a symplectic transformation $\Delta: \cV^0 \to \cV^s$. Following and generalizing~\cite{Chiodo:2008hv,Chiodo:2012qt,Guere:2016ab}, we define the modification factor
\begin{equation}
   \label{eq:85}
    M^s_{\{k_\gamma\}}(z) = \prod_{j=1}^N \exp \left(  -\sum_{0 \leq m < \lfloor \sum_{\gamma \in G} \theta^{\gamma}_j
      k_\gamma + q_j\rfloor} s^{(j)}\left(\left(\langle \sum_{\gamma \in G} \theta^{\gamma}_j
      k_\gamma + q_j \rangle + m\right)z\right)\right) \,,
\end{equation}
where the $\theta_j^\gamma \in [0,1)$ are the phases of an element $\gamma \in G$ and where we use the functions $s^{(j)}(x)$ given in~\eqref{eq:71}. Then the family of functions $u \mapsto I^s(u,-z)$ defined by
\begin{equation}
  \label{eq:84}
  I^s(u,z) =  \sum_{ \{k_\gamma \geq 0 \mid \gamma \in G \}}
  M^s_{\{k_\gamma\}}(z)  J^0_{\{k_\gamma\}}(u,z)
\end{equation}
lies on the cone $\cL^s$. This formula is a natural generalization of the one given in~\cite{Chiodo:2008hv,Chiodo:2012qt} to admissible arbitrary groups $G$ and a derivation within the setting of FJRW theory will be given elsewhere. Note that we have changed the notation for the parametrization of $\cH^{\ext}_{\FJRW}$ from $t$ to $u= \sum_{\gamma \in G} u^\gamma e_\gamma$ since for $s\not=0$, these variables do not satisfy~\eqref{eq:96}. The fact that $I^s(u,-z)$ lies on the $\cL^s$ implies that it contains the same information as $J^s$. Therefore, we only need to bring it into the form~\eqref{eq:96}. We will do this in the next subsection, after we have set $s=\lambda$.

\subsection{The central charge in the FJRW formalism}
\label{sec:central-charge-fjrw}

In this subsection, we specialize the formula~\eqref{eq:84} for the $I$-function to the equivariant case and evaluate it in the non-equivariant limit. In particular, we consider Landau--Ginzburg orbifolds of Calabi--Yau type with $\widehat c = 3$ and $\age(J) = \sum_{j=1}^N q_j = 1$. Furthermore, we also discuss the Gamma class. 

For the specialization $s=\lambda$ we obtain the so-called $e_T$--twisted $J$-function $J_T(t,-z;\lambda) := J^s(t,-z)|_{s=\lambda}$
\begin{equation}
  \label{eq:81}
    J_T(t,-z;\lambda) = - z 1_J + t + \sum_{n=2}^\infty \sum_{b=0}^\infty
  \sum_{\gamma,\gamma' \in G} \frac{1}{n!} \langle
  t,\dots,t,\tau_b(e_\gamma)\rangle^{\ext,s}_{T,0,n+1}
  \eta_T^{\gamma\gamma'} (-z)^{-1-b}e_{\gamma'},
\end{equation}
where
\begin{equation}
  \label{eq:82}
  \eta_{T,\gamma\gamma'} = \eta_{T} (e_\gamma,e_{\gamma'}) = \frac{1}{|G|} \prod_{j \in I^{\gamma }}  \lambda_j \delta_{\gamma (\gamma')^{-1}}
\end{equation}
and $\langle t,\dots,t,\tau_b(e_\gamma)\rangle^{\ext,s}_{T,0,n+1}$ are the  $e_T$--twisted FJRW invariants obtained from the Euler class in~\eqref{eq:70}. Note that
\begin{equation}
  \lim_{\lambda_j \to 0} \langle
  \tau_{a_1}(e_{\gamma(1)}),\dots,
  \tau_{a_n}(e_{\gamma(n)})\rangle_{T,0,n} = \langle
  \tau_{a_1}(e_{\gamma(1)}),\dots,
  \tau_{a_n}(e_{\gamma(n)})\rangle^{ext}_{0,n} \,.
  \label{eq:91}
\end{equation}

The specialization of~\eqref{eq:85} is
\begin{equation}
  \label{eq:83}
      M^s_{\{k_\gamma\}}(z) |_{s=\lambda} 
     = \prod_{j=1}^N z^{\lfloor \sum_{\gamma \in G} \theta^{\gamma}_j
      k_\gamma + q_j\rfloor} \frac{\Gamma(\frac{\lambda_j}{z}-\langle \sum_{\gamma \in G} \theta^{\gamma}_j
      k_\gamma + q_j\rangle +1)}{\Gamma(\frac{\lambda_j}{z}-\sum_{\gamma \in G} \theta^{\gamma}_j
      k_\gamma - q_j +1)} \,.
\end{equation}
At this point, it is useful to introduce the grading operator $\pG\pr$ of~\cite{Chiodo:2012qt}:
\begin{equation}
  \label{eq:86}
  \pG\pr(e_\gamma) = \tfrac{1}{2} \deg(e_\gamma) e_\gamma\,,
\end{equation}
where ``$\deg$'' was defined in~\eqref{eq:87}. Furthermore, we introduce the set $G^{(2)}$ labeling the sectors containing the deformation classes, i.e. in the present situation the elements $\alpha \in \cH ^{\ext}_{\FJRW}$ with $\deg \alpha = 2$ (cf.~\eqref{eq:101}):
\begin{equation}
  \label{eq:88}
  G^{(2)} = \{ \gamma \in G \mid \age(\gamma) = 2\}. 
\end{equation}
Then, we restrict $t$ from the big phase space to the small phase space $\cH^{\ext,(1,1)}_{\FJRW} = \bigoplus_{\gamma \in G^{(2)}} \cH_\gamma$ by setting
\begin{equation}
  \label{eq:89}
  \begin{aligned}
    u^\gamma &=0, \gamma \not \in G^{(2)}\,,\\
    u &= \sum_{\gamma \in G^{(2)}} u^\gamma e_\gamma \,.
  \end{aligned}
\end{equation}
Note that we always have $J^2 \in G^{(2)}$. In~\cite{Chiodo:2008hv,Chiodo:2012qt,Guere:2016ab} one--parameter families along the direction of $e_{J^2}$ were studied. If one of the weights $q_i$ is $1/2$, however, then $\cH_{J^2}$ is a broad sector and $\dim \cH_{J^2} = 0$. The corresponding state $e_{J^2}$ was only artificially introduced in the definition~\eqref{eq:74} of the extended state space. From the discussion in Section~\ref{sec:cc-lgo} we have learned that in such situations there is a change of variables such that the fake deformation along $e_{J^2}$ can be traded for a genuine deformation along a specific state $e_\gamma \in \cH_{narrow}$ with $\deg e_\gamma$. We will return to this point after the discussion of the $J$-function below.

The reason to restrict to the degree two elements is the following. First, we notice that the invariants with only even degree insertions span the entire FJRW theory. This follows~\cite{Fan:2007ba} from~\eqref{eq:93} and~\eqref{eq:97}, which for $\widehat{c}=3$ yields
\begin{equation}
  \label{eq:94}
  2n - \sum_{i=1}^n \deg \alpha_i - 2\sum_{i=1}^n a_i = 0 \,,
\end{equation}
and the fact that $F_0$ satisfies the string equation~\eqref{eq:100}. The string equation implies that every FJRW invariant containing an entry of the form $\tau_0(e_J)$ can be expressed in terms of the remaining invariants unless it is of the form $\langle \tau_0(e_J) \tau_0(\alpha_2) \dots \tau_0(\alpha_n)\rangle_{g,n}$. One finds
\begin{equation}
  \label{eq:95}
  \langle \tau_0(e_J) \tau_0(\alpha_2) \dots \tau_0(\alpha_n)\rangle_{g,n} =
  \begin{cases}
    \eta_{\gamma(2),\gamma(3)} & (g,n) = (0,3) \\
    0 & \text{otherwise}.
  \end{cases}
\end{equation}
Now, we see from~\eqref{eq:94} that either $\deg \alpha_i = 2$ and $a_i=0$ for all $i$, or $\deg \alpha_i = 2$ and $a_i=0$ for all but one $i$, and the remaining entry being $\tau_1(e_J)$. It therefore suffices to compute the $J$-function for $t \in \cH_{\FJRW}^{p,q}$ with $p+q = 2$ only.  

With these preparations, we define the $e_T$-twisted $I$-function by\footnote{We do not multiply $I^s$ with a factor of $u^J$ as in~\cite{Chiodo:2008hv,Chiodo:2012qt} since we require the $e_J$ component $I_{T,J}$ of $I_T$ to be the form $I_{T,J} = 1 + O(u)$.} $I_T(u,z;\lambda) = I^s(u,z)|_{s=\lambda}$ and obtain the following expression 
\begin{equation}
  \label{eq:29}  
  I_T(u,z;\lambda) = z^{-\pG\pr} z
    \sum_{\{k_\gamma \geq 0 \mid \gamma \in G^{(2)}\}} \prod_{\gamma\in G^{(2)}}
    \frac{\left(u^{\gamma}\right)^{k_\gamma}}{k_\gamma!}  \prod_{j=1}^N \frac{\Gamma(\frac{\lambda_j}{z}-\langle \sum_{\gamma \in G^{(2)}} \theta^{\gamma}_j k_\gamma + q_j\rangle +1)}{\Gamma(\frac{\lambda_j}{z}-\sum_{\gamma \in G^{(2)}} \theta^{\gamma}_j 
      k_\gamma - q_j +1)} e_{\widehat{\gamma}(k)}.
\end{equation}
Finally, in this form it is convenient to decompose the sum according to the twisted sectors given by $e_\gamma$:
\begin{equation}
  I_T( u,z;\lambda) = \sum_{\gamma' \in G} I_{T,\gamma'}( u,z;\lambda)  e_{\gamma'},
  \label{eq:30}
\end{equation}
with
\begin{equation}
    I_{T,\gamma'}( u,z;\lambda) = z^{1-\frac{1}{2}\deg e_{\gamma'}}
    \sum_{\substack {\{k_\gamma \geq 0 \mid \gamma \in G^{(2)}\}\\\prod_{\gamma \in G^{(2)}} \gamma^{k_\gamma} = \gamma' }} 
    \prod_{\gamma\in G^{(2)}}
    \frac{\left(u^{\gamma}\right)^{k_\gamma}}{k_\gamma!}  \prod_{j=1}^N \frac{\Gamma(\frac{\lambda_j}{z}-\langle \sum_{\gamma \in G^{(2)}} \theta^{\gamma}_j k_\gamma + q_j\rangle +1)}{\Gamma(\frac{\lambda_j}{z}-\sum_{\gamma \in G^{(2)}} \theta^{\gamma}_j 
      k_\gamma - q_j +1)}. 
  \label{eq:31}  
\end{equation}
Since the coefficient of $z$ in  $I_T(u,z;\lambda)$ is $I_{T,J}$, the fact that $I_T(u,-z;\lambda)$ lies on $\cL^\lambda$, together with the uniqueness of $J_T(t,-z;\lambda)$, implies that 
\begin{equation}
  \label{eq:103}
  J_T(t,-z;\lambda) = \frac{I_T(u,-z;\lambda)}{I_{T,J}( u,-z;\lambda) } \,,
\end{equation}
and hence
\begin{equation}
  \label{eq:104}
  t^\gamma(u) = \frac{I_{T,\gamma}(u,-z;\lambda)}{I_{T,J}( u,-z;\lambda) }, \qquad \gamma \in G^{(2)} \,.
\end{equation}
This is the analogous statement to the ones in Sections~\ref{sec:frame} and~\ref{sec:flat}.  We finally define
\begin{equation}
  I_{W,G}(u,z) = \lim_{\lambda\to 0} I_T(u,z;\lambda), \quad J_{W,G}(t,z) = \lim_{\lambda\to 0} J_T(t,z;\lambda).
\end{equation}

The functions $I_T(u,z;\lambda)$ and $I_{W,G}(u,z)$ have a number of interesting properties. First, by~\eqref{qmardef} and~\eqref{eq:22} we immediately have
\begin{equation}
  \label{eq:138}
  I_{W,G}(u,-1) = I_{LG}(u)\,.
\end{equation}
where we use $q=q^{LG}$ in the definition of $I_{LG}(u)$. Second, from the discussion in Section~\ref{sec:gkz} it follows that $I_{W,G}(u,z)$ is a generating function for a subspace of solutions to a GKZ system near a point with finite monodromy which is defined in terms of $W$ and $G$ only. This subspace has dimension $\dim \cH_{narrow}$ and is closed under the monodromy action. It is straightforward to see that this holds more generally for $I_T(u,z;\lambda)$. Finally, in concrete examples, the sums over $k$ in~\eqref{eq:31} can be rewritten by choosing a fundamental domain for the action of $G$ on $\{k_\gamma \geq 0 \mid \gamma \in G^{(2)}\}$, so that the entire expression admits a much nicer form. This is done in Section~\ref{sec:apps}.

The procedure to determine the FJRW invariants is then as follows. Given $(W,G)$, one first sets up the GKZ system and determines its solutions which give the $I$-function. Then one computes the change of variables~\eqref{eq:104}, inverts the corresponding power series to obtain $u^\gamma$ as functions of $t^{\gamma'}$. After substituting $u = u(t)$ in~\eqref{eq:103} the FJRW invariants can be read off from the expansion~\eqref{eq:81} after taking the limit $\lambda \to 0$ as in~\eqref{eq:91}. 
This is very reminiscent of the standard procedure in mirror symmetry. In fact, the $I$-function has the interpretation of period integral on the mirror Landau-Ginzburg model $(W\spcheck,G\spcheck)$. In that context, the change of variables between the $I$- and $J$-function is then nothing but the mirror map, see e.g.~\cite{Huang:2006hq}. In the present context, however, no input from the B-model is needed.

Now,  we return to the deformations $e_\gamma \in \cH_{\FJRW}^{\ext,(1,1)}$ which are not in the narrow sector $\cH_{narrow}$, for example $e_{J^2}$ if one of the weights is $1/2$.  Then $I_{T,J^2} = 0$, and hence $t^{J^2}=0$. This is in agreement with the vanishing of the extended invariants~\eqref{eq:105} with insertions not in $\cH_{narrow}$. However, there are still nonvanishing FJRW invariants coming from genuine deformations $e_\gamma \in \cH_{narrow}^{(1,1)}$.

In the geometric framework, one can define a $J$-function associated to the anticanonical bundle $K_{\mP_{\Sigma(\Delta)}}\spcheck$ of the toric ambient variety $\mP_{\Sigma(\Delta)}$ of the Calabi-Yau $X$ \cite{Cox:2000vi}. Naturally, this only depends on $\tH^*_{amb}(X,\mathbb{C})$. The Landau-Ginzburg analogue is the extended $J$-function that only sees the narrow sectors of the genuine FJRW $J$-function.

The other ingredient, besides the $I$-function, we need for the D-brane central charge is the Gamma class. The $\widehat{\Gamma}$-integral structure in orbifold GW theory was introduced in~\cite{Iritani:2009ab,Katzarkov:2008hs}. The generalization to the case of FJRW theory for $(\mC^N, W, \langle J\rangle)$ was given in~\cite{Chiodo:2012qt}. For arbitrary LG orbifolds $(W,G)$ we define the Gamma class $\widehat{\Gamma}_{W,G} \in \End \cH_{\FJRW}(W,G)$ as
\begin{equation}
  \widehat{\Gamma}_{W,G} = \bigoplus_{\gamma \in G} \prod_{j \in I_\gamma}
  \Gamma(1- \theta_j^{\gamma} ) \id_{\cH_{\FJRW,\gamma}}\,,
  \label{eq:28}
\end{equation}
where the product is taken to be 1 if $I_\gamma =\emptyset$. 
Note that we have excluded the broad sectors here. If we are replacing $\cH_{\FJRW}(W,G)$ by the extended state space $\cH^{\ext}_{\FJRW}(W,G)$ in~\eqref{eq:74} we can let the product run from $j=1$ to $N$, since then the Gamma class acts by the identity on the auxiliary sectors. We also define the conjugate Gamma class as
\begin{equation}
  \widehat{\Gamma}^*_{W,G} = \bigoplus_{\gamma \in G}\prod_{j \in I_\gamma}
  \Gamma(1-\theta_j^{\gamma^{-1}}) \id_{\cH_{\FJRW,\gamma}}\,.
  \label{eq:11}
\end{equation}
Equivalently, if we define the map $\inv: \cH_{\FJRW}(W,G) \to \cH_{\FJRW}(W,G)$
induced from the natural isomorphism $\varepsilon: \cH_{\FJRW,\gamma} \cong
\cH_{\FJRW,\gamma^{-1}}$, then
\begin{equation}
  \widehat{\Gamma}^*_{W,G} = \inv^*\widehat{\Gamma}_{W,G}
  \label{eq:16} 
\end{equation}
The Gamma class and its conjugate satisfy
\begin{equation}
    \widehat \Gamma _{W, G} \circ \widehat \Gamma^*_{W,
      G} = \bigoplus_{\gamma \in G}
    \prod_{j \in I_\gamma} \left( \frac{2\pi i\e{i\pi\theta_j^\gamma}}{\e{2\pi i \theta_j^\gamma}-1}  \right)  \id_{\cH_{\FJRW,\gamma}}\,.
  \label{eq:27}
\end{equation}
The right-hand side looks like a Todd class of an orbifold~\cite{Iritani:2009ab}. It seems that the presence of a Landau-Ginzburg potential is reflected in the restriction to $j\in I_{\gamma}$. This was first observed in \cite{Walcher:2004tx}.

We are now going to show that the definition of the operator $\widehat \Gamma^*_{W,G}$ in~\eqref{lggamma} agrees with the definition of the conjugate Gamma class given in~\eqref{eq:11}. This follows immediately from the isomorphism~\eqref{eq:124} and from~\eqref{eq:126},
\begin{equation}
  \label{eq:139}
  \theta_j^\gamma = \theta_j^{F([k])} = \left\langle (k^Tq)_j \right\rangle, \qquad j=1,\dots,N. 
\end{equation}
We have already shown in Section~\ref{sec:lggamma} that the expression on the right hand side only depends on $\gamma$. 
By~\eqref{eq:78} the eigenvalue in the sector $\cH_{\FJRW,\gamma}$ equals the eigenvalue in the sector $\cH^{(a,c)}_{\gamma^{-1}J}$. The shift by $J$ is accounted for by the shift by $q_j$ in~\eqref{lggamma}. 


%% file: section5.tex
\section{The hemisphere partition function}
\label{sec:hpf}
One of the main cases of interest where Landau-Ginzburg orbifold models arise is when they can be found at special points of the quantum K\"ahler moduli space of Calabi-Yau manifolds. In this section we will use the hemisphere partition function \cite{Hori:2013ika} for GLSMs to compute central charges of B-branes on Landau-Ginzburg orbifold phases and show that it reproduces the results obtained from our proposed formula. To leading order this has already been shown in \cite{Hori:2013ika}.
\subsection{GLSM and hemisphere partition function}
\label{sec:glsm-hemisph-part}
We briefly summarize the necessary definitions of the GLSM and the hemisphere partition function.

A GLSM can be specified by the quadruple $(\mathsf{G},\mathsf{W},\rho_{\pV},\pR)$, where $\pG$ is the gauge group (we will eventually take $\pG=U(1)^{h}$), $\rho_\pV:\pG\rightarrow GL(\pV)$ is the matter representation, with the vector space $\pV$ the space of chiral multiplets. We restrict ourselves to the case where $\rho_\pV:\pG\rightarrow SL(\pV)$ i.e. where the axial R-charge $U(1)_{A}$ is non-anomalous. Identifying $\pV \cong \mC^{\dim \pV}$ we denote the coordinates on $\pV$ by $z$ which are identified with the scalar components of the chiral multiplets. $\pW\in\mathrm{Sym}(\pV^{*})$ is the superpotential, and $\pR:U(1)_V\rightarrow GL(\pV)$ denotes the vector $R$-charge. We require $\pW$ to have weight $2$ under $\pR$ and we also require charge integrality, i.e. $\pR(e^{i\pi})=\rho_\pV(\pJ)$ for some $\pJ\in \pG$. Denote by $\zeta$ and $\theta$ the FI-parameters and $\theta$-angles, respectively. We define $\pt=\zeta-i\theta$ with $e^{\pt}\in\mathrm{Hom}(\pi_1(\pG),\mathbb{C}^{\ast})^{\pi_{0}(\pG)}$. Note that $\pt$ is different from the flat coordinate $t$ in Sections~\ref{sec:cc-scft} to~\ref{sec:fjrw}. 

A B-type D-brane\footnote{Due to shortage of fonts, we denote GLSM B-branes with the same letter as B-branes in the CFT in section \ref{sec:cc-scft}.} $\mathcal{B}$ in the GLSM is characterized by the data $\mathcal{B}=(\pM,\pQ,\rho_{\pM},{\pr}_{*})$ and a contour $\gamma\subset \mathfrak{t}_{\mathbb{C}}:= \mathrm{Lie}(\pT)\otimes_{\mathbb{R}} \mathbb{C}$ (where $\pT$ denotes the maximal torus of $\pG$). The space $\pM=\pM^0\oplus \pM^1$ is a $\mathbb{Z}_2$-graded free $\mathrm{Sym}(\pV^{\ast})$-module -- the Chan-Paton space. $\pQ\in \mathrm{End}^1(\pM)$ is a matrix factorization of $\pW$. The representations $\rho_{\pM}:\pG\rightarrow GL(\pM)$ and ${\pr}_{*}: \mathfrak{u}(1)_V\rightarrow\mathfrak{gl}(\pM)$ are defined by the conditions that $\pQ$ is gauge invariant and has $\pR$-charge $1$:
\begin{eqnarray}
  \rho _{\pM} (g)^{-1}\pQ(gz)\rho_{\pM}(g)&=&\pQ(z)\\
  \lambda^{\pr_{*}}\pQ(\lambda^{\pR} z) \lambda^{-\pr_{*}}&=&\lambda \pQ(z),
  \end{eqnarray}
for all $g\in \pG$ and $\lambda\in U(1)_V$. For the conditions on the contour $\gamma$ we refer the reader to \cite{Hori:2013ika} (see also \cite{Knapp:2016rec} for a summary). We will not need many details about $\gamma$. It will suffice to have in mind that $\gamma$ is a continuous deformation of $\mathrm{Lie}(\pT) \subset \mathfrak{t}_{\mathbb{C}}$ in the region where $Z_{D^{2}}(\mathcal{B})$ (defined below) is convergent. 

Given this data, we can now give a definition for the hemisphere partition function in the Calabi-Yau case:
\begin{equation}
Z_{D^{2}}(\mathcal{B})=C\int_{\gamma} d^{\mathrm{rk}\pG}\sigma \prod_{\alpha>0}\alpha(\sigma)\sinh(\pi\alpha(\sigma))\prod_{j=1}^{\mathrm{dim}\pV}\Gamma\left(iQ_{j}(\sigma)+\frac{R_{j}}{2}\right)e^{i\pt(\sigma)}f_{\mathcal{B}}(\sigma),
\end{equation}
where $\alpha>0$ are the positive roots of $\pG$, $C$ is a normalization constant and $Q_{i}$ and $R_{i}$ denote the weights of $\rho_{\pV}$ and $\pR$, respectively. The information about the brane $\mathcal{B}$ is encoded in the ``brane factor''
\begin{equation}
  f_{\mathcal{B}}(\sigma)=\mathrm{tr}_\pM\left(e^{i\pi \pr_{\ast}}e^{2\pi\rho_{\pM}(\sigma)}\right) = \sum_{\mu=1}^{\dim \pM}e^{i\pi r^{\mu}}e^{2\pi\sum_\alpha w_{\alpha}^{\mu}\sigma_\alpha},
\end{equation}
where $w_\alpha^\mu$ and $r^\mu$ denote the weights of $\rho_{\pM}$ and ${\bf r}_*$, respectively.
In the following we fix
\begin{equation}
\pG=U(1)^{h},
\end{equation}
and denote by 
\begin{equation}
\pC \in\mathrm{Mat}_{h\times \mathrm{dim}\pV} (\mathbb{Z})
\end{equation}
the charge matrix of the GLSM. We identify the weights $Q_{i}$ with the columns of $\pC$.

\subsubsection*{Landau-Ginzburg phases}
\label{sec:land-ginzb-phas}
Whether or not a GLSM has a Landau-Ginzburg phase is in general an open question. A sufficient criterion on the gauge charges and superpotential was proposed in \cite{Herbst:2008jq} and was later proved in \cite{Clarke:2010ep}. In a nutshell, it is a criterion on the matrix $\pC$ and the superpotential $\pW$. The condition for $\pC$ is that there exists a subset of $h$ linearly independent columns of $\pC$ such that the remaining columns lie in a negative cone $\cC$ of the chosen $h$ columns\footnote{In \cite{Clarke:2010ep} the more general case of $\mathrm{rk}\pC<h$ is considered.}. This is equivalent to saying that there exists a cone $\cC$ of the secondary fan such that the symplectic quotient (i.e. a solution to the D-term equations) $Y=\mu^{-1}(\zeta)/U(1)^{h}$ is isomorphic to $\mathbb{C}^{\mathrm{dim}\pV-h}/\Gamma$ for some finite group\footnote{This is shown in \cite{Clarke:2010ep} and $\mu$ denotes the moment map associated to the action of $\pG$ on $\pV$.} $\Gamma\subset \pG$ for any value of $\zeta$ in the interior of such a cone. (This is the cone $\cC$ that appeared in Section~\ref{sec:gkz}.) Then a Landau-Ginzburg phase exists for $\zeta$ deep in the interior of such a cone if the restricted superpotential $\pW:Y\rightarrow \mathbb{C}$ has an isolated critical point at the origin. The action of $\pG$ on $z$ can be written as
\begin{equation}
z_{j}\mapsto e^{i(\pC^{T}\lambda)_{j}}z_{j}\qquad \lambda\in\mathbb{R}^{h}.
\end{equation}
If the criterion on $\pC$ is satisfied this means there exists a basis where $\pC$ takes the block form
\begin{equation}
\pC=(L\ S)\qquad L\in\mathrm{Mat}_{h\times h}(\mathbb{Z}),S\in \mathrm{Mat}_{h\times N}(\mathbb{Z}),
\end{equation}
where $N=\mathrm{dim}\pV-h$ and $L$ is invertible (over $\mathbb{Q}$) and formed by the $h$ linearly independent columns mentioned above. It coincides with the matrix $L$ associated to $q^{\mathrm{geom}}$ we defined in Section \ref{sec:cc-lgo} up to a change of basis\footnote{\label{fn:LG-GLSM}It would be interesting to study the GLSMs one gets from $q^{\mathrm{LG}}$ and $q^{\mathrm{ext}}$. For a closely related discussion, see~\cite{Halverson:2013eua,Sharpe:2013bwa}.}. Using the Smith normal form of $L^T$ (cf.~(\ref{eq:5})) it is clear that the elements
\begin{equation}\label{embedGG}
e^{2\pi iL^{-T}m}\in \pG=\prod_{\alpha=1}^{h}U(1)_{\alpha}\qquad m\in\mathbb{Z}^{h}
\end{equation}
define an embedding of $G_{\mathrm{orb}}$ into $\pG$ where $G_{\mathrm{orb}}$ denotes the finite subgroup defined by the elementary divisors of $L$, cf.~(\ref{eq:122}). Recall that $G_{\mathrm{orb}}$ is isomorphic to the Landau-Ginzburg orbifold group $G$. In the following we will not distinguish between $G$ and $G_{\mathrm{orb}}$. By writing 
\begin{equation}
q=L^{-1}\pC
\end{equation}
the action of $G_{\mathrm{orb}}$ on $\pV$ is nontrivial only on the fields $z_{h+1},\ldots, z_{h+N}$, which we denote as $\phi_{j}$, $j=1,\ldots,N$ (to make the connection with section \ref{sec:cc-lgo}) and is given by
\begin{equation}
\phi_{j}\mapsto e^{2\pi i(q^{T}m)_{j}}\phi_{j}\qquad m\in\mathbb{Z}^{h}.
\end{equation}
The group $G_{\mathrm{orb}}$ is the unbroken gauge group deep in the phase defined by the matrix $\pC$. However, there can be, in general, situations where we are forced to use F-term and D-term equations simultaneously in order to break the gauge group $\pG$ to a finite subgroup. An example of this is presented in section \ref{sec:two-param-example2}, where we have a Landau-Ginzburg phase, even though $\pC$ do not satisfy the aforementioned criterion of \cite{Clarke:2010ep}. In that example, we can still decompose $\pC=(L \ S)$ but the columns of $S$ do not lie in the negative cone of the columns of $L$. We also cannot exclude the possibility of Landau-Ginzburg phases occurring nonperturbatively as strongly coupled phases in nonabelian GLSMs, in an analogous way as some geometric phases are known to be non-perturbatively realized in non-abelian models. In the context of mirror symmetry of non-abelian GLSMs, Landau-Ginzburg models have recently made an appearance in \cite{Gu:2018fpm}. However we will not consider such scenarios in this work. 

In general, for Landau-Ginzburg phases realized at weakly coupled points, the F-term and D-term equations fix nonzero VEVs for exactly $h$ of the fields, say $\{z_1,\ldots,z_h\}$ which breaks $\pG$ to $G$ as described above and the superpotential in the Landau-Ginzburg phase can be taken to be
\begin{equation}
  W(\phi_1,\ldots,\phi_{N}):=\pW(1,\ldots,1,z_{h+1},\ldots,z_{h+N})\qquad \phi_{j}:=z_{h+j}.
\end{equation}
The R-charge assignments for the massless fields can be obtained directly from the fact that the fields that acquire nonzero VEVs must be assigned R-charge zero. This determines the R-charges of $\phi_{1},\ldots,\phi_{N}$ uniquely by imposing that $W$ is quasi-homogeneous of weight $2$. Denote these R-charges by $R_{j}\in (0,2)$, then
\begin{equation}
  W(\lambda^{R_{j}}\phi_{j})=\lambda^{2}W(\phi_{j}),
\end{equation}
so we can identify $\frac{R_{j}}{2}$ with the values of the left R-charges $q_{j}$ of Section \ref{sec:cc-lgo} which we assume to hold from now on. 

A GLSM brane data $\mathcal{B}$ reduces to a Landau-Ginzburg brane $\overline{\mathcal{B}}=(\overline{M},\sigma,\overline{Q},\overline{\rho},\overline{R})$ with
\begin{equation}
  \label{lgbranedef}
  \begin{aligned}
  \overline{M}&=\pM|_{z_a=1},\\
  \overline{Q}(\phi_1,\ldots,\phi_N)&=\pQ(1,\ldots,1,\phi_1,\ldots,\phi_N),\\
  \overline{\rho}&=\rho_{\pM}\vert_{G},\\
  \overline{R}&=e^{i\pi {\bf r}_{*}}|_{R_a=0,R_{h+j}=2q_j}, 
\end{aligned}
\end{equation}
where $a=1,\dots,h$ and $j=1,\dots,N$. 
Moreover, $\sigma=e^{i\pi {\bf r}_{*}}\rho_{\pM}(\pJ)$ with the restrictions on $\pM$, $\rho_{\pM}$ and $e^{i\pi {\bf r}_*}$ indicated in~(\ref{lgbranedef}) and $\pJ \in \pG$ is an element satisfying the charge integrality condition for $\rho_{\pV}$ and $\pR$. The module $\overline M = \bigoplus_{\mu=1}^{\dim M} \overline M_\mu$ corresponds to the module $\pM$ over the specialization $\overline{S} = \mC[\phi_1,\dots,\phi_N]$ of $\pS=\mC[z_1,\dots,z_{\dim \pV}]$ at $z_a=1$, $a=1,\dots,h$, $z_{a+j}=\phi_j$, $j=1,\dots,N$.

Let us now consider the hemisphere partition function and evaluate it in the Landau-Ginzburg phase. Restricting to the abelian case, one gets
\begin{equation} Z_{D^2}(\mathcal{B})=C\int_{\gamma}d^h\sigma\prod_{a=1}^h\Gamma\left(i\sum_{\alpha=1}^hL_{\alpha a}\sigma_{\alpha}\right)\prod_{j=1}^{N}\Gamma\left(i\sum_{\alpha=1}^hS_{\alpha j}\sigma_\alpha+{q_{j}}\right)e^{i\pt(\sigma)}\sum_{\mu=1}^{\dim \pM}e^{i\pi r^{\mu}}e^{2\pi\sum_\alpha w_{\alpha}^{\mu}\sigma_\alpha},
\end{equation}
where we identified $R_{j}/2=q_{j}$. If $\pC$ satisfies the criterion described above, i.e. when $\zeta=\Re(\pt)$ is in the interior of a cone (of the secondary fan) describing a Landau-Ginzburg orbifold phase, it is easy to see that we can deform $\gamma$ to enclose the poles of the term $\prod_{a=1}^h\Gamma\left(i\sum_{\alpha=1}^hL_{\alpha a}\sigma_{\alpha}\right)$. All of them are simple and located at 
\begin{equation}
  \label{simple}
\sigma_\alpha=i\sum_aL^{-1}_{a \alpha}k_{a},\qquad k\in\mathbb{Z}^{h}_{\geq 0}.
\end{equation}
We expect this to hold when $\Re(\pt)$ is in a cone describing a Landau-Ginzburg orbifold phase. More precisely, we expect that a decomposition of the form $\pC=(L\ S)$ is always possible, where $L$ corresponds to the fields acquiring nonzero VEVs in the Landau-Ginzburg orbifold phase and that the poles are all simple. Even when the criterion of~\cite{Clarke:2010ep} is not satisfied, and the convergence of the integral is not obvious, we expect this to be true. We observe this in all our examples. The mechanism at work is that the factor $f_{\mathcal{B}}$ can cancel some non-simple poles, thus forbidding terms polynomial in $\pt$ in the result. Assuming that this is always the case, we proceed to take the residue of the simple poles at (\ref{simple}):
\begin{equation}
  \begin{aligned}
  Z_{D^2}^{LG}(\mathcal{B})&=\frac{C(2\pi)^h}{|\det L|}\sum_{k_1,\dots,k_h=0}^{\infty}\frac{(-1)^{\sum_{a=1}^h k_a}}{\prod_{a=1}^{h}\Gamma(k_a+1)}\prod_{j=1}^N\Gamma\left(-\sum_{a=1}^hq_{a,h+j}k_a+q_{j}\right)\\
  &\phantom{=} \cdot \e{-\sum_{a}\pt'_ak_a}\sum_{\mu=1}^{\dim \pM}\e{i\pi r^{\mu}}\e{2\pi i\sum_{\alpha,a}w^{\mu}_{\alpha}L^{-1}_{a\alpha}k_a}.
  \end{aligned}
\end{equation}
where we defined
\begin{align}
  \pt'_a &=\sum_{\alpha=1}^h L^{-1}_{a\alpha}\pt_{\alpha}.
  \label{eq:tprime}       
\end{align}
The superscript in $Z_{D^2}^{LG}(\mathcal{B})$ indicates that we have evaluated $Z_{D^{2}}$ in a Landau-Ginzburg phase. By (\ref{embedGG}) and (\ref{lgbranedef}), we can write in a diagonal basis 
\begin{equation}
  \label{eq:129}
  \begin{aligned}
    \left. e^{i\pi r_{\mu}} \right|&=\sigma_{\mu}\overline{\rho}(J^{-1})_{\mu},\\
     \left. \e{2\pi i\sum_{\alpha,a}w^{\mu}_{\alpha}L^{-1}_{a\alpha}k_a} \right|&=\overline{\rho}(\gamma)_{\mu},
  \end{aligned}
\end{equation}
where $(\dots)|$ denotes the restrictions indicated in~\eqref{lgbranedef}. To obtain the second equation, we associate an element $\gamma =F([k]) \in G$ to $[k] \in G_{\mathrm{orb}}$ by~\eqref{eq:124}. Moreover, we change the representative $k$ of $[k] \in G_{\mathrm{orb}}$ using a choice for the matrix $U$ that appears in the Smith normal form $UL^{T}V=D$.  To do this, we proceed as in~\eqref{eq:127} and define
\begin{equation}
  k'=Uk.
\end{equation}
This allows us to write $\gamma = \prod_{a=1}^h g_a^{k'_a}$ for the canonical generators $g_a$ of $\mZ_{d_a} \subset G_{\mathrm{orb}}$. As in Section~\ref{sec:lggamma} we choose $U$ such that $F(\langle J \rangle) = \mZ_{d_1}$ and the fundamental domain $\mathcal{F}$ as in (\ref{eq:19}). 
The matrix $V$ then defines (the weights of) the representation $\overline \rho : G_{\mathrm{orb}} \to GL(\overline{M})$ by
\begin{equation}
  \label{eq:130}
  w'=Vw.
\end{equation}
Different choices of $U$ and $V$ yield equivalent elements $\gamma$ and representations $\overline \rho$, respectively.
Hence,
\begin{equation}
  \sum_{\mu=1}^{\dim \pM} \left.\e{i\pi r^{\mu}}\e{2\pi i\sum_{\alpha,a}w^{\mu}_{\alpha}L^{-1}_{a\alpha}k_a} \right\vert =\mathrm{str}_{\overline M}(\overline{\rho}(J^{-1}\gamma)) .
\end{equation}
Then the formula reads
\begin{equation}
   Z_{D^2}^{LG}(\mathcal{B})=\frac{C(2\pi)^h}{|\det L|}\sum_{k_1,\dots,k_h=0}^{\infty}\frac{(-1)^{\sum_{a=1}^h k_a}}{\prod_{a=1}^{h}\Gamma(k_a+1)}\prod_{j=1}^N\Gamma\left(-\sum_{a=1}^hq_{a,h+j}k_a+q_{j}\right)
\e{-\pt'\cdot k}\mathrm{str}_{\overline M}(\overline{\rho}(J^{-1}\gamma)).
\end{equation}
It has been shown in \cite{Hori:2013ika}, that the leading order term
of $Z_{D^2}^{LG}$ indeed reproduces that RR-charge of the
Landau-Ginzburg brane $\overline\cB$. In the following we will show
that the subleading terms combine into the Landau-Ginzburg central
charge function discussed in Section~\ref{sec:centr-charge-form}.

For this purpose, we apply the reflection formula for the Gamma function in the form
\begin{equation}
 \label{eq:1}
 \Gamma(z) = \frac{2\pi i \e{-i\pi z}}{1-\e{-2\pi i z}}\frac{1}{\Gamma(1-z)}
\end{equation}
to the numerator. This yields
\begin{equation}
 \label{eq:2}
 \begin{aligned}
   Z_{D^2}^{LG}(\mathcal{B})&=-\frac{C(2\pi)^h(2\pi i)^{N}}{|G|}\sum_{k_1,\dots,k_h=0}^{\infty}\frac{(-1)^{\sum_{a=1}^h k_a}}{\prod_{a=1}^{h}\Gamma(k_a+1)}\frac{1}{\prod_{j=1}^N\Gamma\left(1+\sum_{i=1}^hq_{a,h+j}k_a+q_{j}\right)} \\
   &\phantom{=} \cdot \prod_{j=1}^N \frac{\e{i\pi(\sum_{a}q_{a,h+j}k_a+q_j)}}{1-\e{2\pi i (\sum_{a}q_{a,h+j}k_a+q_j)}}\e{-\pt'\cdot k} \str_{\overline M}\overline\rho(J^{k'_1-1} \prod_{i=2}^{h}
g_i^{k'_i}) 
 \end{aligned}
\end{equation}
The relation~\eqref{eq:48} yields $\prod_{j=1}^N
\e{i\pi(\sum_{a}q_{a,h+j}k_a+q_j)} = -(-1)^{\sum_{a=1}^h k_a}$. So the final expression reads
\begin{equation}
  \label{eq:51}
  \begin{aligned}
    Z_{D^2}^{LG}(\mathcal{B})&=-\frac{C(2\pi)^{h}(2\pi i)^N}{|G|} \sum_{k_1,\dots,k_h=0}^{\infty}\frac{1}{\prod_{a=1}^{h}\Gamma(k_a+1)} \frac{1}{\prod_{j=1}^N\Gamma\left(1+\sum_{a=1}^hq_{a,h+j}k_a+q_j\right)} \\
    &\phantom{=} \cdot \frac{1}{\prod_{j=1}^N \left(1-\e{2\pi i (\sum_{a}q_{a,h+j}k_a+q_j)}\right)}\e{-\pt'\cdot k} \str_{\overline M}\overline\rho(J^{k'_1-1} \prod_{m=2}^{h}
g_m^{k'_m}) 
  \end{aligned}  
\end{equation}
The comparison with the formula~\eqref{eq:25} for the central charge
in the Landau-Ginzburg theory yields the identification
\begin{equation}
  Z_{LG}(\overline{\cB},u) = Z_{D^2}^{LG}(\cB,\pt)
  \label{eq:38}
\end{equation}
for $C=(2\pi)^{-h}$ and $u_a=\e{-\pt'_a}$, $a=1,\dots,h$.


%% file: section6.tex
\section{Examples}
\label{sec:apps}
In the following we provide several examples for which we show that the proposed formula for the central charge matches with the hemisphere partition function. Our main focus is on Landau-Ginzburg phases. Section \ref{sec:quintic} is devoted to the quintic. Most of the results presented there can be found in the literature, and it is easy to show that that everything works as expected. While this article is mostly concerned with Landau-Ginzburg orbifolds, we use the results of \cite{Hori:2013ika} and \cite{Chiodo:2012qt} to show that the central charge formula also holds in geometric phases, thus providing evidence that it is indeed universal all over the moduli space. Section \ref{sec:two-param-example1} is devoted to a well-studied two-parameter family. In section \ref{sec:two-param-example2} we consider another two-parameter family where the Landau-Ginzburg potential is not Fermat. Here we show in particular that the methods to determine the matrix $q$ also apply to the non-Fermat case, which extends the scope of a criterion to determine Landau-Ginzburg phases \cite{Herbst:2008jq,Clarke:2010ep}. There are some interesting subtleties related to the fact that for the GLSM of this orbifold the D-term equations are not enough to determine the Landau-Ginzburg phase. Our final example, presented in section \ref{sec:4-parameter-example}, is a four-parameter orbifold. In this case we consider two Landau-Ginzburg orbifolds. The first has broad sectors and would thus lie outside the validity of the proposed central charge formula. We then show that, by quotienting by an additional group, one gets a Landau--Ginzburg orbifold with the same Hodge numbers which has only narrow sectors. This is the Landau-Ginzburg realization of an approach in geometry to associate to a model where some moduli are not torically realized another model where all moduli are accounted for by the toric geometry of the ambient space \cite{Berglund:1995gd}. 
\subsection{Quintic}
\label{sec:quintic}
Most of the results presented here can be found in \cite{Candelas:1990rm,Walcher:2004tx,Fan:2007ba,Hori:2013ika}. Here we show that everything matches up in the central charge formula.
\subsubsection{Landau-Ginzburg orbifold}
We start with the Fermat polynomial
\begin{equation}
  \label{eq:113}
  W=\phi_1^5+\phi_2^5+\phi_3^5+\phi_4^5+\phi_5^5.
\end{equation}
We have $d=5$ and $q_i=\frac{1}{5}$ for all $i=1,\ldots,5$. Following the discussion of section \ref{sec-lgoclosed}, we define a $5\times 5$ matrix $M=5\cdot\mathrm{id}_{5\times 5}$. Note that it is already in its Smith normal form. The full automorphism group is $\mathrm{Aut}(W)\simeq\mathbb{Z}_5^5$. 
In the following we will consider the Landau-Ginzburg orbifold\footnote{For other quintic Calabi-Yaus and different quotients by subgroups of $\mathrm{Aut}(W)$ in the context of the Berglund-H\"ubsch-Krawitz mirror construction, see \cite{Doran:2011pw}.} $(W,\langle J\rangle)$.

The state space is $\mathcal{H}=\bigoplus_{\gamma\in \langle J\rangle}\mathcal{H}_{\gamma}$. When choosing a concrete basis we have to be careful whether we use the labeling of the $(a,c)$-ring, the one of the $(c,c)$-ring which is associated to the mirror B-model, or the one in FJRW theory. In this case we have the following relation in accordance with (\ref{eq:64}) and (\ref{eq:78}):
\begin{equation}
  \label{orbbas1}
  \mathbf{e}^{(a,c)}_{J\gamma}=\mathbf{e}^{(c,c)}_{\gamma}=e_{\gamma^{-1}}.
  \end{equation}
We further use $\gamma=J^{\ell}$, $\ell=0,\ldots,4$ and abbreviate ${e}_{\gamma}\equiv {e}_{\ell}$. The sector associated to $\mathbf{e}_{1}^{(a,c)}=\mathbf{e}_0^{(c,c)}$ is $204$-dimensional and thus a broad sector. The sectors $\ell=1,\ldots,4$ only contain the vacuum as a ground state. Hence, these sectors are narrow. Restricting to narrow sectors, the definition (\ref{lgpairing}) yields $(\mathbf{e}^{(c,c)}_{\gamma},\mathbf{e}^{(c,c)}_{\gamma^{-1}})=(\mathbf{e}^{(a,c)}_{J\gamma},\mathbf{e}^{(a,c)}_{J\gamma^{-1}})=(e_{\gamma^{-1}},e_{\gamma})=\frac{1}{5}$. In the following we consistently use the FJRW-basis, because it provides the most intuitive labeling, and only write the final results in terms of the $(a,c)$-rings.

To define the $I$-function and the Gamma class, we need the $q$-matrix:
\begin{equation}
  q=(\begin{array}{rrrrrr}1&-\frac{1}{5}&-\frac{1}{5}&-\frac{1}{5}&-\frac{1}{5}&-\frac{1}{5}\end{array}).
\end{equation}
Related to this, we also define matrices $\pC=(L\ S)$
  so that $q=L^{-1}\pC$. Here we have
  \begin{equation}
\pC=(\begin{array}{rrrrrr}-5&1&1&1&1&1\end{array}).
  \end{equation}
  The matrix $L$ encodes the equivalence relations of $G_{\mathrm{orb}}$. Using definition (\ref{eq:125}) and choosing the fundamental domain (\ref{eq:19}), we define $k\in\mathbb{Z}$ satisfying $k\sim k+5m$. Furthermore we choose $\ell=k+1\:\mathrm{mod}\:5$. It is convenient to introduce states $e_{[k]} \in \cH_{\FJRW}$ for classes $[k] \in G_{\mathrm{orb}}$. The only sector containing marginal deformations is $\gamma=J^2$ ($\ell=2$). We denote the corresponding coordinate by $u$.

  Inserting into the definition (\ref{lggamma}) of the Gamma class restricted to the narrow sectors we get 
  \begin{equation}
    \widehat{\Gamma}=\bigoplus_{\ell=1}^4\Gamma\left(1-\left\langle -\frac{\ell}{5}\right\rangle\right)^5=\bigoplus_{\ell=1}^4\Gamma\left(\frac{\ell}{5}\right)^5.
  \end{equation}
  
  Next, we can define the $I$-function (\ref{eq:22}), after a shift $k\rightarrow k+1$:
  \begin{eqnarray}
    I_{LG}(u)&=&
    -\sum_{k=1}^{\infty}(-1)^{5\left\langle\frac{k}{5}\right\rangle}\frac{u^{k-1}}{\Gamma(k)}\frac{\Gamma\left(\left\langle-\frac{k}{5}\right\rangle\right)^5}{\Gamma\left(1-\frac{k}{5}\right)^5}{e}_{[k]}\nonumber\\
    &\stackrel{k=5n+\ell}{=}&
    -\sum_{n=0}^{\infty}\sum_{\ell=1}^4(-1)^{\ell}\frac{u^{5n+\ell-1}}{\Gamma(5n+\ell)}\frac{\Gamma\left(\left\langle-\frac{5n+\ell}{5} \right\rangle\right)^5}{\Gamma\left(1-\frac{5n+\ell}{5}\right)^5}e_{\ell}.
  \end{eqnarray}
  In the next steps we use standard identities for Gamma functions and well-known results for the quintic. These are summarized in appendix \ref{app-quintic}. Applying the reflection formula the expression can be rewritten as
\begin{equation}
   I_{LG}(u)=-\sum_{\ell=1}^4\frac{(-1)^{\ell}}{\pi^5}\Gamma\left(\left\langle-\frac{\ell}{5} \right\rangle\right)^5\sin^5\frac{\pi \ell}{5}\sum_{n=0}^{\infty}\frac{(-1)^{5n}u^{5n+\ell-1}\Gamma\left(n+\frac{\ell}{5}\right)^5}{\Gamma(5n+\ell)}e_{\ell}.
  \end{equation}
Now we use the reflection formula again in combination with the observation that $\langle x+n\rangle=\langle x\rangle$ and $\langle -x\rangle=1-x$ for $0<x<1$ and $n\in\mathbb{Z}$. 
Upon the identification $u=-5\psi$ we can write the result in terms of the well-known Landau-Ginzburg periods (\ref{d5-basis2}):
\begin{equation}
  I_{LG}(u)=\sum_{\ell=1}^4\frac{1}{\Gamma\left(\frac{\ell}{5}\right)^5}\hat{\varpi}_\ell\:e_{\ell}\equiv \sum_{\ell=1}^4I_{1-\ell}\:\mathbf{e}^{(a,c)}_{1-\ell}.
\end{equation}
Note that the inverse of the Gamma class appears as a normalization factor. In our identification with $\hat{\varpi}_{\ell}$ we have absorbed a factor $\frac{1}{u}$ in the definition. While $\hat{\varpi}_{\ell}$ was not defined this way in \cite{Candelas:1990rm}, it is necessary from the point of view of D-brane masses that at least one of the LG periods starts with a constant term. Otherwise all D-branes would be massless at the LG point $u=0$ contradicting the fact that there is no singularity at the LG point. 

To give a complete definition of the central charge we have to pick a Landau-Ginzburg brane $\overline{\mathcal{B}}=(\overline{M},\sigma,\overline{Q},\overline{R},\overline{\rho}(g))$. Since all the twisted sectors only contain the vacuum, the formula (\ref{rr-walcher}) for the Chern character reduces to $\mathrm{ch}(\overline{Q})=\mathrm{str}(\overline{\rho}(J^{\ell}))\,e_{\ell}$. Putting everything together, the central charge is
\begin{equation}
  \label{d5-zgen}
  Z_{LG}(\overline{\mathcal{B}})=\frac{1}{5}\sum_{\ell=1}^4 \mathrm{str}(\overline{\rho}(J^{-1})^\ell)\hat{\varpi}_\ell.
\end{equation}

It is instructive to discuss an explicit example of a Landau-Ginzburg brane. Consider the matrix factorization\footnote{A further interesting example is the ``canonical'' matrix factorization $\overline{Q}=\sum_{i=1}^5\phi_i\eta_i+\phi_i^4\bar{\eta}_i$ that has also been used in~\cite{Chiodo:2012qt}.} of~\eqref{eq:113}
\begin{equation}
\label{d0orb}
\overline{Q}=(\phi_1+\phi_2)\eta_1+(\phi_1^4-\phi_1\phi_2^3+\phi_1^2\phi_2^2-\phi_1\phi_2^3+\phi_2^4)\bar{\eta}_1+\sum_{i=3}^5\phi_i\eta_{i-1}+\phi_i^4\bar{\eta}_{i-1},
\end{equation}
where $\eta_i,\bar{\eta}_i$ satisfy the Clifford algebra $\{\eta_i,\bar{\eta}_j\}=\delta_{ij}$, and all other anticommutators are zero. To this matrix factorization we can associate five matrices $\overline{\rho}_k(J)$ ($k=0,\ldots,4$) satisfying (\ref{def-orbmat}) and $\overline{\rho}_k^5=\mathrm{id}$ and $\mathrm{str}(\overline{\rho}_k(J^{-1}))^\ell=J^{k \ell}(-1+J^\ell)^4$, where $J=e^{\frac{2\pi i}{5}}$. The data $\overline{\mathcal{B}}_k\equiv(\overline{M}_k,\sigma,\overline{Q},\overline{R},\overline{\rho}_k)$ describes five branes in a $\mathbb{Z}_5$-orbit. The significance of our choice of matrix factorization is that it describes a set of branes of minimal charge that generate the K-theory lattice~\cite{Ashok:2004zb,Brunner:2005fv}. One of these branes is the analytic continuation of the D0-brane to the Landau-Ginzburg point \cite{Herbst:2008jq}. Furthermore one can show that $\overline{\mathcal{B}}_k\rightarrow \overline{\mathcal{B}}_{k+1}$ under Landau-Ginzburg monodromy. Let us consider the case $k=0$. Making use of (\ref{d5-basrel}) and (\ref{d5-basis1}) the central charge is
\begin{equation}
  Z_{LG}(\overline{\mathcal{B}}_0)=\frac{1}{5}\sum_{\ell=1}^4(-1+J^\ell)^4\hat{\varpi}_\ell=-(2\pi i)^4\varpi_0.
  \end{equation}
Similarly, one computes $Z(\overline{\mathcal{B}}_k)=-(2\pi i)^4\varpi_k$. 

As we have seen, two sets of periods, $\hat{\varpi}_\ell$ and $\varpi_\ell$, appear in the discussion. The former transforms diagonally under monodromy transformations. Further note that $\left.\hat{\varpi}_\ell\right\vert_{u\rightarrow 0}= 0$ for $\ell\neq 1$. Since there are no massless branes at the Landau-Ginzburg point, the $\hat{\varpi}_{\ell}$, in contrast to the $\varpi_\ell$, do not have an interpretation as the quantum corrected central charge of some physical brane. 
\subsubsection{GLSM}
We proceed to show that the hemisphere partition function of the quintic evaluated in the LG phase yields the result (\ref{d5-zgen}) for the central charge. This calculation has already been done in \cite{Hori:2013ika}. We repeat it for the readers' convenience. The orbifold has gauge group $\pG=U(1)$ and one FI-theta parameter $\pt=\zeta-i\theta$. The maximal torus of the gauge group is parameterized by the scalar $\sigma$. The chiral matter content is 
\begin{equation}
\begin{array}{c|rc|c}
&p&\phi_1,\ldots,\phi_5&\text{FI}\\
\hline
U(1)&-5&1&\zeta\\
\hline
R&0&\frac{2}{5}&-
\end{array}
\end{equation}
Here we have chosen the GLSM R-charges to match with the LG R-charges. The GLSM superpotential is $\pW=p\cdot G_5(x)$, where $G_5$ is homogeneous of degree $5$. When we talk about branes and matrix factorizations we will choose $G_5$ to be the Fermat quintic. The hemisphere partition function for the quintic is then
\begin{equation}
  \label{d5-zd2def}
Z_{D^2}(\mathcal{B})=C\int d\sigma\: \Gamma(-5i\sigma)\Gamma\left(i\sigma+\frac{1}{5}\right)^5 e^{i\pt\sigma}f_{\mathcal{B}}(\sigma).
\end{equation}
To evaluate (\ref{d5-zd2def}) in the LG phase at $\zeta\ll0$ we have to close the integration contour in the direction of negative imaginary axis, i.e. the contour is oriented clockwise. The first Gamma-factor has a first order pole at $\sigma=-\frac{i}{5}k$ with $k\in\mathbb{Z}_{\geq 0}$. This is easily evaluated
\begin{equation}
Z_{D^2}^{LG}(\mathcal{B})=\frac{2\pi C}{5}\sum_{k=0}^{\infty}\frac{\Gamma\left(\frac{1}{5}(1+k)\right)^5}{\Gamma(1+k)}(-1)^k e^{\pt\frac{k}{5}}f_{\mathcal{B}}\left(-i\frac{k}{5}\right),
\end{equation}
where the brane factor is $f_{\mathcal{B}}(\sigma)=\sum_{\mu}e^{i\pi r^{\mu}}e^{2\pi w^{\mu}\sigma}$. Inserting this explicitly and rewriting the sum using $k=5n+\ell-1$ with $n\in\mathbb{Z}_{\geq 0}$ and $\ell=1,\ldots,5$ we get
\begin{equation}
  Z_{D^2}^{LG}(\mathcal{B})=\frac{2\pi C}{5}\sum_{n=0}^{\infty}\sum_{\ell=1}^5\frac{\Gamma\left(n+\frac{\ell}{5}\right)^5}{\Gamma(5n+\ell)}(-e^{\frac{\pt}{5}})^{5n+\ell-1}\sum_{\mu}e^{i\pi r^{\mu}}e^{-\frac{2\pi i}{5}w^{\mu}(\ell-1)}.
\end{equation}
With $J=e^{\frac{2\pi i}{5}}$ we have the following identification for the brane data in the Landau-Ginzburg phase:
\begin{equation}
e^{i\pi r_{\mu}}=\sigma_{\mu}\overline{\rho}(J^{-1})_{\mu}\qquad e^{-\frac{2\pi i}{5}{w^{\mu}}(\ell-1)}=\overline{\rho}(J^{1-\ell})_{\mu},
\end{equation}
where by $\mu$ we denote the $\mu$th diagonal component of the corresponding matrix, and $\sigma_{\mu}$ is the $\mathbb{Z}_2$-grading in the boundary Landau-Ginzburg orbifold. Putting everything together, the brane factor reduces to the Chern character in the Landau-Ginzburg phase:
\begin{equation}
  \sum_{\mu}e^{i\pi r_{\mu}}e^{-\frac{2\pi i}{5}w^{\mu}(\ell-1)}=\sum_{\mu}\sigma_{\mu}\overline{\rho}(J^{-\ell})_{\mu}=\mathrm{str}(\overline{\rho}(J^{-1})^{\ell}).
  \end{equation}
Comparing with the definition (\ref{d5-basis1}), we immediately recover (\ref{d5-zgen}) upon identifying $u=e^{\frac{\pt}{5}}$ and choosing $C=(2\pi)^{-1}$. Note that we do not have to restrict to the narrow sector since the Chern character for $\ell=5$ corresponding to the broad sector vanishes trivially. Our result is also in agreement with \cite{Hori:2013ika}.
\subsubsection{GLSM and geometry}
In \cite{Hori:2013ika,hori2019notes} it was shown for geometric phases of GLSMs that the hemisphere partition function computes the quantum corrected central charge of a D-brane. Here, we repeat this discussion for the quintic to show that the central charge formula (\ref{ccfinalf}) is consistent with these results. In geometric phases the $I$-function has made an appearance in the context of supersymmetric localization in GLSMs in \cite{Bonelli:2013mma,Ueda:2016wfa,Kim:2016jye,Gerhardus:2018zwb,Goto:2018bci}. 

We start off with (\ref{d5-zd2def}) and change coordinates to $\sigma=in+\frac{z}{2\pi}$. In this section we choose an R-charge assignment such that $p$ has charge $2$ and the $\phi_i$ have charge $0$ in order to be in agreement with the R-charges in the large volume phase. Evaluation in the large volume phase results in 
\begin{equation}
  Z^{LV}_{D^2} = 
  \sum_{n=1}^\infty \oint_0
  \frac{\diff{}{z}}{2\pi} \Gamma(5n+\tfrac{5z}{2\pi i} +1) \Gamma(-n
  -\tfrac{z}{2\pi i})^5  \e{-t n+\frac{i}{2\pi}t z} f_{\mathcal{B}}(in+\tfrac{z}{2\pi}).
\end{equation}
Using $e^{2\pi q_j(in)}=1$ on the brane factor and
\begin{equation}
  \Gamma(-n-\tfrac{z}{2\pi i})=\frac{2\pi i(-1)^{n+1} \e{-\frac{z}{2}}}{\Gamma(1+n+\frac{z}{2\pi i})(1-\e{-z})},
\end{equation}
by means of the reflection formula we obtain
\begin{equation}
  Z^{LV}_{D^2}= (2\pi i)^5 
  \sum_{n=1}^\infty \oint_0
  \frac{\diff{}{z}}{2\pi} \frac{\Gamma(5n+\tfrac{5z}{2\pi i} +1)}{ \Gamma(1+n
    +\tfrac{z}{2\pi i})^5}  \frac{(-1)^{5n+5} \e{-\frac{5z}{2}}}{(1-\e{-z})^5}\e{-t n+\frac{i}{2\pi}t z} f_{\mathcal{B}}(\tfrac{z}{2\pi}).
\end{equation}
Making use of the identity $\frac{z^5}{5z}(1-\e{5z}) = -\frac{z^5}{5z}\e{5z}(1-\e{-5z}) $ and taking into account the theta angle shift at large volume \cite{Herbst:2008jq} by defining  $t' = t -5\pi i$ we rewrite this further as
\begin{equation}
    Z^{LV}_{D^2} = (2\pi i)^5 
    \sum_{n=1}^\infty \oint_0 \frac{\diff{}{z}}{2\pi}\frac{5z}{z^5}
    \frac{\Gamma(5n+\tfrac{5z}{2\pi i} +1)}{ \Gamma(1+n
      +\tfrac{z}{2\pi i})^5} \frac{z^5 (1-\e{-5z})
    }{5z(1-\e{-z})^5}\e{-t' n-\frac{z}{2\pi i}t '} \frac{f_{\mathcal{B}}(\tfrac{z}{2\pi})}{(1-\e{5z})}.
\end{equation}
Now we use the following relation between the brane factor and the Chern character of the brane $\mathcal{B}^{LV}$ in the large volume phase \cite{Hori:2013ika}:
\begin{equation}
  \chern(\mathcal{B}^{LV}) = \frac{f_{\mathcal{B}}(\tfrac{z}{2\pi})}{(1-\e{5z})}.
\end{equation}
Further, we use the following identity for any formal power series $g$ in $z$
\begin{equation}
  \int_X g(H) = \int_{\mP^4} 5H g(H) = \oint_0 \frac{\diff{}{z}}{2\pi
    i} \frac{5z}{z^5} g(z), 
\end{equation}
where $X$ is the quintic. Then the hemisphere partition function can be rewritten as
\begin{equation}
  \begin{aligned}
    Z^{LV}_{D^2}(\cB,t;H) &= (2\pi i)^5 
    \sum_{n=1}^\infty \int_X \frac{\Gamma(1+5n+\tfrac{5H}{2\pi i})}{
      \Gamma(1+n +\tfrac{H}{2\pi i})^5} \frac{H^5 (1-\e{-5H})
    }{5H(1-\e{-H})^5}\e{-t' n-\frac{H}{2\pi i}t'}
    \chern(\mathcal{B}^{LV})\\
    &= (2\pi i)^5 
    \int_X \chern(\mathcal{B}^{LV}) \e{-\frac{H}{2\pi i}t'} \sum_{n=1}^\infty \e{-t'n} \frac{\Gamma(1+5n+\tfrac{5H}{2\pi i})}{
      \Gamma(1+n +\tfrac{H}{2\pi i})^5} \todd(X). \\
  \end{aligned}
\end{equation}
In the second line we used the definition of the Todd class for the quintic
\begin{equation}
  \todd(X) = \frac{H^5(1-\e{-5H})}{5H(1-\e{-H})^5}.
\end{equation}
Now we connect this result to the central charge formula. We have already recovered the Chern character of the brane. We still need to identify the pairing, the $I$-function and the Gamma class. The pairing in $\tH^*(X,\mC)$ is given by $\langle \alpha, \beta\rangle = \int_X \alpha \wedge \beta$ for $\alpha,\beta \in \tH^*(X,\mC)$. The Gamma class for the quintic and its conjugate are
\[
  \widehat{\Gamma}_X = \frac{\Gamma(1+\frac{H}{2\pi i})^5}{\Gamma(1+\frac{5H}{2\pi
      i})}, \qquad \widehat{\Gamma}^*_X = \frac{\Gamma(1-\frac{H}{2\pi
      i})^5}{\Gamma(1-\frac{5H}{2\pi i})}. 
\]
Using the reflection formula and the definition of the Todd class above, they satisfy the relation $\todd(X)=\widehat{\Gamma}_X \widehat{\Gamma}^*_X$. The definition for the $I$-function can be found for instance in~\cite{Cox:2000vi,Chiodo:2012qt}. Specializing to the quintic and suitably choosing the parameters, it can be written as
\begin{equation}
  I_{X}=e^{-\frac{H}{2\pi i}t'}\frac{\Gamma\left(1+\frac{H}{2\pi i}\right)^5}{\Gamma\left(1+\frac{5H}{2\pi i}\right)}\sum_{n\geq 0}e^{-t'n}\frac{\Gamma\left(1+5n+\frac{5H}{2\pi i}\right)}{\Gamma\left(1+n+\frac{H}{2\pi i}\right)^5}.
\end{equation}
Then we can write
\begin{equation}
   Z^{LV}_{D^2}(\cB,t;H) = (2\pi i)^5 
   \langle  \chern(\mathcal{B}^{LV}) , \widehat{\Gamma}^*_X  I_{X}(-t',\tfrac{H}{2\pi i})
   \rangle.
   \label{eq:8}  
\end{equation}
We thus have verified that the central charge formula holds with $\Gamma_{LV} = \widehat \Gamma^*_X$. 
\subsection{Two-parameter family 1}
\label{sec:two-param-example1}
Our next example is well-studied in the mirror symmetry literature. This model has been discussed in detail in \cite{Candelas:1993dm}. 
\subsubsection{Landau-Ginzburg orbifold}
Consider the Fermat polynomial
\begin{equation}
  W=\phi_1^8+\phi_2^8+\phi_3^4+\phi_4^4+\phi_5^4.
\end{equation}
The automorphism group is $\mathrm{Aut}(W)=\mathbb{Z}_8^2\times\mathbb{Z}_4^3$. We discuss the Landau-Ginzburg orbifold $(W,\langle J\rangle)$ where $\langle J\rangle$ acts on the $\phi_i$ with weights $q=\theta^{J}=\left(\frac{1}{8},\frac{1}{8},\frac{1}{4},\frac{1}{4},\frac{1}{4}\right)$.

The labeling conventions of the state space are as in (\ref{orbbas1}) for the quintic. Choosing $J=e^{\frac{2\pi i}{8}}$, we label the sectors by $\gamma=J^{\ell}$ with $\ell=0,\ldots,7$. This corresponds to the FJRW labeling convention with basis vectors $e_{\gamma}\equiv e_{\ell}$. The sectors labelled by $\ell=0$ and $\ell=4$ are broad and have dimensions $168$ and $6$, respectively. They contribute to the odd part of $\cH_{\FJRW}$. The remaining sectors have the vacuum as the only Ramond-Ramond ground state, hence they are narrow. The sectors contributing to the marginal deformations are $\gamma\in G^{(2)}=\{J^2,J^5\}$. The coordinates are $u_1,u_2$, respectively. The orbifold is the Landau-Ginzburg description of the degree $8$ Calabi-Yau hypersurface in weighted $\mathbb{P}^4_{11222}$ with Hodge numbers $(h^{1,1},h^{2,1})=(2,86)$.

To determine the matrix $q$ we follow the steps indicated in Section~\ref{sec:deformations}. The computation is straightforward for this case, so we refrain from giving details. We refer the reader to the subsequent examples which are slightly less trivial. The result for $q$ is 
\begin{equation}
\label{d8-cgmatrix}
  q=\left(\begin{array}{rrrrrrr}
    1&0&-\frac{1}{4}&-\frac{1}{4}&-\frac{1}{4}&-\frac{1}{8}&-\frac{1}{8}\\
    0&1&0&0&0&-\frac{1}{2}&-\frac{1}{2}
    \end{array}\right).
\end{equation}
From this, we can extract the matrix
\begin{equation}
  L=\left(\begin{array}{rr}-4&1\\0&-2\end{array}\right),
\end{equation}
so that there is a matrix $\pC=L\cdot q$ that will be identified with the GLSM charge matrix. 
It is convenient to introduce states $e_{[k_1,k_2]} \in \cH_{\FJRW}$ for classes $[k_1,k_2] \in G_{\mathrm{orb}}$ determined by the equivalence relations encoded in $L$. Concretely, we have 
\begin{equation}
  \label{d8periodicities}
  (k_1+4,k_2)\sim (k_1,k_2+1)\qquad (k_1,k_2+2)\sim (k_1,k_2).
\end{equation}
Given this, one can assign a state $e_{\ell}\leftrightarrow e_{[k_1,k_2]}$ with fixed $(k_1,k_2)$ to each sector $\ell$ of the Landau-Ginzburg orbifold, as we will show below. 
In order to make contact with the results of~\cite{Candelas:1993dm} we will not use the Smith decomposition as discussed in Section~\ref{sec:cc-lgo}. Rather, we choose the following decomposition:
\begin{equation}
  k_1=4n+r-1,\quad k_2=2m+s,\qquad r=1,\ldots,4, \quad s=0,1.
  \end{equation}
Inserting into the definition (\ref{lggamma}) of the Gamma class we get for the six narrow sectors:
\begin{equation}
  \widehat{\Gamma}=\bigoplus_{r=1}^3\Gamma\left(\frac{r}{4}\right)^3\Gamma\left(\frac{r}{8}\right)^2\oplus \bigoplus_{r=1}^3\Gamma\left(\frac{r}{4}\right)^3\Gamma\left(\frac{1}{2}+\frac{r}{8}\right)^2.
\end{equation}
The assignment to the twisted sectors is discussed below in (\ref{twopartwist}). 
For the $I$-function (\ref{eq:22}) we get
\begin{eqnarray}
  I_{LG}(u)&=&-
  \sum_{m,n,r,s}u_1^{4n+r-1}u_2^{2m+s}\frac{(-1)^{G(k,q)}}{\Gamma(4n+r)\Gamma(1+2m+s)}\nonumber\\
  &&\cdot\frac{\Gamma\left(\left\langle-n-\frac{r}{4}\right\rangle\right)^3\Gamma\left(\left\langle-\frac{n}{2}-\frac{r}{8}-m-\frac{s}{2}\right\rangle\right)^2}{\Gamma\left(1-\left(n+\frac{r}{4}\right)\right)^3\Gamma\left(1-\left(\frac{n}{2}+\frac{r}{8}+m+\frac{s}{2}\right)\right)^2}e_{[4n+r,2m+s]},
\end{eqnarray}
where we have chosen the abbreviation $G(k,q)$ for the sign appearing in (\ref{eq:21}). 
To further evaluate this expression we have to distinguish between cases where $n+s$ is even or odd. Using the reflection formula this can be rewritten as
\begin{eqnarray}
  \label{2parifct}
  I_{LG}(u)&=&-
  \sum_{m,n,r,s}(-1)^{G(k,q)}u_1^{4n+r-1}u_2^{2m+s}\frac{(-1)^s\Gamma\left(n+\frac{r}{4}\right)^3\Gamma\left(m+\frac{r}{8}+\frac{n+s}{2}\right)^2}{\Gamma(4n+r)\Gamma(1+2m+s)}e_{4n+r,2m+s}\nonumber\\
  &&\cdot\left\{\begin{array}{cl}\frac{1}{\Gamma\left(\frac{r}{4}\right)^3\Gamma\left(\frac{r}{8}\right)^2}&n+s\:\:\textrm{even} \\ -\frac{1}{\Gamma\left(\frac{r}{4}\right)^3\Gamma\left(\frac{r}{8}+\frac{1}{2}\right)^2}&n+s\:\:\textrm{odd}.\end{array} \right.
  \end{eqnarray}
Further details on the calculation can be found in Appendix \ref{app-d8}. Restricting to $r=1,2,3$ amounts to restricting to the narrow sector. We note that there is a relative minus sign between between the two choices of $n+s$. This relative sign is removed by $(-1)^{G(k,q)}$. Using the properties of $\langle x\rangle$ we find
\begin{equation}
  G(k,q)=3\left\langle\frac{r}{4}\right\rangle+2\left\langle\frac{r}{8}+\frac{n+s}{2} \right\rangle=\left\{\begin{array}{cl}r&n+s\:\:\textrm{even} \\ r+1&n+s\:\:\textrm{odd}.\end{array} \right. 
  \end{equation}
We also observe the appearance of the inverse of the Gamma class in the definition of the $I$-function. 
To make contact with the periods, we recall the definitions in Appendix \ref{app-d8}. We identify
\begin{equation}
  u_1=(-2^{12}\psi^4)^{\frac{1}{4}}\qquad u_2=2\phi,
\end{equation}
where $\phi$ and $\psi$ parameterize the complex structure deformations of the mirror Landau-Ginzburg potential away from the Fermat point, see (\ref{d8-wlg}). Given the definition of the periods of the mirror in~\cite{Candelas:1993dm} it is useful explicitly divide the contributions from $n+s$ even into $n\in2\mathbb{Z},s=0$ and $n\in2\mathbb{Z}+1,s=1$ and those from  $n+s$ odd into $n\in2\mathbb{Z}+1,s=0$ and $n\in2\mathbb{Z},s=1$. Then one can define periods $\hat{\varpi}_{r}^{ev}$ and $\hat{\varpi}_r^{od}$ whose explicit expressions are
\begin{align}
  \hat{\varpi}_r^{ev}=&(-1)^{r+1}\sum_{n\in 2\mathbb{Z}_{\geq0}}\frac{\Gamma\left(n+\frac{r}{4}\right)^4}{\Gamma(4n+r)}(-1)^{n+\frac{r-1}{4}}(2^{12}\psi^4)^{n+\frac{r-1}{4}}\sum_{m}\frac{\Gamma\left(m+\frac{n}{2}+\frac{r}{8}\right)^2}{\Gamma\left(n+\frac{r}{4}\right)\Gamma(2m+1)}(2\phi)^{2m}\nonumber\label{d8-perev}\\
 & +(-1)^{r}\sum_{n\in 2\mathbb{Z}_{\geq0}+1}\frac{\Gamma\left(n+\frac{r}{4}\right)^4}{\Gamma(4n+r)}(-1)^{n+\frac{r-1}{4}}(2^{12}\psi^4)^{n+\frac{r-1}{4}}\sum_{m}\frac{\Gamma\left(m+\frac{n}{2}+\frac{r}{8}+\frac{1}{2}\right)^2}{\Gamma\left(n+\frac{r}{4}\right)\Gamma(2m+2)}(2\phi)^{2m+1},\nonumber\\
\end{align}
\begin{align}
  \hat{\varpi}_r^{od}=&(-1)^{r+1}\sum_{n\in 2\mathbb{Z}_{\geq0}+1}\frac{\Gamma\left(n+\frac{r}{4}\right)^4}{\Gamma(4n+r)}(-1)^{n+\frac{r-1}{4}}(2^{12}\psi^4)^{n+\frac{r-1}{4}}\sum_{m}\frac{\Gamma\left(m+\frac{n}{2}+\frac{r}{8}\right)^2}{\Gamma\left(n+\frac{r}{4}\right)\Gamma(2m+1)}(2\phi)^{2m}\nonumber\label{d8-perod}\\
 & +(-1)^{r}\sum_{n\in 2\mathbb{Z}_{\geq0}}\frac{\Gamma\left(n+\frac{r}{4}\right)^4}{\Gamma(4n+r)}(-1)^{n+\frac{r-1}{4}}(2^{12}\psi^4)^{n+\frac{r-1}{4}}\sum_{m}\frac{\Gamma\left(m+\frac{n}{2}+\frac{r}{8}+\frac{1}{2}\right)^2}{\Gamma\left(n+\frac{r}{4}\right)\Gamma(2m+2)}(2\phi)^{2m+1}.\nonumber\\
\end{align}
In terms of the expressions $\xi_r,\eta_r$ that are defined in~\cite{Candelas:1993dm} (see Appendix \ref{app-d8}) we find the relation
\begin{equation}
  \hat{\varpi}_r^{ev}=-\frac{2\pi i}{u_1} \frac{\alpha^{-2r}}{\alpha-1} (\xi_r+\eta_r),\qquad
  \hat{\varpi}_r^{od}=-\frac{2\pi i}{u_1} \frac{\alpha^{-2r}}{\alpha+1} (\xi_r-\eta_r),
\end{equation}
where $\alpha$ is a primitive eighth root of unity.

In summary, we get the following expression for the full $I$-function
\begin{equation}
  I_{LG}(u)=
  \sum_{r=1}^3\frac{1}{\Gamma\left(\frac{r}{4}\right)^3\Gamma\left(\frac{r}{8}\right)^2}\hat{\varpi}_{r}^{ev}e_r+\frac{1}{\Gamma\left(\frac{r}{4}\right)^3\Gamma\left(\frac{1}{2}+\frac{r}{8}\right)^2}\hat{\varpi}_{r}^{od}e_{r+4}.
\end{equation}
As for the quintic, we have absorbed an additional factor $\frac{1}{u_1}$ in the definition of the periods so that $\hat{\varpi}_{1}^{ev}$ starts with a constant term. As promised above, we have defined $e_{\ell}$ associated to the $\ell$-th twisted sector of the Landau-Ginzburg orbifold by identifying
\begin{equation}
  \label{twopartwist}
  \ell=r+4[(n+s)\:\mathrm{mod}\:2].
\end{equation}
One can show that the periods $\hat{\varpi}_{r}^{ev}$ and $\hat{\varpi}_r^{od}$ satisfy the Picard-Fuchs equations (\ref{twoparpf}) at the Landau-Ginzburg point. Furthermore the monodromy matrix for these periods is diagonal. As for the quintic, the inverse of the Gamma class appears in the definition of the $I$-function. The calculation of the Chern character works as for the quintic. Using the pairing (\ref{lgpairing}) we obtain
\begin{equation}
  Z_{LG}(\overline{\mathcal{B}})=\frac{1}{8}\sum_{r=1}^3\left[\mathrm{str}(\overline{\rho}(J^{-1})^r)\hat{\varpi}_r^{ev}+\mathrm{str}(\overline{\rho}(J^{-1})^{r+4})\hat{\varpi}_r^{od}\right].
\end{equation}
\subsubsection{GLSM}
Let us consider the $\pG=U(1)^2$ GLSM associated to the two-parameter family. The charges are
\begin{equation}
  \begin{array}{r|rr|rrrrr|c}
    &p&\phi_6&\phi_3&\phi_4&\phi_5&\phi_1&\phi_2&\mathrm{FI}\\
    \hline
  U(1)_1&-4&1&1&1&1&0&0&\zeta_1\\
  U(1)_2&0&-2&0&0&0&1&1&\zeta_2\\
  \hline
  U(1)_a&-8&0&2&2&2&1&1&2\zeta_1+\zeta_2\\
  \hline
  R&0&0&\frac{1}{2}&\frac{1}{2}&\frac{1}{2}&\frac{1}{4}&\frac{1}{4}&-
  \end{array}
\end{equation}
The first two lines correspond to the matrix $\pC=(L\ S)$. By $U(1)_a$ we denote the combination of the $U(1)$-charges that is broken to $\mathbb{Z}_8$ in the Landau-Ginzburg phase. For a suitable choice of complex structure parameters the GLSM superpotential can be written as
\begin{equation}
  \pW=p\left(\left(\phi_1^8+\phi_2^8\right)\phi_6^4+\phi_3^4+\phi_4^4+\phi_5^4\right).
\end{equation}
The generic form is $\pW=p\cdot G_{(4,0)}(\phi_1,\ldots,\phi_6)$ where $G_{(4,0)}$ is a homogeneous polynomial of degree $(4,0)$. The orbifold has four phases. The Landau-Ginzburg phase sits at $\zeta_2<0$ and $2\zeta_1+\zeta_2<0$ where the fields $p$ and $\phi_6$ acquire a non-zero VEV. The geometric phase is at $\zeta_1>0,\zeta_2>0$ and corresponds to the smooth hypersurface $G_{(4,0)}=0$ in the ambient toric variety defined by the two $U(1)$s. There is also an orbifold phase corresponding to a singular degree $8$ hypersurface in $\mathbb{P}^4_{11222}$, and a hybrid phase.

The hemisphere partition function is
\begin{equation}
Z_{D^2}(\mathcal{B})=C\int d^2\sigma\:\Gamma\left(i\sigma_1+\frac{1}{4}\right)^3\Gamma\left(i\sigma_2+\frac{1}{8}\right)^2\Gamma(i\sigma_1-2i\sigma_2)\Gamma(-4i\sigma_1)e^{i\pt_1\sigma_1+i\pt_2\sigma_2}f_{\mathcal{B}}(\sigma).
\end{equation}
To evaluate it in the Landau-Ginzburg phase we follow the general discussion in Section \ref{sec:hpf}. It is convenient to change the coordinates to $\rho=L^T(i\sigma)$, i.e. $\rho_1=-4i\sigma_1,\rho_2=i{\sigma_1}-2i\sigma_2$. The partition function transforms into
\begin{equation}
  Z_{D^2}(\mathcal{B})=
  -\frac{C}{8}\int d^2\rho\: \Gamma\left(-\frac{1}{4}\rho_1\right)^3\Gamma\left(-\frac{1}{8}(4\rho_2+\rho_1)\right)^2\Gamma(\rho_2)\Gamma(\rho_1)e^{\frac{1}{2}\rho_2(-\pt_2)+\frac{1}{8}\rho_1(-2\pt_1-\pt_2)}f_{\mathcal{B}}(\rho).
\end{equation}
Note that this change of coordinates amounts to transforming the charge matrix of the GLSM into the $q$-matrix (\ref{d8-cgmatrix}). 

In the Landau-Ginzburg phase the first order poles at $\rho_1=-l$ and  $\rho_2=-k$ with $k,l\in\mathbb{Z}_{\geq 0}$ contribute. After further changing the summation variables to $l=4n+r-1$ and $k=2m+s$ with $n,m\in\mathbb{Z}_{\geq 0}$ and $r=1,2,3,4$, $s=0,1$ one ends up with
\begin{align}
  Z_{D^2}^{LG}(\mathcal{B})&=
  \frac{(2\pi)^2C}{8}\sum_{m,n,r,s}\frac{\Gamma\left(n+\frac{r}{4}\right)^3\Gamma\left(m+\frac{n}{2}+\frac{r}{8}+\frac{s}{2}\right)^2}{\Gamma(4n+r)\Gamma(2m+s+1)}\nonumber\\
  &\quad \cdot (-1)^{r+s-1}e^{\frac{1}{2}(2m+s)\pt_2}e^{\frac{1}{8}(4n+r-1)(2\pt_1+\pt_2)}f_{\mathcal{B}}(m,n,r,s).
\end{align}
We still have to take into account the brane factor $f_{\mathcal{B}}=\sum_{\mu\in M}e^{i\pi r^{\mu}}e^{2\pi(w_1^{\mu}\sigma_1+w_2^{\mu}\sigma_2)}$. Going through all the changes of coordinates and summation indices one gets for one summand of fixed $(w_1^{\mu},w_2^{\mu})$:
\begin{equation}
  e^{2\pi(w_1^{\mu}\sigma_1+w_2^{\mu}\sigma_2)}=e^{-\frac{2\pi i}{8}(r-1)(2w_1^{\mu}+w_2^{\mu})}e^{-i\pi w_2^{\mu}(n+s)}.
\end{equation}
We insert this into the hemisphere partition function, together with the identifications $e^{\frac{2\pt_1+\pt_2}{8}}=u_1$ and $e^{\frac{\pt_2}{2}}=u_2$, or
\begin{equation}
  e^{\pt_1+\frac{\pt_2}{2}}=-(8\psi)^4\qquad e^{\pt_2}=(2\phi)^2.
\end{equation}
Then we get 
\begin{eqnarray}
  Z_{D^2}^{LG}(\mathcal{B})&=&-
  \frac{(2\pi)^2C}{8}\sum_{n,r}(-1)^{r}\sum_{\mu}e^{i\pi r_{\mu}}e^{-\frac{2\pi i}{8}(r-1)(2w^{\mu}_1+w^{\mu}_2)}e^{-i\pi w_2^{\mu}(n+s)}\frac{\Gamma\left(n+\frac{r}{4}\right)^4}{\Gamma(4n+r)}\nonumber\\ && \cdot(-1)^{n+\frac{r-1}{4}}(2^{12}\psi^4)^{n+\frac{r-1}{4}}\sum_{m,s}\frac{\Gamma\left(m+\frac{n}{2}+\frac{r}{8}+\frac{s}{2}\right)^2}{\Gamma\left(n+\frac{r}{4}\right)\Gamma(2m+s+1)}(-1)^s(2\phi)^{2m+s}.
\end{eqnarray}
The brane factor is not completely independent of the summation index $n$ of the infinite sum. However, the contribution only depends on whether $(n+s)$ is even or odd. In the even case, we get $e^{-\frac{2\pi i}{8}(2w^{\mu}_1+w^{\mu}_2)(r-1)}$ with $e^{\frac{2\pi i}{8}(2w^{\mu}_1+w^{\mu}_2)}\in\mathbb{Z}_8$. In complete analogy to the quintic the brane factor reduces to $\overline{\rho}(J^{-r})$. The contributions with $(n+s)$ odd effectively extend the range of $r$ to $r=1,\ldots,8$. To see this, note for instance that $e^{-\frac{2\pi i}{8}(2w_1^{\mu}+w_2^{\mu})\cdot 5}=e^{-\frac{2\pi i}{8}(2w_1^{\mu}+w_2^{\mu})}e^{-\frac{2\pi i}{8}(8w_1^{\mu}+4w_2^{\mu})}=e^{-\frac{2\pi i}{8}(2w_1^{\mu}+w_2^{\mu})}e^{-i\pi w_2^{\mu}}$. In the last step we have used that the brane charges $w_{\alpha}^{\mu}$ are integers. We have obtained a contribution to the brane factor with odd $(n+s)$. Dividing the terms in the hemisphere partition function accordingly, we recover the definitions (\ref{d8-perev}) and (\ref{d8-perod}) of the periods $\hat{\varpi}_r^{ev}$ and $\hat{\varpi}_r^{od}$. In this way we can write the hemisphere partition function as
\begin{equation}
  Z_{D^2}^{LG}(\mathcal{B})=\frac{(2\pi)^2C}{8}\sum_{r=1}^3\mathrm{str}(\overline{\rho}(J^{-r})\hat{\varpi}_r^{ev}+\mathrm{str}(\overline{\rho}(J^{-(r+4)})\hat{\varpi}_r^{od}.
\end{equation}
This confirms the result of the central charge formula with $C=(2\pi)^{-2}$.

\subsubsection{FJRW invariants}
Since we have all the necessary ingredients, we also compute the FJRW invariants for this orbifold. The invariants given below remain conjectural, as we have not checked them by independent methods. To our knowledge this is the first time FJRW invariants have been computed for a multi-parameter orbifold.

First we note that $\{e_{J^2}, e_{J^5}\}\in\mathcal{H}^{(1,1)}_{\FJRW}$ and we write $t=t_1e_{J^2}+t_2e_{J^5}$. The $J$-function (\ref{eq:81}) then takes the following form
\begin{equation}
  \begin{aligned}
  J(t,-z)&=1(-z)e_J+t_1e_{J^2}+t_2e_{J^5}+\\
  &\phantom{=}+(-z)^{-1}\sum_{n_1,n_2}\langle \left(e_{J^2}\right)^{n_1-1} \left(e_{J^5}\right)^{n_2} e_{J^2}\rangle_{0,n_1+n_2}\frac{t_1^{n_1-1}t_2^{n_2}}{(n_1-1)! n_2!}\eta^{J^2J^6}e_{J^6}\\
  &\phantom{=}+(-z)^{-1}\sum_{n_1,n_2}\langle \left(e_{J^2}\right)^{n_1}\left(e_{J^5}\right)^{n_2-1} e_{J^5}\rangle_{0,n_1+n_2}\frac{t_1^{n_1}t_2^{n_2-1}}{n_1!(n_2-1)!}\eta^{J^5J^3}e_{J^3}\\
  &\phantom{=}+(-z)^{-2}\sum_{n_1,n_2}\langle \left(e_{J^2}\right)^{n_1}\left(e_{J^5}\right)^{n_2} \tau_1(e_{J^1})\rangle_{0,n_1+n_2+1}\frac{t_1^{n_1}t_2^{n_2}}{n_1!n_2!}\eta^{J^1J^7}e_{J^7}\\
  &= 1(-z)e_J+t_1e_{J^2}+t_2e_{J^5}+ (-z)^{-1} \left(\cF_{t_1} e_{J^6} + \cF_{t_2} e_{J^3} \right) + (-z)^{-2} \cF_0 e_{J^7}\,,
  \end{aligned}
  \label{2parj}
\end{equation}
where we set $z=1$. Now we use the connection (\ref{eq:103}) to the $I$-function. First we read off the mirror map (\ref{eq:104}):
\begin{equation}
  \begin{aligned}
  t_1(u) & = u_{{1}}+{\frac {3}{128}}\,u_{{1}}{u_{{2}}}^{2}+{\frac {301}{98304}}\,u
_{{1}}{u_{{2}}}^{4}-{\frac {9}{163840}}\,{u_{{1}}}^{5}u_{{2}}+{\frac {
      32677}{62914560}}\,u_{{1}}{u_{{2}}}^{6}
  + O(u^8)\\
  t_2(u) & = u_{{2}}+{\frac {11}{192}}\,{u_{{2}}}^{3}-{\frac {1}{1536}}\,{u_{{1}}}^
{4}+{\frac {15}{2048}}\,{u_{{2}}}^{5}-{\frac {11}{98304}}\,{u_{{1}}}^{
  4}{u_{{2}}}^{2}+{\frac {1549}{1310720}}\,{u_{{2}}}^{7}
+ O(u^8).
  \end{aligned}  
  \label{eq:115}
\end{equation}
Inverting these series and inserting into (\ref{eq:103}) we can read off the invariants from (\ref{2parj}). There are only two types of invariants for which we define the following abbreviations. 
\begin{align}
  \mathrm{FJRW}_{n_1,n_2}&:=\langle\left(e_{J^2}\right)^{n_1}\left(e_{J^5}\right)^{n_2} \rangle_{0,n_1+n_2} \nonumber\\
\mathrm{FJRW}^0_{n_1,n_2}&:=\langle\left(e_{J^2}\right)^{n_1}\left(e_{J^5}\right)^{n_2} \tau_1(e_{J^1}) \rangle_{0,n_1+n_2+1}. 
\end{align}
For both types of invariants the selection rule (\ref{eq:76}) leads to the constraint $n_1+4n_2=6\mod 8$. The invariants are given in Tables \ref{tab1} and \ref{tab2}, respectively. One can verify that there is an $\mathcal{F}$ such that the special geometry relation $\mathcal{F}_0=2\mathcal{F}-t_1\mathcal{F}_{t_1}-t_2\mathcal{F}_{t_2}$ is satisfied, where $\mathcal{F}_{t_i}=\frac{\partial\mathcal{F}}{\partial t_i}$. 
\begin{table}[!h]
 \def\arraystretch{1.3}
  \begin{center}
  \begin{small}
  $
  \begin{array}[h]{c|cccccccccc}
    n_1\backslash n_2&0&1&2&3&4&5&6&7&8&9
    \\ \hline
    2&0&\frac{1}{8}&0&{\frac {1}{256}}&0&{\frac {7}{4096}}&0&{\frac {273}{131072}}&0&{\frac {5027}{1048576}}\\
    6&{\frac {3}{512}}&0&{\frac {9}{8192}}&0&{\frac {243}{262144}}&0&{\frac {3717}{2097152}}&0&{\frac {398709}{67108864}}&0\\
    10&0&{\frac {1143}{262144}}&0&{\frac {44559}{8388608}}&0&{\frac {1821915}{134217728}}&0&{\frac {250010901}{4294967296}}&0&{\frac {6399635103}{17179869184}}
  \end{array}
  $
  \end{small}
  \caption{The invariants $\mathrm{FJRW}_{n_1,n_2}$}\label{tab1}
\end{center}
\end{table}

\begin{table}[!h]
  \def\arraystretch{1.3}
  \begin{center}
  \begin{small}
    $
      \begin{array}[h]{c|cccccccccc}
n_1\backslash n_2&0&1&2&3&4&5&6&7&8&9
\\  \hline 
    2&0&\frac{1}{8}&0&{\frac {3}{256}}&0&{\frac {35}{4096}}&0
&{\frac {1911}{131072}}&0&{\frac {45243}{1048576}}
\\ 6&{\frac {3}{128}}&0&{\frac {27}{4096}}&0&{
\frac {243}{32768}}&0&{\frac {18585}{1048576}}&0&{\frac {1196127}{
16777216}}&0\\ 10&0&{\frac {10287}{262144}}&0&{
\frac {490149}{8388608}}&0&{\frac {23684895}{134217728}}&0&{\frac {
3750163515}{4294967296}}&0&{\frac {108793796751}{17179869184}}
    \end{array}
    $
  \end{small}
  \caption{The invariants $\mathrm{FJRW}^0_{n_1,n_2}$}\label{tab2}
  \end{center}
\end{table}

\subsection{Two-parameter family 2}
\label{sec:two-param-example2}
Our next example is a Landau-Ginzburg orbifold which is not a Gorenstein singularity, i.e. the weights $w_j = q_j d$ do not divide $d$ for all $j$. Such Landau-Ginzburg orbifolds have been studied in the context of mirror symmetry and LG/CY-correspondence in~\cite{Hosono:1993qy,Candelas:1994bu,Berglund:1995gd}. The present example was considered in the latter two references. The superpotential $W$ is still invertible and consists of Fermat and chain type polynomials. The hypersurface $W=0$ in $\mP^4(w_1,\dots,w_N)$, however, has non-Gorenstein singularities inherited from the ambient weighted projective space. In~\cite{Hosono:1993qy,Candelas:1994bu,Berglund:1995gd} it is proposed to pass to a birationally equivalent hypersurface $\widetilde W = 0$ in a Gorenstein toric variety $\mP_{\Delta^*}$ for a reflexive polytope $\Delta^*$. It was shown that this is possible for all Landau-Ginzburg orbifolds with $\widehat c=3$ and $N=4,5$. The toric variety $\mP_{\Delta^*}$ is the natural setting in which the LG/CY-correspondence can be studied with the gauged linear sigma model. We will show that in the context of the latter there are some new features as opposed to the Fermat hypersurfaces which are probably remnants of the fact that one starts with a non-Gorenstein singularity. The outcome is that the $I$-functions as defined in Section~\ref{sec:cc-lgo} in terms of the matrix $q$ associated to $(W,G)$, and as defined in Section~\ref{sec:fjrw} in terms of Givental's formalism agree. In particular, the $I$-functions take the same form independent of whether $(W,G)$ is Gorenstein or not. In this subsection we focus on determining the matrix~$q$. The calculation of the $I$-function and the evaluation of the hemisphere partition function are completely analogous to the previous example. We point out, however, that due to the fact that $W$ is not Gorenstein, there are contributions from non-concave insertions in the computation of the FJRW invariants. Hence, one needs to work with the more general virtual class~\eqref{nonconc} and use the results of~\cite{Guere:2016ab}.

\subsubsection{Landau-Ginzburg orbifold}
We consider the following orbifold:
\begin{equation}
  W=\phi_1^7+\phi_2^7+\phi_3^7+\phi_4^3\phi_2+\phi_5^3\phi_3,
\end{equation}
where the weights of the $\phi_i$ are $q=\left(\frac{1}{7},\frac{1}{7},\frac{1}{7},\frac{2}{7},\frac{2}{7}\right)$, and we choose $G=\langle J\rangle \cong \mZ_7$. This orbifold describes a Calabi-Yau threefold with Hodge numbers $(h^{1,1},{h^{2,1}})=(2,95)$. The untwisted sector is broad and $192$-dimensional. The six twisted sectors only contain the vacuum, i.e. they are narrow.

Let us discuss the $q$ matrix in some detail.   From
$W$ we obtain the matrix of exponents
\begin{equation}
  M=
  \begin{pmatrix}
    7 & 0 & 0 & 0 & 0 \\
    0 & 7 & 0 & 1 & 0 \\
    0 & 0 & 7 & 0 & 1 \\
    0 & 0 & 0 & 3 & 0 \\
    0 & 0 & 0 & 0 & 3 \\

  \end{pmatrix}.
\end{equation}
From the Smith normal form of $M$ we get $\Aut(W) =\mZ_7 \times
\mZ_{21} \times \mZ_{21}$ whose generators can be read off from the
rows of $M^{-1}$ in (\ref{eq:minverse})
\begin{equation}
  M^{-1}=
  \begin{pmatrix}
    \frac{1}{7} & 0 & 0 & 0 & 0\\
    0 & \frac{1}{7} & 0 & -\frac{1}{21} & 0\\
    0 & 0 & \frac{1}{7} & 0 & -\frac{1}{21}\\
    0 & 0 & 0 & \frac{1}{3} & 0\\
    0 & 0 & 0 & 0 &\frac{1}{3}\\
  \end{pmatrix}.
  \end{equation}
  We denote the generators of $\Aut(W)$ by $g_1,\ldots,g_5$. Note that $g_2^7=g_4^{-1}$ and $g_3^7=g_5^{-1}$ so that $\Aut(W)=\langle g_1,g_2,g_3\rangle\simeq\mathbb{Z}_7\times\mathbb{Z}_{21}\times\mathbb{Z}_{21}$. The grading element is
  \begin{equation}
    J=g_1\cdot\ldots\cdot g_5=((V^{-1})^T\cdot(1,1,1,1,1)^T)=\frac{1}{7}(1,1,1,2,2)^T,
  \end{equation}
  where, by abuse of notation we denote $\rho_m(g)\equiv g$. The columns of $M^{-1}$ generate $\Aut(W^T)$ and we denote the generators by $g_1^{\vee},\ldots,g_5^{\vee}$. They satisfy the relations $g_2^{\vee}=(g_4^{\vee})^{-3}$ and $g_3^{\vee}=(g_5^{\vee})^{-3}$. We choose as minimal generators $\{g_1^{\vee},g_4^{\vee},g_5^{\vee}\}$. The grading element of $\Aut(W^T)$ is
  \begin{equation}
    J^{\vee}=g_1^{\vee}\cdot\ldots\cdot g_5^{\vee}=(V^{-1}\cdot(1,1,1,1,1)^T)=\frac{1}{21}(3,2,2,7,7)^T. 
    \end{equation}
  Recall furthermore that $\Aut(W^T)\simeq\mathrm{Hom}(\Aut(W),\mC^{\ast})\simeq \Aut(W)$.

  In order to determine $G^{\vee}$, it is useful to have $J^{\vee}$ among the generators of $G^{\vee}$. We observe that
  \begin{eqnarray}
    J^{\vee}&=&g_1^{\vee}(g_4^{\vee})^{-2}(g_5^{\vee})^{-2}\in SL(5,\mC),\\
    g^{\vee}&:=&g_4^{\vee}(g_5^{\vee})^{-1}=\frac{1}{21}(0,20,1,7,14)\in SL(5,\mC).
  \end{eqnarray}
  With $g_4^{\vee}=((g^{\vee})^2(J^{\vee})^{-1}g_1^{\vee})^{16}$ and $g_5^{\vee}=((g^{\vee})^2J^{\vee}(g_1^{\vee})^{-1})^5$, we can write $\mathrm{Hom}(\Aut(W),\mC^{\ast})=\langle J^{\vee},g^{\vee},g_1^{\vee}\rangle$. Since $g_1^{\vee}\notin SL(5,\mC)$ we have $G^{\vee}\cap SL(5,\mathbb{C})=\langle J^{\vee},g^{\vee}\rangle$. The vectors $v\in \mathcal{A}_{\mathrm{ext}}$ 
  have to satisfy $J^{\vee}\cdot v=1$ and $g^{\vee}\cdot v=0\mod\mathbb{Z}$.  The solutions can be arranged into the columns of the following matrix:
  \begin{equation}
  M\spcheck =
  \begin{pmatrix}
 1 & 3 & 7 & 0 & 0 & 0 & 0 \\
 1 & 3 & 0 & 7 & 0 & 0 & 0 \\
 1 & 3 & 0 & 0 & 7 & 0 & 0 \\
 1 & 0 & 0 & 1 & 0 & 3 & 0 \\
 1 & 0 & 0 & 0 & 1 & 0 & 3 \\
  \end{pmatrix}
\end{equation}
where the first two columns form the submatrix $M'$. Finally, the
matrix $q$ is
\begin{equation}
  q=\left(\begin{array}{rrrrrrr}1&0&-\frac{1}{7}&-\frac{1}{7}&-\frac{1}{7}&-\frac{2}{7}&-\frac{2}{7}\\
0&1&-\frac{3}{7}&-\frac{3}{7}&-\frac{3}{7}&\frac{1}{7}&\frac{1}{7}.
  \end{array}\right)
  \end{equation}
Note that the bottom right entries are positive. Such situations have not been discussed in \cite{Clarke:2010ep}, so this is an example, alluded to in section \ref{sec:hpf}, where the columns of $S$ do not lie in the cone spanned by the negative columns of $L$. The matrix $L$ is
\begin{equation}
  L=\left(\begin{array}{rr}-3&1\\-1&-2 \end{array}\right).
  \end{equation}
\subsubsection{GLSM}
Let us briefly comment on the GLSM of this orbifold. The gauge group is $\pG=U(1)^2$ and the matter content is
\begin{equation}
  \begin{array}{c|rr|rrrrr|c}
    &p&\phi_6&\phi_4&\phi_5&\phi_1&\phi_2&\phi_3&\mathrm{FI}\\
    \hline
  U(1)_1&-3&1&1&1&0&0&0&\zeta_1\\
  U(1)_2&-1&-2&0&0&1&1&1&\zeta_2\\
  \hline
  U(1)_a&-7&0&2&2&1&1&1&2\zeta_1+\zeta_2\\
  \hline
  R&0&0&\frac{4}{7}&\frac{4}{7}&\frac{2}{7}&\frac{2}{7}&\frac{2}{7}&-
  \end{array}
\end{equation}
One can consider the GLSM potential
\begin{equation}
  \label{d7-glsmpot}
  \pW=p\left(\phi_6^3\left(\phi_1^7+\phi_2^7+\phi_3^7\right)+\phi_4^3\phi_2+\phi_5^3\phi_3 \right).
\end{equation}
Finding the Landau-Ginzburg point of this GLSM is more difficult than for the GLSMs related to Fermat-type Landau-Ginzburg potentials. The Landau-Ginzburg phase is at $2\zeta_1+\zeta_2<0$ and $-\zeta_1+3\zeta_2<0$, where in both cases $\zeta_2<0$. The associated D-terms are
\begin{align}
  -|p|^2-2|\phi_6|^2+\sum_{i=1}^3|\phi_i|^2=&\zeta_2\nonumber\\
  3\sum_{i=1}^3|\phi_i|^2-(|\phi_4|^2+|\phi_5|^2)-7|\phi_6|^2=&-\zeta_1+3\zeta_2\nonumber\\
  -7|p|^2+\sum_{i=1}^3|\phi_i|^2+2(|\phi_4|^2+|\phi_5|^2)=&2\zeta_1+\zeta_2.
\end{align}
In the Landau-Ginzburg phase we have $p\neq0,\phi_6\neq 0$. However, this cannot be concluded from the D-terms alone since $\phi_6=0$ is not excluded. To see that $\phi_6\neq 0$, one also has to take into account the F-term equations. If one sets $\phi_6=0$ the F-terms imply that also $\phi_1,\ldots,\phi_5$ are zero, which is disallowed by the D-terms. Hence one concludes that $\phi_6\neq 0$.

This phenomenon is also reflected in the hemisphere partition function
\begin{equation}
Z_{D^2}=C\int d^2\sigma\:\Gamma\left(i\sigma_1+\frac{2}{7}\right)^2\Gamma\left(i\sigma_2+\frac{1}{7}\right)^3\Gamma(i\sigma_1-2i\sigma_2)\Gamma(-3i\sigma_1-i\sigma_2)e^{i\pt_1\sigma_1+i\pt_2\sigma_2}f_{\mathcal{B}}(\sigma).
\end{equation}
After changing the coordinates to $\rho=L^T(i\sigma)$, we get
\begin{align}
  Z_{D^2}&=-\frac{C}{7}\int d^2\rho\: \Gamma\left(\frac{1}{7}(-2\rho_1+\rho_2+2)\right)^2\Gamma\left(\frac{1}{7}(-\rho_1-3\rho_2+1)\right)^3\Gamma(\rho_2)\Gamma(\rho_1)\nonumber\\
  &\quad\cdot e^{\frac{\rho_1}{7}(-2\pt_1-\pt_2)+\frac{\rho_2}{7}(-\pt_1+3\pt_2)}f_{\mathcal{B}}(\rho).
\end{align}
The integral converges in the Landau-Ginzburg phase for $\rho_1\in\mathbb{Z}_{\leq0},\rho_2\in\mathbb{Z}_{\leq 0}$. Apart from the expected first order poles there seem to be additional poles from $\Gamma\left(\frac{1}{7}(-2\rho_1+\rho_2+2)\right)^2$. We claim that these will always be cancelled by the brane factor. To gather some evidence, consider matrix factorizations of (\ref{d7-glsmpot}). If one considers matrix factorizations such as $\pQ=\phi_4\eta_1+\phi_5\eta_2+p\phi_4^2\phi_2\bar{\eta}_1+p\phi_5^2\phi_3\bar{\eta}_2+...$, where the monomials containing $\phi_4,\phi_5$ are factorized individually, the brane factor will always contain a factor $(1-e^{2\pi k\sigma_1})(1-e^{2\pi l\sigma_1})$ with $k,l\in\{1,2\}$. This will cancel the unwanted second-order poles. One could for instance avoid this by constructing a matrix factorization $\pQ=(\phi_4+\alpha \phi_5)\eta_1+\ldots$, which would only give a factor $(1-e^{2\pi \sigma_1})$ in the brane factor, thus leaving an extra first order pole. However, this is not possible for generic $\phi_2,\phi_3$. Also other standard types of matrix factorizations such as $\pQ=p\eta+G_{(3,1)}(\phi_1,\ldots,\phi_6)\bar{\eta}$ do not change this, since the associated hemisphere partition function is zero. The fact that one needs the brane factor to see that only the first order poles contribute is consistent with the observation that one needs the F-terms to see the Landau-Ginzburg phase: the brane factor is the only datum where F-term information enters.

Once the issue of the poles has been clarified, the evaluation of the hemisphere partition function in the Landau-Ginzburg phase is the analogous to the two-parameter family of Section~\ref{sec:two-param-example1}, so we omit the details. We end with the FJRW invariants in Table~\ref{tab:FJRWd7}.
\begin{table}[!h]
  \def\arraystretch{1.3}
  \centering
    $
    \begin {array}{c|ccccccccccccc}
      (n_1,n_2) & (2,1)&(0,4)&(5,0)&(3,3)&(1,6)&(6,2)&
(4,5)&(2,8)&(9,1)\\ 
   \hline
\FJRW_{n_1,n_2} &\frac{1}{7}&{\frac {27}{49}}&{\frac {5}{
343}}&-{\frac {3}{343}}&{\frac {9612}{16807}}&{\frac {414}{117649}}&-{
\frac {6096}{117649}}&{\frac {3365820}{823543}}&{\frac {1692}{823543}}
\\ 
\FJRW^0_{n_1,n_2}&\frac{1}{7}&{\frac {54}{49}}&{\frac {15}{343}}&-{\frac {
12}{343}}&{\frac {48060}{16807}}&{\frac {2484}{117649}}&-{\frac {6096}
{16807}}&{\frac {26926560}{823543}}&{\frac {13536}{823543}}
\end {array} 
$
\caption{The first few non-zero  FJRW invariants
  $\mathrm{FJRW}_{n_1,n_2} =\left\langle \left(e_{J^2}\right)^{n_1}\left(e_{J^4}\right)^{n_2} \right\rangle_{0,n_1+n_2}$ and $
  \mathrm{FJRW}^0_{n_1,n_2}=\left\langle \left(e_{J^2}\right)^{n_1}\left(e_{J^4}\right)^{n_2} \tau_1\left(e_J\right)\right\rangle_{0,n_1+n_2+1}$.
}  
\label{tab:FJRWd7}
\end{table}

\subsection{4-parameter family and broad sectors}
\label{sec:4-parameter-example}
Out final set of examples shows a connection between an family with broad sectors and a related family where all the moduli are encoded in narrow sectors. 
\subsubsection{Landau-Ginzburg orbifolds}
We consider two Landau-Ginzburg orbifolds $(W,G_1)$ and $(W,G_2)$ with Fermat superpotential
\begin{equation}
  \label{d9-fermat}
  W=\phi_1^9+\phi_2^9+\phi_3^9+\phi_4^3+\phi_5^3.
\end{equation}
The automorphism group is $\Aut(W)=\mathbb{Z}_{9}^3\times\mathbb{Z}_3^2$. The orbifold groups are $G_1 = \langle J\rangle \cong \mZ_9$ and $G_2 = \langle J, g\rangle \cong \mZ_9 \times \mZ_3$, where the generators $J$ and $g$ are specified in terms of their phases as
\begin{equation}
  \begin{aligned}
    \theta^J &=\frac{1}{9}(1,1,1,3,3)\\
    \theta^g&=\frac{1}{3}(0,0,0,1,2).
  \end{aligned}
  \label{eq:50}
\end{equation}
Analyzing the two orbifolds with PALP one finds that in both cases $(h^{1,1},h^{2,1})=(4,112)$. The way this is encoded in the twisted sectors is however very different, see Appendix~\ref{sec:palp-landau-ginzburg} for details on both cases.

Let us start with $(W,G_1)$. We label the sectors by $\gamma=J^{\ell}$ and write $e_{\gamma}=e_{\ell}$. The $\ell=0$ sector $\cH_{0}$ is broad and has dimension $226$. This accounts for the odd part of $\cH_{\FJRW}$. For the twisted sectors labelled by $\ell=1,\ldots,8$ we find that $\dim \cH_3 = \dim \cH_6 = 2$, i.e.~these sectors are broad but contribute to the even part of $\cH_{\FJRW}$.  On the mirror, the elements of the $(c,c)$-ring corresponding to the RR ground states are $\{\phi_4,\phi_5\}$. The other twisted sectors are narrow, i.e. they are one-dimensional and only contain the vacuum. If we apply our definition of the central charge function $Z_{LG}$ to this orbifold we have to restrict to the narrow sectors. The result therefore only depends on two of the four moduli. 

The second orbifold $(W,G_2)$ provides a way to make all four moduli visible with the methods at hand. We use PALP to analyze the state space. We label the twisted sectors of $\cH_{\FJRW}$ corresponding to $(J^{\ell_1},g^{\ell_2})$ by $(\ell_1,\ell_2)$ with $\ell_1\in\{0,\ldots,8\}$ and $\ell_2\in\{0,1,2\}$, and introduce basis vectors $e_\gamma=e_{\ell_1,\ell_2}$ as in~\eqref{orbbas1}. Now, only the sectors $(0,\ell_2)$, $\ell_2=0,1,2$ are broad and nonzero and contribute to the odd part of $\cH_{\FJRW}$. All the other sectors are either zero or only contain the vacuum state. In particular the sectors with $\ell_1=3,6$ -- that were broad in the $(W,G_1)$ orbifold above -- split up into two narrow sectors labelled by $\ell_2=1,2$, respectively. Therefore all of the even part of $\cH_{\FJRW}$ is accounted for by narrow sectors, and the central charge function $Z_{LG}(u)$ depends on all four marginal directions, as we will now demonstrate.

Let us construct the matrix $q$. The exponent matrix $M=\mathrm{diag}(9,9,9,3,3)$ is already in Smith normal form. From this we can deduce $\mathrm{Aut}(W)\simeq(\mathbb{Z}_9)^3\times(\mathbb{Z}_3)^2\simeq\mathrm{Aut}(W^T)$. As usual we denote the respective generators by $g_1,\ldots,g_5$ and $g_1^{\vee},\ldots,g_5^{\vee}$. Note that $J=g_1\cdot\ldots\cdot g_5$ and $g=g_4g_2^{-1}$. We only consider $G_2$ here. 
To determine $G_2\spcheck$ 
we define
\begin{equation}
  g^{\vee}=g_4^{\vee}(g_5^{\vee})^{-1}, \quad \tilde{g}_1^{\vee}=g_1^{\vee}(g_2^{\vee})^{-1},\quad \tilde{g}_2^{\vee}=g_1^{\vee}(g_3^{\vee})^{-1}.
\end{equation}
Hence, $\mathrm{Hom}(\mathrm{Aut}(W),\mC^{\ast})\cap SL(5,\mC)=\langle J^{\vee},g^{\vee}, \tilde{g}_1^{\vee},\tilde{g}_2^{\vee}\rangle$, where $J\spcheck =  g_1\spcheck\cdot\ldots\cdot g_5\spcheck$. From this we conclude that $\mathrm{Hom}(G_2,\mC^{\ast})=\langle \tilde{g}_1^{\vee},\tilde{g}_2^{\vee}\rangle$. Therefore $\mathcal{A}_{\mathrm{ext}}$ in (\ref{eq:52}) consists of $v\in(\mathbb{Z}_{\geq0})^5$ satisfying
\begin{equation}
  J^{\vee}v=1,\qquad \tilde{g}_{1,2}^{\vee}v=0\mod\mathbb{Z}.
  \end{equation}
This has thirteen solutions:
\begin{equation}
  \label{eq:131}
  \begin{aligned}
      \cA_{\mathrm{ext}} &=\{(0, 0, 0, 0, 3), (0, 0, 0, 1, 2), (0, 0, 0, 2, 1), (0, 0, 0, 3, 
      0), (0, 0, 9, 0, 0), \\
      &\phantom{=} (0, 9, 0, 0, 0), (1, 1, 1, 0, 2), (1, 1, 1, 1, 
      1), (1, 1, 1, 2, 0), (2, 2, 2, 0, 1), \\
      &\phantom{=}
      (2, 2, 2, 1, 0), (3, 3, 3, 0, 0), (9, 0, 0, 0, 0)\}.
  \end{aligned}
\end{equation}
To get $\cA_{\mathrm{geom}}$ in~\eqref{eq:116} we remove each vector that only contains a single zero.
We arrange the vectors $v \in \cA_{\mathrm{geom}}$ into the matrix
\begin{equation}
  M^{\vee}=\left(\begin{array}{ccccccccc}
    1&0&0&3&9&0&0&0&0\\
    1&0&0&3&0&9&0&0&0\\
    1&0&0&3&0&0&9&0&0\\
    1&1&2&0&0&0&0&3&0\\
    1&2&1&0&0&0&0&0&3
  \end{array}\right).
\end{equation}
The sectors corresponding to the deformations of the polynomial $W^T$ are labelled by
\begin{equation}
  \label{4pardefgeom}
  \gamma\in\{J^2,g,g^2,J^4 \}.
  \end{equation}
The kernel of this matrix, with the normalization determined by~\eqref{eq:120}, determines the matrix $q$ 
\begin{equation}
  \label{4parq}
 q\equiv q^{\mathrm{geom}}= \left(
\begin{array}{rrrrrrrrr}
 1 & 0& 0& 0& -\frac{1}{9} & -\frac{1}{9} &  -\frac{1}{9} &  -\frac{1}{3} &
   -\frac{1}{3} \\
 0 & 1& 0 &0 &0 & 0 & 0  & -\frac{1}{3} & -\frac{2}{3} \\
 0 &0 & 1&0 &0 & 0 & 0 &  -\frac{2}{3} & -\frac{1}{3} \\
 0 &0 &0&1 & -\frac{1}{3} & -\frac{1}{3} & -\frac{1}{3} & 0 & 0 \\
\end{array}
\right).
\end{equation}
In this example, $\cA_{\mathrm{LG}}$ is different from $\cA_{\mathrm{geom}}$. The marginal deformation sectors $\cH^{(a,c)}_{\gamma^{-1},(-1,1)}$ in~\eqref{eq:118} correspond to the twisted sectors $\cH_{\FJRW,\gamma}$ with $\gamma \in G^{(2)}$ where
\begin{equation}
  \label{4parg2}
  G^{(2)}=\{ J^2, J^3g, J^3g^2,  J^4 \}
  \end{equation}
  in~\eqref{eq:88}. By~\eqref{eq:133}, these correspond to the vectors
\begin{equation}
  (1,1,1,1,1),(3,3,3,0,0), (2,2,2,0,1), (2,2,2,1,0) \in \cA_{\mathrm{ext}}.
  \label{eq:134}
\end{equation}
Together with the column vectors of the matrix $M^T$, these are precisely the vectors in the set $\cA_{\mathrm{LG}}$ in~\eqref{eq:135}.

Given (\ref{4parq}) we have 
\begin{equation}
  L=\left(\begin{array}{rrrr}
    -3&2&2&1\\
    0&-2&1&0\\
    0&1&-2&0\\
    0&0&0&-3
  \end{array}\right).
\end{equation}
Note that this choice of $L$ is not unique since $q$ has more than one $4\times 4$ minor of value $\frac{1}{27}$. It is however easy to show that all the other choices of $L$ are related by similarity transformations. Erasing the second and third rows and columns $L$ and second and third rows and last two columns in $q$ corresponds to the data for the orbifold $(W,\langle J\rangle)$. 
Again we introduce representatives states $e_{[(k_1,k_2,k_3,k_4)]} \in \cH_{\FJRW}$ for classes $[(k_1,k_2,k_3,k_4)]$. The equivalence relations as encoded in $L$ are:
\begin{equation}
  \begin{aligned}
   (k_1,k_2+2,k_3+2,k_4+1)&\sim (k_1+3,k_2,k_3,k_4)& (k_1,k_2,k_3+1,k_4)&\sim (k_1,k_2+2,k_3,k_4)\\
    (k_1,k_2,k_3+2,k_4)&\sim  (k_1,k_2+1,k_3,k_4) &(k_1,k_2,k_3,k_4)&\sim (k_1,k_2,k_3,k_4+3).
    \end{aligned}
\end{equation}
In order to evaluate the central charge formula in the Landau-Ginzburg phase we use the Smith decomposition. One choice for decomposing $L^T$ is given by
\begin{equation}
  \label{4par-smith}
  U=\left(
\begin{array}{rrrr}
 1 & 0 & 0 & 3 \\
 0 & 1 & 2 & -6 \\
 0 & 0 & 1 & -2 \\
 0 & 0 & 0 & 1 \\
\end{array}
\right)
\quad
S=\left(
\begin{array}{rrrr}
 9 & 0 & 0 & 0 \\
 0 & 3 & 0 & 0 \\
 0 & 0 & 1 & 0 \\
 0 & 0 & 0 & 1 \\
\end{array}
\right)\quad
V=\left(
\begin{array}{rrrr}
 -3 & 0 & 0 & 1 \\
 -6 & -2 & 1 & 0 \\
 -6 & -1 & 0 & 0 \\
 -1 & 0 & 0 & 0 \\
\end{array}
\right).
\end{equation}
We define $k'=U\cdot k$:
\begin{equation}
  k_1=k_1'-3k_4', \quad k_2=k_2'-2k_3'+2k_4', \quad k_3=k_3'+2k_4', \quad k_4=k_4' .
\end{equation}
The primed basis explicitly exhibits the periodicities of the orbifold $(W,G_2)$. Therefore it makes sense to define
\begin{eqnarray}
  k_1'&=&9n_1+\ell_1-1\qquad \ell_1=1,\ldots,9, \quad n_1\in\mathbb{Z}_{\geq 0},\\
  k_2'&=&3n_2+\ell_2\qquad \ell_2=0,1,2,\quad n_2\in\mathbb{Z}_{\geq 0}.
\end{eqnarray}

Let us collect the ingredients that enter the central charge formula. Mapping the states $e_{[(k_1,k_2,k_3,k_4)]}$ to the twisted sectors $e_{(\ell_1,\ell_2)}$ is obvious in the $k'$-basis:
\begin{equation}
  e_{[(k_1,k_2,k_3,k_4)]}=e_{[\ell_1-1+9n_1-3k_4',\ell_2+3n_2-2k_3'+2k_4',k_3'+2k_4',k_4']}= e_{(\ell_1,\ell_2)}.
  \end{equation}
In other words, a fixed $(\ell_1,\ell_2)$ with any $(n_1,n_2,k_3',k_4')$ contributes to the same sector.

The Gamma class reduces to\footnote{Note that the labeling $\widehat{\Gamma}_{\ell_1,\ell_2}$ does not coincide with the labeling of (\ref{lggamma}) which has been defined using the labeling of the $(a,c)$-ring.} 
\begin{equation}
  \widehat{\Gamma}=\bigoplus_{\ell_1,\ell_2}\Gamma\left(1-\left\langle-\frac{\ell_1}{9}\right\rangle\right)^3\Gamma\left(1-\left\langle-\frac{\ell_1}{3}-\frac{\ell_2}{3}\right\rangle\right)\Gamma\left(1-\left\langle -\frac{\ell_1}{3}-\frac{2\ell_2}{3}\right\rangle\right)=\bigoplus_{\ell_1,\ell_2}\widehat{\Gamma}_{\ell_1,\ell_2}.
\end{equation}
Further, we get for the sign in (\ref{eq:21}):
\begin{equation}
  (-1)^{G(k,q)}=3\left\langle\frac{\ell_1}{9}\right\rangle+\left\langle\frac{\ell_1}{3}+\frac{\ell_2}{3}\right\rangle+\left\langle\frac{\ell_1}{3}+\frac{2\ell_2}{3}\right\rangle.
\end{equation}
Next, we compute the $I$-function. We use the coordinates $u_1,\ldots,u_4$ corresponding to the elements in (\ref{4pardefgeom}), respectively. Following the discussion in Section~\ref{sec:deformations} these are not the ``natural'' coordinates describing the deformations from the point of view of the Landau-Ginzburg description as the sectors $\cH_g$ and $\cH_{g^2}$ are actually zero. The ``natural'' coordinates are instead $u^{\mathrm{LG}}_a,a=1,\dots,4$ corresponding to the genuine marginal deformation sectors labelled by~\eqref{4parg2}. As discussed in Section~\ref{sec:gkz} there is a change of variables $u^{\mathrm{LG}}_a= u^{\mathrm{LG}}_a(u_1,\dots,u_4),a=1,\dots,4$ in terms of rational functions. In the present case, they turn out to be rather involved, so we refrain from displaying them here.
We now continue our discussion using $q^{\mathrm{geom}}$. In Appendix \ref{app-4par} we give the FJRW invariants for this family, which we compute using $q^{\mathrm{LG}}$. 
Inserting into (\ref{eq:22}), the $I$-function is:
\begin{align}
  I_{LG}(u)=&-
  \sum_{n_1,n_2,k_3',k_4'}\frac{(-1)^{G(k,q)} u_1^{\ell_1-1+9n_1-3k_4'}u_2^{\ell_2+3n_2-2k_3'+2k_4'}u_3^{k_3'+2k_4'}u_4^{k_4'}}{\Gamma(\ell_1+9n_1-3k_4')\Gamma(1+\ell_2+3n_2-2k_3'+2k_4')\Gamma(1+k_3'+2k_4')\Gamma(1+k_4')}\nonumber\\
  &\cdot \frac{\Gamma\left(\left\langle-\frac{\ell_1}{9}\right\rangle\right)^3\Gamma\left(\left\langle-\frac{\ell_1}{3}-\frac{\ell_2}{3}\right\rangle\right)}{\Gamma\left(1-\left(\frac{\ell_1}{9}+n_1\right)\right)^3\Gamma\left(1-\left(\frac{\ell_1}{3}+3n_1+\frac{\ell_2}{3}+n_2+k_4'\right)\right)}\nonumber\\
 & \cdot\frac{\Gamma\left(\left\langle-\frac{\ell_1}{3}-\frac{2\ell_2}{3}\right\rangle\right)}{\Gamma\left(1-\left(\frac{\ell_1}{3}+3n_1+\frac{2\ell_2}{3}+2n_2-k_3'+k_4'\right)\right)}e_{(\ell_1,\ell_2)}.
\end{align}
Note that the powers of the exponents of the $u_i$ are all positive, because they coincide with $k_i\geq 0$. The same holds for the arguments of the denominator of the Gamma functions in the first line. 
It is convenient to rewrite this by applying the reflection formula to the numerator and the denominator in the second and third line.
Reflecting the numerator immediately gives the inverse of the Gamma class. 
Let us consider the $\sin$-factors on gets after applying the reflection formula:
\begin{eqnarray}
  &&\frac{\sin^3\pi\left(\frac{\ell_1}{9}+n_1\right)\sin\pi\left(\frac{\ell_1}{3}+3n_1+\frac{\ell_2}{3}+n_2+k_4'\right)\sin\pi\left(\frac{\ell_1}{3}+3n_1+\frac{2\ell_2}{3}+2n_2-k_3'+k_4'\right)}{\sin^3\pi\left(\left\langle -\frac{\ell_1}{9}\right\rangle\right)\sin\pi\left(\left\langle -\frac{\ell_1}{3}-\frac{\ell_2}{3}\right\rangle\right)\sin\pi\left(\left\langle -\frac{\ell_1}{3}-\frac{2\ell_2}{3}\right\rangle\right)}\nonumber\\
  &=&(-1)^{9n_1+3n_2-k_3'+2k_4'}\frac{\sin^3\pi\left(\frac{\ell_1}{9}\right)\sin\pi\left(\frac{\ell_1}{3}+\frac{\ell_2}{3}\right)\sin\pi\left(\frac{\ell_1}{3}+\frac{2\ell_2}{3}\right)}{\sin^3\pi\left(\left\langle -\frac{\ell_1}{9}\right\rangle\right)\sin\pi\left(\left\langle -\frac{\ell_1}{3}-\frac{\ell_2}{3}\right\rangle\right)\sin\pi\left(\left\langle -\frac{\ell_1}{3}-\frac{2\ell_2}{3}\right\rangle\right)}
\end{eqnarray}
This expression is well-defined in all the narrow sectors. We will now argue that this expression, in combination with $-(-1)^{G(k,q)}$, produces the sign $(-1)^{\ell_1-1-9n_1+\ell_2+3n_2-k_3'+2k_4'}$ that we will also see in the hemisphere partition function below.

Whenever $\ell_2=0$ and $\ell_1$ corresponds to a narrow sector the numerator and the denominator cancel without a sign. To see this, note first that for $|x|<1$ and $x>0$
\begin{equation}
  \sin\pi\langle -x\rangle=\sin\pi(1-x)=-\sin\pi(-x)=\sin\pi x.
\end{equation}
Further note that there are no additional signs from the $\sin^3$-terms since the argument can never become greater than $1$ in the narrow sectors. There can be signs from the other $\sin^1$-factors, but for $\ell_2=0$ these will be the same in both $\sin^1$-factors in the numerator and thus cancel.

Finally let us consider the cases with $\ell_2=1,2$, which we technically only have to consider for $\ell_1=3,6$. For $\ell_2=2$ the third $\sin$-factor in the numerator gives an additional minus sign.

Now let us consider $(-1)^{G(k,q)}$. Also here we immediately see that for $\ell_2=0$ and $\ell_1$ narrow we get that $G(k,q)=\ell_1$, up to some shifts by even numbers that do not matter. For the narrow sectors with $\ell_2\neq 0$ there is a sign discrepancy between $G(k,q)$ and $\ell_1+\ell_2$ whenever $\ell_2=2$. This is precisely cancelled by the excess minus signs we got when we have rewritten the $I$-function. 

In the end we find that the $I$-function formally has the form
\begin{equation}
  I_{LG}(u)=\frac{1}{\widehat{\Gamma}_{\ell_1,\ell_2}}\sum_{\ell_1,\ell_2}\hat{\varpi}_{\ell_1,\ell_2}e_{(\ell_1,\ell_2)},
\end{equation}
where the explicit form of the periods $\hat{\varpi}_{\ell_1,\ell_2}$ is
\begin{eqnarray}
  \hat{\varpi}_{\ell_1,\ell_2}&=&\sum_{n_1,n_2,k_3',k_4'}\frac{(-1)^{\ell_1-1+9n_1+\ell_2+3n_2-k_3'-2k_4'}u_1^{\ell_1-1+9n_1-3k_4'}u_2^{\ell_2+3n_2-2k_3'+2k_4'}u_3^{k_3'+2k_4'}u_4^{k_4'}}{\Gamma(\ell_1+9n_1-3k_4')\Gamma(1+\ell_2+3n_2-2k_3'+2k_4')\Gamma(1+k_3'+2k_4')\Gamma(1+k_4')}\nonumber\\
  &&\cdot \Gamma\left(\frac{\ell_1+9n_1}{9}\right)^3\Gamma\left(\frac{\ell_1+9n_1}{3}+\frac{\ell_2+3n_2}{3}+k_4'\right)\nonumber\\
  &&\cdot\Gamma\left(\frac{\ell_1+3n_1}{3}+\frac{2(\ell_2+3n_2)}{3}-k_3'+k_4'\right).
\end{eqnarray}
 One can show that they satisfy the GKZ differential equations of the mirror Calabi-Yau at the Landau-Ginzburg point and they transform diagonally under Landau-Ginzburg monodromy. Further details on these differential operators can be found in appendix \ref{app-4par}.

 Since we only have narrow sectors, the Landau-Ginzburg central charge (\ref{rr-walcher}) for a Landau-Ginzburg brane $(\overline{Q},\overline{\rho}(J,g))$ reduces to $\mathrm{ch}_{\ell_1,\ell_2}(\overline{Q})=\mathrm{str}(\overline{\rho}(J^{\ell_1}g^{\ell_2})e_{\ell_1,\ell_2}$. This is only non-zero in the narrow sectors. Using the standard pairing (\ref{lgpairing}) the central charge can be written as
\begin{equation}
  Z_{LG}=\frac{1}{27}\sum_{\ell_1,\ell_2}\mathrm{str}(\overline{\rho}(J^{\ell_1-1}g^{\ell_2})\hat{\varpi}_{\ell_1,\ell_2}.
\end{equation}
Given the $I$-function we can also compute the FJRW invariants. The results can be found in Appendix \ref{app-4par}.
\subsubsection{Relation between $(W,G_1)$ and $(W,G_2)$}
A natural question to ask is whether this four-parameter orbifold is equivalent to the four-parameter orbifold with broad sectors. One criterion is that the chiral rings should isomorphic as rings and not only as vector spaces. This is the case in geometry: For the geometry $X_2$ associated to $(W,G_2)$ all the divisor classes are induced from divisor classes of the ambient variety and the intersection ring $H^*(X_2)$ can be computed by standard methods. For the geometry $X_1$ associated to $(W,G_1)$ only two of the four divisor classes come from the ambient variety. The other two are primitive classes. Using the methods described in \cite{Hosono:1993qy,Mavlyutov:2000dj} we can show that $H^*(X_1) \cong H^*(X_2)$ as rings.

For a Landau-Ginzburg orbifold the ring structure constants of the state space $\cH_{\FJRW}$ are the FJRW invariants. Since FJRW theory only sees the narrow sectors, one does not get the full set of structure constants for $(W,G_1)$. Therefore it is currently not possible to compare the rings of the two orbifolds. One way to show the equivalence would be to proceed via analytic continuation to geometry.

Let us also briefly comment on the matrix factorizations of the two orbifolds. Let us start with $(W,G_1)$. In \cite{Caviezel:2005th} the matrix factorizations that account for the full RR-charge lattice have been identified as
  \begin{align}
    \label{z9-mf}
    \overline{Q}=&(\phi_1-\beta_1 \phi_2)\eta_1+\prod_{j=1}^{8} (\phi_1-\beta_1^{2j+1} \phi_2)\bar{\eta}_1+\phi_3\eta_2+\phi_3^{8}\bar{\eta}_2\nonumber\\
  &+(\phi_4-\beta_2 \phi_5)\eta_3+\prod_{j=1}^{2} (\phi_4-\beta_2^{2j+1} \phi_5)\bar{\eta}_3\qquad \beta_1=e^{\frac{i\pi}{9}},\beta_2=\{e^{\frac{i\pi}{3}},-1\}.
  \end{align}
  The two choices of the parameter $\beta_2$ means that one needs two matrix factorizations to see the full charge lattice. Using (\ref{rr-walcher}) for either of the two matrix factorizations, one finds non-zero contributions to the RR-charge from all twisted sectors, including the two contributions from $\{\phi_4,\phi_5\}$ in the sectors $\ell=3,6$. Note that this is not the case for the ``canonical'' matrix factorization
\begin{equation}
  \overline{Q}_{can}=\sum_{i}\phi_i\eta_i+\frac{1}{w_i}\frac{\partial W}{\partial \phi_i}\bar{\eta}_i,
  \end{equation}
where the contribution to the RR-charge in the sectors $\ell=3,6$ is zero.

On the other hand, if we consider the orbifold $(W,G_2)$  (\ref{z9-mf}) is not a valid matrix factorization because it is not equivariant under the $\mathbb{Z}_3$ action. However, if we consider the canonical matrix factorization, each element of the narrow sector contributes to the RR-charge. It would be interesting to further compare the D-brane categories of these two orbifolds. 
\subsubsection{GLSM}
Consider a GLSM with $\pG=U(1)^4$ and the following matter content:
\begin{equation}
  \label{4parglsm}
  \begin{array}{c|rrrr|rrrrr|c}
    &p&\phi_6&\phi_7&\phi_8&\phi_1&\phi_2&\phi_3&\phi_4&\phi_5&\mathrm{FI}\\
    \hline
    U(1)_1&-3&2&2&1&0&0&0&-1&-1&\zeta_1\\
    U(1)_2&0&-2&1&0&0&0&0&0&1&\zeta_2\\
    U(1)_3&0&1&-2&0&0&0&0&1&0&\zeta_3\\
    U(1)_4&0&0&0&-3&1&1&1&0&0&\zeta_4\\
    \hline
    U(1)_a&-9&0&0&0&1&1&1&3&3&3\zeta_1+6\zeta_2+6\zeta_3+\zeta_4\\
    U(1)_b &0&-3&0&0&0&0&0&1&2&2\zeta_2+\zeta_3\\
    \hline
    R&0&0&0&0&\frac{2}{9}&\frac{2}{9}&\frac{2}{9}&\frac{2}{3}&\frac{2}{3}&-
    \end{array}
\end{equation}
This GLSM can also be obtained by following \cite{Berglund:1995gd} where it was shown in examples how to modify a lattice polytope associated to a Calabi-Yau with non-toric moduli to obtain a Calabi-Yau with the same Hodge data where all the moduli are torically realized. As an interesting side remark, we point out that this GLSM exhibits a second Landau-Ginzburg phase by giving a VEV to the fields $p$, $\phi_4$, $\phi_5$, $\phi_8$. This yields $q=(\frac{1}{9}, \frac{1}{9}, \frac{1}{9}, \frac{1}{3}, \frac{1}{3})$ and orbifold group $\langle J \rangle$, hence corresponds to $(W',G_1)$ with $W'$ consisting of Fermat and loop terms.

The top left block in (\ref{4parglsm}) encodes the matrix $L$. Indeed, this orbifold has a Landau-Ginzburg phase where $U(1)_a$ and $U(1)_b$ are broken to $\mathbb{Z}_9\times\mathbb{Z}_3$. The GLSM that leads to the Landau-Ginzburg orbifold with only the $\mathbb{Z}_9$-orbifold is obtained from (\ref{4parglsm}) by removing $U(1)_2$ and $U(1)_3$ and the matter fields $\phi_6$ and $\phi_7$.

The Fermat superpotential (\ref{d9-fermat}) lifts to the following GLSM potential
\begin{equation}
  \pW=p\left((\phi_1^9+\phi_2^9+\phi_3^9)\phi_8^3+\phi_4^3\phi_6\phi_7^2+\phi_5^3\phi_6^2\phi_7 \right).
\end{equation}
The hemisphere partition function is 
\begin{align}
  Z_{D^2}(\mathcal{B})=&C\int d^4\sigma\: \Gamma(-3i\sigma_1)\Gamma(2i\sigma_1-2i\sigma_2+i\sigma_3) \Gamma(2i\sigma_1+i\sigma_2-2i\sigma_3)\Gamma(i\sigma_1-3i\sigma_4)\nonumber\\
  &\cdot\Gamma\left(i\sigma_4+\frac{1}{9}\right)^3\Gamma\left(-i\sigma_1+i\sigma_3+\frac{1}{3}\right)\Gamma\left(-i\sigma_1+i\sigma_2+\frac{1}{3}\right)e^{i\sum_j \pt_j\sigma_j}f_{\mathcal{B}}(\sigma).
\end{align}
Using $L$, we change the coordinates to
\begin{equation}
  \rho_1=-3(i\sigma_1)\quad \rho_2=2(i\sigma_1)-2(i\sigma_2)+(i\sigma_3)\quad \rho_3=2(i\sigma_1)+(i\sigma_2)-2(i\sigma_3)\quad\rho_4=(i\sigma_1)-3(i\sigma_4),
\end{equation}
which results in
\begin{align}
  Z_{D^2}(\mathcal{B})=&\frac{C}{27}\int d^4\rho\Gamma(\rho_1)\Gamma\left(\rho_2\right)\Gamma\left(\rho_3\right)\Gamma(\rho_4)\Gamma\left(-\frac{\rho_1}{9}-\frac{\rho_4}{3}+\frac{1}{9}\right)^3\Gamma\left(-\frac{\rho_1}{3}-\frac{\rho_2}{3}-\frac{2\rho_3}{3}+\frac{1}{3}\right)\nonumber\\
  &\cdot \Gamma\left(-\frac{\rho_1}{3}-\frac{2\rho_2}{3}-\frac{\rho_4}{3}+\frac{1}{3}\right) e^{-\frac{1}{9}\rho_1(3\pt_1+6\pt_2+6\pt_3+\pt_4)}e^{-\frac{1}{3}\rho_2(2\pt_2+\pt_3)}e^{-\frac{1}{3}\rho_3(\pt_2+2\pt_3)}e^{-\frac{1}{3}\rho_4\pt_4}f_{\mathcal{B}}(\rho).
\end{align}
In the Landau-Ginzburg phase the poles at $\rho_i=-k_i$ with $k_i\geq 0$ contribute. Again, we use the Smith decomposition (\ref{4par-smith}). Defining
\begin{equation}
  w^{\mu}_1=\bar{w}^{\mu}_4 \quad w^{\mu}_2=\bar{w}^{\mu}_3 \quad w^{\mu}_3=-\bar{w}^{\mu}_2-2\bar{w}^{\mu}_3 \quad w^{\mu}_4=-\bar{w}^{\mu}_1+6\bar{w}^{\mu}_2+6\bar{w}^{\mu}_3-3\bar{w}^{\mu}_4,
\end{equation}
the exponent of $e^{2\pi\sum_iw^{\mu}_i\sigma_i}$ becomes
\begin{equation}
  \frac{2\pi i}{9}\left(\bar{w}^{\mu}_1(\ell_1-1)+\bar{w}^{\mu}_2\ell_2+9\bar{w}^{\mu}_1n_1+9\bar{w}^{\mu}_2n_2+9\bar{w}^{\mu}_3k_3'+9\bar{w}^{\mu}_4k_4'\right)\sim \frac{2\pi i}{9}\left(\bar{w}^{\mu}_1\ell_1+\bar{w}^{\mu}_2\ell_2\right),
\end{equation}
where we have used that $\bar{w}^{\mu}_i$ and $k_i'$ are integer. We have implemented the same coordinate changes as for the $I$-function.  With
\begin{equation}
  \bar{w}^{\mu}_1=-3w^{\mu}_1-6w^{\mu}_2-6w^{\mu}_3-w^{\mu}_4,\qquad \bar{w}^{\mu}_2=-2w^{\mu}_2-w^{\mu}_3,
  \end{equation}
the whole brane factor reduces to $\mathrm{ch}_{\ell_1,\ell_2}(\overline{Q})$.

With $z_i=e^{-\pt_i}$ we expect the Landau-Ginzburg periods to depend on the following variables expressed in terms of the $z_i$:
\begin{equation}
  \label{d9-lglvvar}
  u_1=
  z_1^{-\frac{1}{3}}z_2^{-\frac{2}{3}}z_3^{-\frac{2}{3}}z_4^{-\frac{1}{9}}\quad u_2=z_2^{-\frac{2}{3}}z_3^{-\frac{1}{3}}\quad u_3=z_2^{-\frac{1}{3}}z_3^{-\frac{2}{3}}\quad u_4=z_4^{-\frac{1}{3}}.
\end{equation}
This combines into 
\begin{equation}
e^{it(\sigma)}=u_1^{k_1'}u_2^{k_2'}\left(\frac{u_3}{u_2^2}\right)^{k_3'}\left(\frac{u_2^2u_3^2u_4}{u_1^3}\right)^{k_4'}.
\end{equation}
This is the same combination that appears in the $I$-function. The Gamma terms in the primed basis before applying any reflection formulas or performing the integral are
\begin{eqnarray}
  &&  \Gamma\left(-k_1'+3k_4'\right)\Gamma\left(-k_2'+2k_3'-2k_4'\right)\Gamma\left(-k_3'-2k_4' \right)\Gamma\left(-k_4'\right)\nonumber\\
  &&\cdot \Gamma\left(\frac{1}{9}+\frac{k_1'}{9}\right)^3\Gamma\left(\frac{1}{3}+\frac{k_1'}{3}+\frac{k_2'}{3}+k_4'\right)\Gamma\left(\frac{1}{3}+\frac{k_1'}{3}+\frac{2k_2'}{3}-k_3'+k_4'\right).
\end{eqnarray}

While it is not obvious in the primed basis, all the poles come from the first four Gamma factors. When we evaluate the residue we apply the reflection formula on them, which will lead to sign contributions. Putting everything together, we arrive at the following result for the hemisphere partition function.
\begin{align}
  Z_{D^2}^{LG}=&\frac{(2\pi)^4C}{27}\sum_{\ell_{1,2},n_{1,2},k_{3,4}'}\frac{(-1)^{\ell_1-1+9n_1+\ell_2+3n_2-k_3'-2k_4'}u_1^{\ell_1-1+9n_1-3k_4'}u_2^{\ell_2+3n_2-2k_3'+2k_4'}u_3^{k_3'+2k_4'}u_4^{k_4'}}{\Gamma(\ell_1+9n_1-3k_4')\Gamma(1+\ell_2+3n_2-2k_3'+2k_4')\Gamma(1+k_3'+2k_4')\Gamma(1+k_4')}\nonumber\\
  &\cdot \Gamma\left(\frac{\ell_1+9n_1}{9}\right)^3\Gamma\left(\frac{\ell_1+9n_1}{3}+\frac{\ell_2+3n_2}{3}+k_4'\right)\Gamma\left(\frac{\ell_1+3n_1}{3}+\frac{2(\ell_2+3n_2)}{3}-k_3'+k_4'\right)\nonumber\\
  &\cdot\sum_Me^{i\pi r}e^{\frac{2\pi i}{9}(\bar{w}^{\mu}_1(\ell_1-1)+\bar{w}^{\mu}_2\ell_2)}\nonumber\\
  =&\frac{(2\pi)^4C}{27}\sum_{\ell_1,\ell_2}\mathrm{str}(\overline{\rho}(J^{\ell_1-1}g^{\ell_2}))\hat{\varpi}_{\ell_1,\ell_2}.
\end{align}
This matches precisely with the central charge formula. It would be interesting to study D-branes and D-brane transport in this GLSM beyond the level of brane charges.
\subsubsection{Scope of the construction}
 We have demonstrated that it is consistent to replace a Landau-Ginzburg orbifold with a broad sector by a different Landau-Ginzburg orbifold, related by an orbifold, that only has narrow sectors so that the FJRW and GLSM technology can be applied.

Normally, orbifolding changes a theory significantly, so we cannot expect that our construction works for all orbifolds with broad sectors. The example we have given, however, is not a mere coincidence, and there are further ones in this class. To understand this, we consider the reflexive polytopes that encode the toric data of these orbifolds and analyze how the additional orbifold acts on them.

We observe that the geometry $X_1$ associated to $(W,\langle J\rangle)$ is a genus one fibration with a $3$-section. There is an additional $\mathbb{Z}_3$-action permuting the components of the $3$-section. The geometry of $X_2$ associated to $(W,\langle J,g\rangle)$ is an elliptic fibration. The idea is that performing the orbifold with respect to this $\mathbb{Z}_3$-action on $X_1$, one should get $X_2$. 

The fibers of genus one fibered Calabi-Yau hypersurfaces in a toric variety are characterized by a two-dimensional reflexive section of the lattice polytope associated to the toric variety. 

\begin{figure}
  \begin{center}
    \includegraphics{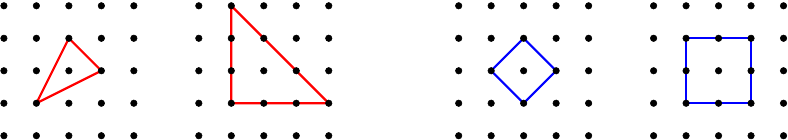}
\end{center}
    \caption{Two-dimensional reflexive lattice polytope related via lattice refinements corresponding to a $\mathbb{Z}_3$ (left) and a $\mathbb{Z}_2$ (right) orbifold.}\label{fig-polytopes}
\end{figure}

Orbifolding is equivalent to a lattice refinement in the point lattice of the polytope. Among the $16$ reflexive lattice polytopes in two dimensions there are two pairs that are related via a lattice refinement. This is depicted in figure \ref{fig-polytopes}.
The corresponding orbifold actions are a $\mathbb{Z}_3$-action with weights $(1,2)$ and a $\mathbb{Z}_2$-action with weights $(1,1)$. We suspect that whenever the polytope associated to our orbifold contains these structures, it is possible to resolve some of the broad sectors into narrow ones by orbifolding. Of course this only works if the RR ground states in the broad sectors are built of the fiber coordinates of the elliptic fibration. Broad sectors that do not have this property cannot simply be resolved into narrow ones by orbifolding.

A further family where this works is the $\mathbb{Z}_{12}$ orbifold of the Fermat Landau-Ginzburg orbifold with weights $w=(1,1,1,3,6)$. This corresponds to a Calabi-Yau with Hodge numbers $(h^{1,1},h^{2,1})=(3,164)$. The twisted sectors $k=4,8$ are one-dimensional, but broad, since the RR ground state is not the identity. It is easy to see that the corresponding geometry is an elliptic fibration and one can show, using PALP, that a $\mathbb{Z}_2$-orbifold with weights $(0,0,0,1,1)$ does not change the Hodge numbers. One can further show that all the twisted sectors of the resulting theory are narrow. The calculation of the central charge in the Landau-Ginzburg orbifold and using the hemisphere partition function is completely analogous to the four-parameter families presented here. In geometry these families have been discussed in detail in \cite{Berglund:1995gd}. There, the connection between the two families has been established via modifications of the corresponding lattice polytopes.

Going through the list of Calabi-Yaus related to $A$-type Gepner orbifolds (see for instance \cite{Caviezel:2005th}) one easily finds further orbifolds that fit into this class. A full classification should be straightforward.

An example where this construction does not work is the $\mathbb{Z}_{12}$-orbifold of the Fermat Landau-Ginzburg orbifold with weights $w=(1,1,3,3,4)$. The Calabi-Yau has Hodge numbers $(h^{1,1},h^{2,1})=(5,168)$. The broad sectors correspond to the twisted sectors $k=4,8$, whose space of RR ground states is three-dimensional. An attempt to resolve this into narrow sectors by a $\mathbb{Z}_4$-orbifold fails, as any such operation changes the Hodge numbers. This is consistent with the fact that the toric geometry is not an elliptic fibration. 


%% file: section7.tex
\section{Outlook}
In this work we have proposed a formula for the exact central charge of a B-type D-brane that we conjecture to hold in all regions of the K\"ahler moduli space of a Calabi-Yau. For Landau-Ginzburg orbifolds, we proposed explicit expressions for the mathematical objects that enter the central charge formula. Our results are in agreement with the GLSM and FJRW theory. There are various directions for further research.

The obvious direction is to consider regions in the moduli space that are neither geometric, nor Landau-Ginzburg orbifolds. The majority of phases of GLSMs with abelian and non-abelian gauge groups is of this more general type. The methods we have applied in this article, in particular FJRW theory, should generalize to such situations.

Another challenging direction of research concerns broad sectors. Our proposal for the central charge a priori only applies to narrow sectors. In multi-parameter models broad sectors will typically contribute to the central charge. In an example, we have worked around this issue by modifying the orbifold so that the broad sectors turn into narrow ones, but this approach will not always work. In order to honestly take into account broad sectors one should include them into the general formalism.

Furthermore it would be desirable to get a better physics understanding of the $I$-function and the Gamma class. From a CFT perspective it is not obvious why the Gamma class plays a central role. It would be interesting to see the Gamma class arise from a CFT argument. A similar issue concerns the $I$-function, which encodes all the quantum corrections in the central charge formula. While one can give a mathematical definition and have we proposed an explicit expression that works for a large class of examples, it is not clear from a physics point of view why this particular object is of such central importance.  

One interesting issue concerns the differential equations satisfied by the $I$-function. It can be shown that it satisfies the GKZ system. However, special geometry implies that it also satisfies the Picard-Fuchs differential equations. The solutions spaces differ by those solutions that are projected out by the action of $G$.  One should therefore be able to express the group action explicitly in terms of additional differential operators. These operators are expected to constrain the GKZ system to the Picard-Fuchs system. 

Another gap that should be filled is to give a detailed version of the path integral derivation of the correlation functions in the A-twisted Landau-Ginzburg orbifold coupled to topological gravity that we only sketched in Section~\ref{sec:brief-guide-fjrw}, resulting in the FJRW virtual class. 
Further aspects that we have not covered are the generalization to non-abelian Landau-Ginzburg orbifolds and to $\hat c \not =3$.


%% file: appendix.tex
\appendix
\label{sec:appendix}
\section{Alternative derivation of $q^{\mathrm{ext}}$}
\label{app-qq}
In this appendix we recall the mirror map for Landau-Ginzburg
orbifolds following~\cite{Kreuzer:1994np,Krawitz:2010ab} 
which suggests why our definition of $q^{\mathrm{ext}}$ could have a direct interpretation in terms of the $(a,c)$-ring elements. 

Consider an element in $\mathcal{H}^{(a,c)}$ of $(W,G,\bar{\rho}_{m},\mathbb{C}_L^{\ast})$, which we can represent as
\begin{equation}
\prod_{j\in I^{\gamma J^{-1}}}
\phi_{j}^{l_{j}}|0\rangle^{(a,c)}_{\gamma}, \qquad l_j \in \mZ_{\geq
  0}, j=1,\dots N\, .
\end{equation}
We choose a set of generators $\bar{g}_1,\dots,\bar{g}_N$ of
$\Aut(W^T)$ (as in~\eqref{phiaut} with $M$ replaced by $M^T$) and
define the element $\bar{\gamma} \in \Aut(W^{T})$ by
\begin{equation}
\bar{\gamma}:=\prod_{j\in I^{\gamma J^{-1}}}\bar{g}_{j}^{l_{j}+1}.
\end{equation}
By the isomorphism $\Aut(W^T) \cong \Aut(W)$, this element gets mapped
to $\gamma J^{-1} \in
\Aut(W)$. By~\eqref{phiaut} this defines exponents $v_1,\dots, v_N$ via
\begin{equation}
  \label{eq:133}
  \gamma J^{-1}=\prod_{\alpha\in I^{\bar{\gamma}}}g_{\alpha}^{v_{\alpha}+1}.
\end{equation}
Then the mirror map is
\begin{equation}
  \label{mmap}
\prod_{i\in I^{\gamma
    J^{-1}}}\phi_{j}^{l_{j}}|0\rangle^{(a,c)}_{\gamma}\longleftrightarrow
\prod_{\alpha\in I^{\bar{\gamma}}} y_{\alpha}^{v_\alpha}|0\rangle^{(c,c)}_{\bar{\gamma}}.
\end{equation}
A few remarks are in order. This mirror map is shown to be one-to-one
in the case of the so called atomic invertible polynomials
\cite{Kreuzer:1992bi,Krawitz:2010ab} of chain and Fermat type and
there is some ambiguity for the case of loop type. Once this ambiguity
is fixed, the mirror map gives an isomorphism between
$\mathcal{H}^{(a,c)}$ of $(W,G)$ and $\mathcal{H}^{(c,c)}$ of
$(W^{T},G^{T})$, where $G^{T} \equiv G\spcheck$ in~\eqref{eq:57}. 
Then, the map is extended to an isomorphism when $W$ is a sum of atomic invertible polynomials, just by taking the tensor product of the individual maps \cite{Krawitz:2010ab}. More important for us is that this map is an isomorphism even without projecting to gauge invariant states \cite{Kreuzer:1994np,Krawitz:2010ab}, namely it provides an isomorphism between the spaces $H^{(a,c)}$ of $(W,G)$ and $H^{(c,c)}$ of $(W^{T},G^{T})$.

Now we come to the matrix $q$ of Section~\ref{sec:deformations}. When a state belongs to a narrow sector
of $H^{(a,c)}$ it is clear that (\ref{mmap}) maps it to a state in the
untwisted sector $H^{(c,c)}_{e}$ of the form
$\prod_{\alpha=1}^{N}y_{\alpha}^{v_{\alpha}}|0\rangle_{e}^{(c,c)}$.
In particular, the marginal deformations $\cO_{\gamma_a}\in
\cH^{(a,c)}_{(-1,1),\gamma_a}$ (cf.~\eqref{eq:118}
and~\eqref{qmardef}) are mapped to states in
$\cH^{(c,c)}_{e,(1,1)}$ given by a vector $v_a\in\mZ^N$. The condition
that the R-charges (with respect to $J\spcheck\in G\spcheck$) are $1$ is
equivalent to the condition $1=J\spcheck \cdot v_a$, i.e. we can
identify the state with a monomial deformation
$\prod_{\alpha=1}^{N}y_{\alpha}^{v_{\alpha}} $ of
$W^{T}$. Furthermore, since also $g\spcheck \cdot v=0$ for all
$g\spcheck \in G\spcheck$, we find that $v_a \in \cA_{\ext}$. In fact,
one can argue that $v_a \in \cA_{\mathrm{LG}}$ given
in~\eqref{eq:135}. We can repeat the procedure to obtain the matrix
$q$ in~\eqref{eq:120} by replacing the set $\cA_{\mathrm{ext}}$ by
  $\cA_{\mathrm{LG}}$. The resulting $h\times(h+N)$ matrix then agrees
  with the matrix $q^{\mathrm{LG}}$ defined in~(\ref{qmardef}).

Since we will also make a connection to the gauged linear sigma model
and geometry, we have found that it is useful to define an extended
matrix $q^{\ext}$ that also captures deformations which may seem
unnatural from the Landau-Ginzburg point of view. A working hypothesis
is that one should at least include unprojected sectors $H^{(a,c)}_\gamma$ satisfying
\begin{equation}
  F_{R}(\mathcal{O}_{\gamma,\mu})-F_{L}(\mathcal{O}_{\gamma,\mu}) \in \{1,2\}.
\end{equation}
This includes the marginal deformations, both narrow and broad, but
can also include further sectors that may be empty after
projection. These correspond to trivial monomial deformations of
$W^T$. The broad marginal deformations cannot be identified with monomial deformations of $W^{T}$. Ignoring the latter, we can define
an extended matrix $q^{\mathrm{ext}}\in\mathrm{Mat}_{\hat{h}\times (\hat{h}+N)}(\mQ)$ by
\begin{equation}
  \label{eq:132}
  q^{\ext}_{\hat{a},\hat{b}}=\delta_{\hat{a},\hat{b}}, \qquad
  q^{\ext}_{\hat{a},\hat{h}+j}=-\theta^{\gamma^{-1}_{\hat{a}}}_{j}\qquad
  \text{ \ for \ }
  \left\{\begin{array}{c}\hat{a},\hat{b}=1,\ldots,\hat{h} \geq h\\ j=1,\ldots,N. \end{array} \right. 
\end{equation}
We claim that the matrix $q^{\ext}$ defined in~\eqref{eq:119} agrees
with the matrix defined in~\eqref{eq:120}.

\section{Details on examples}
\subsection{Quintic}
\label{app-quintic}
The Landau-Ginzburg periods of the quintic can be found in \cite{Candelas:1990rm}. There are two bases of periods. These can be obtained, for instance, by solving the Picard-Fuchs equation of the mirror quintic characterized by
\begin{equation}
  \phi_1^5+\phi_2^5+\phi_3^5+\phi_4^5+\phi_5^5-5\psi \phi_1\phi_2\phi_3\phi_4\phi_5.
\end{equation}
The Gepner point is at $\psi=0$. This is related to the large complex structure  coordinate $z$ via $z=-(5\psi)^{-5}$. Comparing with the $I$-function we have $u=-5\psi$.

One basis of periods is given by
\begin{equation}
\label{d5-basis1}
\varpi_j=-\frac{1}{5}\sum_{m=1}^{\infty}\omega^{2m}\frac{\Gamma\left(\frac{m}{5}\right)(5\psi)^m}{\Gamma(m)\Gamma\left(1-\frac{m}{5}\right)^4}\omega^{jm}\qquad j=0,\ldots 4,
\end{equation}
with $\omega\equiv J=e^{\frac{2\pi i}{5}}$. Under monodromy around the Gepner point at $\psi=0$ the periods transform as $\varpi_j\rightarrow\varpi_{j+1}$, modulo the relation $\sum_{j=0}^{4}\varpi_j=0$. 

There is a second basis given by
\begin{equation}
\label{d5-basis2}
\hat{\varpi}_k=\sum_{n=0}^{\infty}\frac{\Gamma\left(n+\frac{k}{5}\right)^5}{\Gamma(5n+k)}(5\psi)^{5n+k}\qquad k=1,\ldots,4. 
\end{equation}
The two bases are related via
\begin{equation}
\label{d5-basrel}
\varpi_j=-\frac{1}{5}\frac{1}{(2\pi i)^4}\sum_{k=1}^4\omega^{jk}(-1+\omega^k)^4\hat{\varpi}_k.
\end{equation}
\subsection{Two-parameter example 1}
\label{app-d8}
Two bases of LG periods of the two-parameter degree $8$ example have been discussed in \cite{Candelas:1993dm}. The periods can be obtained by solving the Picard-Fuchs equation of the mirror hypersurface characterized by the equation
\begin{equation}
  \label{d8-wlg}
  \phi_1^8+\phi_2^8+\phi_3^4+\phi_4^4+\phi_5^4-8\psi \phi_1\phi_2\phi_3\phi_4\phi_5-2\phi \phi_1^4\phi_2^4.
\end{equation}
The Gepner point is at $(\psi,\phi)=0$. The relation to coordinates $(z_1,z_2)$ at the large complex structure point is given by 
\begin{equation}
  z_1z_2^{\frac{1}{2}}=-(8\psi)^{-4}\qquad z_2=(2\phi)^{-2}.
  \end{equation}
The Picard-Fuchs operators at the Landau-Ginzburg point are
\begin{eqnarray}
  \label{twoparpf}
  \mathcal{L}_1&=&32\psi^2\theta_{\psi}^2\theta_{\phi}-\phi(\theta_{\psi}-1)(\theta_{\psi}-2)(\theta_{\psi}-3)\nonumber\\
  \mathcal{L}_2&=&16\theta_{\phi}(\theta_{\phi}-1)-\phi^2(4\theta_{\phi}+\theta_{\psi})^2.
\end{eqnarray}
One basis of periods is given by
\begin{equation}
  \label{d8-basis1}
  \varpi_j(\psi,\phi)=-\frac{1}{4}\sum_{m=1}^{\infty}\frac{(-1)^m\alpha^{mj}\Gamma\left(\frac{m}{4}\right)}{\Gamma(m)\Gamma\left(1-\frac{m}{4}\right)^3}(2^{12}\psi^4)^{\frac{m}{4}}u_{-\frac{m}{4}}((-1)^j\phi),
  \end{equation}
with $\alpha=e^{\frac{2\pi i}{8}}$ and
\begin{equation}
  u_{\nu}(\phi)=(2\phi)^{\nu} {}_2F_1\left(-\frac{\nu}{2},-\frac{\nu}{2}+\frac{1}{2};1;\frac{1}{\phi^2} \right).
\end{equation}
Since the Landau-Ginzburg point is at $\phi=0$ we have to analytically continue ${}_2F_1$ to $\phi=0$. This gives a sum of two terms
\begin{equation}
  u_{\nu}(\pm\phi)=\frac{1}{4\pi i}\frac{1-e^{i\pi\nu}}{\Gamma(-\nu)}\sum_{m=0}^{\infty}\frac{\Gamma\left(-\frac{\nu}{2}+m\right)^2}{\Gamma(2m+1)}(2\phi)^{2m}\mp \frac{1}{4\pi i}\frac{1+e^{i\pi\nu}}{\Gamma(-\nu)}\sum_{m=0}^{\infty}\frac{\Gamma\left(-\frac{\nu}{2}+\frac{1}{2}+m\right)^2}{\Gamma(2m+2)}(2\phi)^{2m+1}.
  \end{equation}
There is a second basis given by
\begin{eqnarray}
  \label{d8-basis2}
  \xi_r(\psi,\phi)&=&\sum_{n=0}^{\infty}\frac{\Gamma\left(n+\frac{r}{4}\right)^4}{\Gamma(4n+r)}(2^{12}\psi^4)^{n+\frac{r}{4}}(-1)^nu_{-\left(n+\frac{r}{4}\right)}(\phi)\nonumber \\
  \eta_r(\psi,\phi)&=&\sum_{n=0}^{\infty}\frac{\Gamma\left(n+\frac{r}{4}\right)^4}{\Gamma(4n+r)}(2^{12}\psi^4)^{n+\frac{r}{4}}u_{-\left(n+\frac{r}{4}\right)}(-\phi).
\end{eqnarray}
Upon evaluating the $I$-function one gets four contributions, depending on whether some combinations of the summation variables are even or odd. Let is give some intermediate steps of the calculation. The expression we want to rewrite is
\begin{equation}
  I_{LG}(u)=-
  \sum_{m,n,r,s}\frac{(-1)^{G(k,q)}u_1^{4n+r-1}u_2^{2m+s}}{\Gamma(1+2m+s)\Gamma(4n+r)}\frac{\Gamma\left(\left\langle-n-\frac{r}{4}\right\rangle\right)^3\Gamma\left(\left\langle-\frac{n}{2}-\frac{r}{8}-m-\frac{s}{2}\right\rangle\right)^2}{\Gamma\left(1-\left(n+\frac{r}{4}\right)\right)^3\Gamma\left(1-\left(\frac{n}{2}+\frac{r}{8}+m+\frac{s}{2}\right)\right)^2}\phi_{4n+r,2m+s}
\end{equation}
Applying the reflection formula to the denominator of the second quotient produces a term
\begin{equation}
  \frac{(-1)^n}{\pi^5}\sin^3\frac{r}{4}\sin^2\left(\frac{r}{8}+\frac{n+s}{2}\right)\Gamma\left(n+\frac{r}{4}\right)^3\Gamma\left(m+\frac{r}{8}+\frac{n+s}{2}\right)^2
\end{equation}
Next, we apply the reflection formula to the $\Gamma(\langle\cdot\rangle)$ which simplifies to $\Gamma\left(\left\langle-\frac{r}{4}\right\rangle\right)^3\Gamma\left(\left\langle-\frac{r}{8}-\frac{n+s}{2} \right\rangle\right)^2$. Here we have to distinguish between even and odd $n+s$. With $a\in\mathbb{Z}$ and using 
$\Gamma\left(\langle -\frac{k}{d}\rangle\right) = 1-  \frac{k}{d}$ if $d\in\mZ_{>0}$ and $k\in \{1,\dots,d-1\}$, we get
\begin{itemize}
\item $n+s=2a$
  \begin{equation}
    \frac{(-1)^n}{\pi^5}\sin^3\left(\frac{r}{4}\right)\sin^2\left(\frac{r}{8}+a\right)\Gamma\left(\left\langle-\frac{r}{4}\right\rangle\right)^3\Gamma\left(\left\langle-\frac{r}{8}-a \right\rangle\right)^2=\frac{(-1)^s}{\Gamma\left(\frac{r}{4}\right)^3\Gamma\left(\frac{r}{8}\right)^2}
    \end{equation}
\item $n+s=1+2a$
  \begin{equation}
    \frac{(-1)^n}{\pi^5}\sin^3\left(\frac{r}{4}\right)\sin^2\left(\frac{r}{8}+\frac{2a+1}{2}\right)\Gamma\left(\left\langle-\frac{r}{4}\right\rangle\right)^3\Gamma\left(\left\langle-\frac{r}{8}-a-\frac{1}{2}\right\rangle\right)^2=-\frac{(-1)^s}{\Gamma\left(\frac{r}{4}\right)^3\Gamma\left(\frac{r}{8}+\frac{1}{2}\right)^2}
    \end{equation}
  \end{itemize}
Combining all the expressions, we arrive at the result in the main text.
\subsection{Four-parameter example}
\label{app-4par}
\subsubsection{Differential operators}
The GKZ differential operators at the Landau-Ginzburg point are 
\begin{equation}
  \label{eq:114}
  \begin{aligned}
    \cL_1 &= 9\, {u_{{1}}}^{3} \theta_{{2}}\theta_{{3}}\theta_{{4}} \left( \theta_{{3}}-1 \right) 
    \left( \theta_{{2}}-1 \right) \\
    &\phantom{=} +  {u_{{2}}}^{2
}{u_{{3}}}^{2}u_{{4}}\left( \theta_{{1}}-1
 \right)  \left( \theta_{{1}}-2 \right)  \left( \theta_{{1}}-3
 \right)  \left( \theta_{{2}}+\theta_{{1}}+2\,\theta_{{3}} \right) 
 \left( 2\,\theta_{{2}}+\theta_{{1}}+\theta_{{3}} \right) \\
   \cL_2 &= 3\,  u_{{3}}\theta_{{2}} \left( \theta_{{2}}-1 \right)+{u_{{2}}}^{2}\theta_{{3}}
   \left( 2\,\theta_{{2}}+\theta_{{1}}+\theta_{{3}} \right) \\
   \cL_3 &= 3\, u_{{2}}\theta_{{3}} \left( \theta_{{3}}-1 \right) + {u_{{3}}}^{2}\theta_{{2}}
     \left( \theta_{{2}}+\theta_{{1}}+2\,\theta_{{3}} \right) \\
   \cL_4 &= 729\,\theta_{{4}} \left( \theta_{{4}}-1 \right)  \left( \theta_{{4}}-
2 \right) + {u_{{4}}}^{3} \left( \theta_{{1}}+3\,\theta_{{4}} \right) ^{3}
  \end{aligned}
\end{equation}
where $\theta_i=u_i\frac{\partial}{\partial u_i}$ and the $u_i$ being the coordinates at the Landau-Ginzburg point. It can be shown that the $I$-function satisfy the GKZ equations.
\subsubsection{FJRW invariants}
The twisted sectors $\mathcal{H}_{\gamma}$ corresponding to the marginal deformations are given by $\gamma\in G^{(2)}$ as in (\ref{4parg2}). Hence, there are two types of invariants
\begin{align}
  \mathrm{FJRW}_{n_1,n_2,n_3,n_4}&:=\left\langle \left(e_{J^2}\right)^{n_1}\left(e_{J^3g}\right)^{n_2}\left(e_{J^3g^2}\right)^{n_3}\left(e_{J^4}\right)^{n_4} \right\rangle_{0,n_1+n_2+n_3+n_4}\nonumber\\
  \mathrm{FJRW}^0_{n_1,n_2,n_3,n_4}&:=\left\langle \left(e_{J^2}\right)^{n_1}\left(e_{J^3g}\right)^{n_2}\left(e_{J^3g^2}\right)^{n_3}\left(e_{J^4}\right)^{n_4} \tau_1\left(e_J\right)\right\rangle_{0,n_1+n_2+n_3+n_4+1}.
\end{align}
In Table \ref{tab-fjrw4par} we give the first few non-zero invariants organized in terms of $|n|=\sum_in_i$.
\begin{table}
  \def\arraystretch{1.3}
  \begin{center}
  $\begin{array}{c|c|cc}
    |n|&(n_1,n_2,n_3,n_4)&\mathrm{FJRW}_{n_1,n_2,n_3,n_4}&\mathrm{FJRW}^0_{n_1,n_2,n_3,n_4}\\
    \hline
    3&(0,1,1,1)&\frac{1}{9}&\frac{1}{9}\\
    &(1,0,0,2)&\frac{1}{9}&\frac{1}{9}\\
    \hline
    4&(1,0,3,0)&\frac {2}{27}&{\frac {2}{27}
}\\ &(1,3,0,0)&\frac {2}{27}&{\frac {2}{27}}
\\ \hline
    5&(3,1,1,0)&{\frac {1}{81}}&\frac {2}{27}
\\ \hline
     6&(0,1,1,4)&{\frac {16}{2187}}&{\frac {64}{2187}}
\\ &(1,0,0,5)&{\frac {11}{2187}}&{\frac {44}{2187}}
\\ \hline
     7&(0,1,4,2)&{\frac {76}{19683}}&{\frac {380}{
19683}}\\ 
     &(0,4,1,2)&{\frac {76}{19683}}&{\frac {
380}{19683}}\\ 
     &(1,0,3,3)&{\frac {25}{6561}}&{
\frac {125}{6561}}\\ 
     &(1,3,0,3)&{\frac {25}{6561}}&
{\frac {125}{6561}}\\ 
     &(7,0,0,0)&{\frac {2}{243}}&{
\frac {10}{243}}\\ 
   \hline
   8&(0,1,7,0)&{\frac {368}{59049}}&
{\frac {736}{19683}}\\ 
    &(0,4,4,0)&{\frac {128}{
59049}}&{\frac {256}{19683}}\\ 
    &(0,7,1,0)&{\frac {
368}{59049}}&{\frac {736}{19683}}\\ 
    &(1,0,6,1)&{
\frac {346}{59049}}&{\frac {692}{19683}}\\ 
    &(1,3,3,
1)&{\frac {121}{59049}}&{\frac {242}{19683}}\\ 
   &(1,
6,0,1)&{\frac {346}{59049}}&{\frac {692}{19683}}\\ 
   &(2,2,2,2)&{\frac {130}{59049}}&{\frac {260}{19683}}
\\ &(3,1,1,3)&{\frac {29}{19683}}&{\frac {58}{6561}
}\\ 
   \hline
   9&(0,1,1,7)&{\frac {280}{19683}}&{\frac {1960}{
19683}}\\ 
   &(1,0,0,8)&{\frac {4624}{531441}}&{\frac 
{32368}{531441}}\\ 
   &(2,2,5,0)&{\frac {584}{177147}}
&{\frac {4088}{177147}}\\ 
   &(2,5,2,0)&{\frac {584}{
177147}}&{\frac {4088}{177147}}\\ 
   &(3,1,4,1)&{
\frac {58}{19683}}&{\frac {406}{19683}}\\ 
   &(3,4,1,1
)&{\frac {58}{19683}}&{\frac {406}{19683}}\\ 
  &(4,0,
3,2)&{\frac {4}{2187}}&{\frac {28}{2187}}\\ 
  &(4,3,0
,2)&{\frac {4}{2187}}&{\frac {28}{2187}}\\ 
    \end{array}
    $\caption{The invariants $\mathrm{FJRW}_{n_1,n_2,n_3,n_4}$ and $\mathrm{FJRW}^0_{n_1,n_2,n_3,n_4}$}\label{tab-fjrw4par}
    \end{center}
  \end{table}


%% file: PALP.tex
\section{PALP and Landau--Ginzburg orbifolds}
\label{sec:palp-landau-ginzburg}

The program {\tt poly.x} of software package PALP
\cite{Kreuzer:2002uu} is capable of analyzing Calabi--Yau
Landau-Ginzburg orbifolds. It implements the results of
\cite{Vafa:1989xc,Intriligator:1990ua}. The option {\tt poly.x -L}
provides the information on how the twisted sectors contribute to the
Hodge numbers. Since this option has not been discussed in detail in
the PALP manual~\cite{Braun:2012vh}, we provide a detailed explanation here.

In general, there is no need to specify $W$ only the group $G$ needs to be
entered. If $G=\langle J \rangle$ we simply enter the numbers $d, w_1, \dots,
w_N$ where $q_j=\frac{w_j}{d}$, $j=1,\dots,N$. For illustration,
consider the example from Section~\ref{sec:4-parameter-example}: $W = x_1^9 + x_2^9 + x_3^9 + x_4^3 +
x_5^3$ with $d=9$ and $w=(1,1,1,3,3)$ and $G=\langle J\rangle$.\\
{\tt
  ./poly.x -L \\
type degree and weights  [d  w1 w2 ...]: 9 1 1 1 3 3 \\
sec[0] th= 0  0  0  0  0  QL= 0/9 dQ= 0  q00+=1 q11+=112 q22+=112 q33+=1\\
sec[1] th= 1  1  1  3  3  QL= 0/9 dQ= 3  q03+=1\\
sec[2] th= 2  2  2  6  6  QL= 9/9 dQ= 1  q12+=1\\
sec[3] th= 3  3  3  0  0  QL= 6/9 dQ= 1  q12+=2\\
sec[4] th= 4  4  4  3  3  QL= 9/9 dQ= 1  q12+=1\\
sec[5] th= 5  5  5  6  6  QL=18/9 dQ=-1  q21+=1\\
sec[6] th= 6  6  6  0  0  QL=15/9 dQ=-1  q21+=2\\
sec[7] th= 7  7  7  3  3  QL=18/9 dQ=-1  q21+=1\\
sec[8] th= 8  8  8  6  6  QL=27/9 dQ=-3  q30+=1\\
WittenIndex=-216, Trace=236\\
9 1 1 1 3 3 M:145 5 N:7 5 V:4,112 [-216]\\
}
Here {\tt sec[i]} corresponds to $\cH^{(c,c)}_\gamma$ with $\gamma = J^i$,
{\tt th= i1 i2 ... iN} corresponds to $\theta^\gamma = (
\frac{i_1}{d} , \dots, \frac{i_N}{d})$. 

The value of {\tt QL} corresponds to
$q_+$, the value of {\tt
  dQ} corresponds to $d_\gamma -
2\age(\gamma)$ with the notation as in
Section~\ref{sec:some-details}. Finally, the pair $(i,j)$ in {\tt qij}
corresponds to $(i,\widehat c-j)$ in the sector
$\cH^{(c,c)\, i,j}_\gamma \cong \cH_{\FJRW,\gamma}^{i,\hat c-j}$, and
the value of {\tt qij+=} corresponds to $\dim \cH_\gamma^{(c,c)\,i,j}$. Only
the sectors with {\tt qij > 0} are displayed.

Note that it is easy to spot the broad sectors by looking
for {\tt 0}'s among ${\tt th}$. In this example, there are three broad
sectors, {\tt sec[0]}, {\tt sec[3]}, {\tt sec[6]}. The untwisted
sector has an odd number of zero phases, hence it is odd. While the
$J^3$-- and $J^6$--twisted sectors have an even number of zero phases
and therefore contribute to the even part of $\cH_{\FJRW}$.

The penultimate line gives the Witten index and the sum of all Hodge
numbers, as computed by the Poincar\'e polynomial of the chiral
ring. For comparison, the last line lists the numbers of points and
vertices of the corresponding $M$-- and $N$--lattice polytopes. This
is explained in great detail in~\cite{Braun:2012vh}.

For a bigger group, one needs to add to
{\tt d w1 w2 ... wN} the further generators $g$ in the form {\tt /Zn: k1
  k2 ... kN}. Here, {\tt n} is the order of the
generator, and {\tt k1 k2 ... kN} are related to the phases of $g$ by
$\theta_j^g = \frac{k_j}{n}$. In the example above, we
consider now the group $G = \langle J, g \rangle$ where the generator $g$
acts on $\mC^5$ by $\diag(1,1,1,\zeta_3,\zeta_3^2)$, $\zeta_3^3 = 1$.\\ 
{\tt
./poly.x -L\\
type degree and weights  [d  w1 w2 ...]: 9 1 1 1 3 3 /Z3: 0 0 0 1 2\\
sec[0:0] th= 0  0  0  0  0  QL= 0/9 dQ= 0  q00+=1 q11+=56 q22+=56 q33+=1\\
sec[0:1] th= 0  0  0  3  6  QL= 3/9 dQ= 0  q11+=28 q22+=28\\
sec[0:2] th= 0  0  0  6  3  QL= 3/9 dQ= 0  q11+=28 q22+=28\\
sec[1:0] th= 1  1  1  3  3  QL= 0/9 dQ= 3  q03+=1\\
sec[2:0] th= 2  2  2  6  6  QL= 9/9 dQ= 1  q12+=1\\
sec[3:1] th= 3  3  3  3  6  QL= 9/9 dQ= 1  q12+=1\\
sec[3:2] th= 3  3  3  6  3  QL= 9/9 dQ= 1  q12+=1\\
sec[4:0] th= 4  4  4  3  3  QL= 9/9 dQ= 1  q12+=1\\
sec[5:0] th= 5  5  5  6  6  QL=18/9 dQ=-1  q21+=1\\
sec[6:1] th= 6  6  6  3  6  QL=18/9 dQ=-1  q21+=1\\
sec[6:2] th= 6  6  6  6  3  QL=18/9 dQ=-1  q21+=1\\
sec[7:0] th= 7  7  7  3  3  QL=18/9 dQ=-1  q21+=1\\
sec[8:0] th= 8  8  8  6  6  QL=27/9 dQ=-3  q30+=1\\
WittenIndex=-216, Trace=236\\
9 1 1 1 3 3 /Z3: 0 0 0 1 2 M:67 5 N:13 5 V:4,112 [-216]
}

In this case, {\tt sec[i;j]} corresponds to $\cH^{(c,c)}_\gamma$ with
$\gamma = J^ig^j$. The remaining quantities have the same meaning as
above.

Note that we can also determine $\tS\tL(N,\mC) \cap \Aut^{\diag}(W)$ as follows:\\
{\tt
  ./poly.x -fv | ./cws.x -N\\
Degrees and weights  `d1 w11 w12 ... d2 w21 w22 ...'
  or `\#lines \#columns' \\ (= `PolyDim \#Points' or `\#Points PolyDim'):\\
9 1 1 1 3 3 \\
Type the 20 coordinates as dim=4 lines with \#pts=5 columns: \\
9 1 1 1 3 3 /Z9: 4 8 0 0 6 /Z3: 2 0 0 0 1 /Z3: 1 0 0 2 0\\
}
Hence, we read off that $\tS\tL(N,\mC) \cap \Aut^{\diag}(W) \cong \mu_9 \times \mu_9 \times \mu_3
\times \mu_3$. This works, however, only for Fermat polynomials
$W$. More generally, we can enter the exponent matrix $M$ explicitly as
follows: We remove from $M$ the column (or row) corresponding to the highest
weight. Then we shift the entries by $-1$. Consider the example
$W=x_1^7 + x_2^7 + x_3^7 + x_2x_4^3+ x_3x_5^3$.\\
{\tt
./poly.x -fv | ./cws.x -N\\
Degrees and weights  `d1 w11 w12 ... d2 w21 w22 ...'
  or `\#lines \#columns' \\ (= `PolyDim \#Points' or `\#Points PolyDim'):\\
4 5\\
 6 -1 -1 -1 -1\\
-1  6 -1 -1 -1\\
-1 -1  6 -1 -1\\
-1  0 -1  2 -1\\
Type the 20 coordinates as dim=4 lines with \#pts=5 columns:\\
7 1 1 1 2 2 /Z21: 9 9 0 4 20 /Z7: 5 3 0 6 0 \\
}
